\documentclass[12pt]{article}
\usepackage{graphicx}
\usepackage{amsmath,amsthm,epsfig,latexsym,color}
\usepackage[numbers,square,sort&compress]{natbib}

\def\hybrid{\topmargin 0pt      \oddsidemargin 0pt
        \headheight 0pt \headsep 0pt
       \voffset-1cm
        \textwidth 6.25in       
       \textheight 9.5in       
        \marginparwidth 0.0in
        \parskip 5pt plus 1pt   \jot = 1.5ex}
\catcode`\@=11
\def\marginnote#1{}

\newcount\hour
\newcount\minute
\newtoks\amorpm
\hour=\time\divide\hour by60
\minute=\time{\multiply\hour by60 \global\advance\minute by-\hour}
\edef\standardtime{{\ifnum\hour<12 \global\amorpm={am}%
        \else\global\amorpm={pm}\advance\hour by-12 \fi
        \ifnum\hour=0 \hour=12 \fi
        \number\hour:\ifnum\minute<10 0\fi\number\minute\the\amorpm}}
\edef\militarytime{\number\hour:\ifnum\minute<10 0\fi\number\minute}

\def\draftlabel#1{{\@bsphack\if@filesw {\let\thepage\relax
   \xdef\@gtempa{\write\@auxout{\string
      \newlabel{#1}{{\@currentlabel}{\thepage}}}}}\@gtempa
   \if@nobreak \ifvmode\nobreak\fi\fi\fi\@esphack}
        \gdef\@eqnlabel{#1}}
\def\@eqnlabel{}
\def\@vacuum{}
\def\draftmarginnote#1{\marginpar{\raggedright\scriptsize\tt#1}}

\def\draftlabel#1{{\@bsphack\if@filesw {\let\thepage\relax
   \xdef\@gtempa{\write\@auxout{\string
      \newlabel{#1}{{\@currentlabel}{\thepage}}}}}\@gtempa
   \if@nobreak \ifvmode\nobreak\fi\fi\fi\@esphack}
        \gdef\@eqnlabel{#1}}
\def\@eqnlabel{}
\def\@vacuum{}
\def\draftmarginnote#1{\marginpar{\raggedright\scriptsize\tt#1}}

\def\draft{\oddsidemargin -.5truein
        \def\@oddfoot{\sl preliminary draft \hfil
        \rm\thepage\hfil\sl\today\quad\militarytime}
        \let\@evenfoot\@oddfoot \overfullrule 3pt
        \let\label=\draftlabel
        \let\marginnote=\draftmarginnote
   \def\@eqnnum{(\theequation)\rlap{\kern\marginparsep\tt\@eqnlabel}%
\global\let\@eqnlabel\@vacuum}  }

\def\numberbysection{\@addtoreset{equation}{section}
        \def\theequation{\thesection.\arabic{equation}}}

\def\underline#1{\relax\ifmmode\@@underline#1\else
        $\@@underline{\hbox{#1}}$\relax\fi}

\def\titlepage{\@restonecolfalse\if@twocolumn\@restonecoltrue\onecolumn
     \else \newpage \fi \thispagestyle{empty}\c@page\z@
        \def\thefootnote{\fnsymbol{footnote}} }

\def\endtitlepage{\if@restonecol\twocolumn \else  \fi
        \def\thefootnote{\arabic{footnote}}
        \setcounter{footnote}{0}}  
\relax

\numberbysection
\hybrid

\newfont{\Bbb}{msbm10 scaled 1\@ptsize00}
\newfont{\Bbbb}{msbm7 scaled 1\@ptsize00}
\newcommand{\CC}{\mbox{\Bbb C}}

\newcommand{\DDD}{\raise-1pt\hbox{$\mbox{\Bbbb D}$}}

\newcommand{\UUU}{\raise-1pt\hbox{$\mbox{\Bbbb U}$}}

\newcommand{\ZZ}{\mbox{\Bbb Z}}
\newcommand{\z}{\raise-1pt\hbox{$\mbox{\Bbbb Z}$}}

\def\res{\mathop{\hbox{res}}\limits}

\def\beq{\begin{equation}}
\def\eeq{\end{equation}}
\def\p{\partial}

\newcommand{\bu}{\bar u}
\newcommand{\bv}{\bar v}
\newcommand{\bw}{\bar w}

\begin{document}








\begin{center}
\

\vspace{10mm}
 {\LARGE{Scalar products of Bethe vectors in the 8-vertex model}}
\\
\vspace{20mm} {\large  {N. Slavnov${}^*$}
 \ \ \ \ \ \  {A. Zabrodin${}^*$}
 \ \ \ \ \ \  {A. Zotov${}^*$\footnote{{\rm E-mails:}{\rm\ nslavnov@mi-ras.ru,\ zabrodin@itep.ru,\ zotov@mi-ras.ru}}
}
  }
 \vspace{12mm}

{\small{\rm
 ${}^*$Steklov Mathematical Institute of Russian Academy of Sciences,\\ Gubkina str. 8, Moscow,
119991,  Russia}}

\vspace{4mm}


\vspace{0mm}

\begin{abstract}

We obtain a determinant representation of normalized scalar products of
on-shell and  off-shell Bethe vectors in the inhomogeneous 8-vertex model.
We consider the case of rational anisotropy parameter and use the generalized
algebraic Bethe ansatz approach. Our method is to
obtain a system of linear equations for the scalar products, prove its
solvability and solve it in terms of determinants of explicitly known matrices.

\end{abstract}

 \end{center}

\bigskip
\bigskip

\tableofcontents

\section{Introduction}

The study of low-dimensional strongly correlated systems is of great importance and  interest.
Among the numerous physical low-dimensional models, a special role is played by the 8-vertex model \cite{Sut70,FanW70,Baxter71,Baxter-book}, which is closely related or in some sense
equivalent to the completely anisotropic $XYZ$ Heisenberg magnet \cite{Hei28}.
This model is completely integrable, because the corresponding
matrix of Boltzmann weights  (the $R$-matrix)
satisfies the Yang--Baxter equation, and transfer matrices
commute for any values of the arguments. However, unlike the 6-vertex model, in the
8-vertex model, the total flow through the vertex, generally speaking, is not conserved.
In the language of the $XYZ$ chain, this leads to the fact that the third component of
the total spin is no longer an integral of motion. All this implies significant
difficulties in the study of this model.

The spectral problem for the 8-vertex model was solved by R.~Baxter via the $Q$-operator method \cite{Baxter71a,Baxter72,Baxter73}, which subsequently found wide applications.
Further research on the 8-vertex model took place in several directions.
Besides studying the properties of the $T$--$Q$ equation and the solutions of Bethe equations
\cite{KluZ88,KluZ89,T95,FMC03,FabM05,FabM07,FMC06,Fab07,BazM07,T15},
it is also worth mentioning the study of the underlying quantum algebras \cite{Skl82,FodIJKMY94,Fro97,JimKOS99,BufRT12}. A number of
works were devoted to the study of correlation functions of the 8-vertex model
\cite{JohKM73,JimMN,LasP98,Las02,Shi04,BooJMST05}.
However, there are still many open questions in this direction.

In 1979, the Quantum Inverse Scattering Method (QISM)
\cite{FadST79,FadLH96} was applied to the
$XYZ$ Heisenberg chain in paper \cite{FT79}. This required the development
of some generalization of the algebraic Bethe ansatz, since this method in
its original formulation is not applicable to the $XYZ$ chain. The generalized
algebraic Bethe ansatz allows us to obtain Bethe equations that determine the
spectrum of the Hamiltonian, as well as construct the eigenvectors of the
transfer matrix. The question arises of the applicability of this method to
the calculation of form factors and correlation functions. This way looks very
attractive, since the application of the standard algebraic Bethe ansatz to this
problem has been developed quite well to date. Let us briefly recall the main
features of this approach.

The calculation of correlation functions within the QISM consists of several
stages. At the first stage, it is necessary to derive the action of local
operators on physical states (on-shell Bethe vectors). This is achieved either
within the framework of the composite model \cite{IzeK84}, or by explicitly solving
the quantum inverse problem \cite{KitMT99}. The latter allows us to
express explicitly local spin operators via elements of the monodromy matrix. For the $XYZ$
chain, the quantum inverse problem was solved in \cite{GohK00} (see also \cite{MaiT00}).

At the next step, it is necessary to calculate the arising scalar products
of Bethe vectors.
In these scalar products, one of the vectors is still an eigenvector of the Hamiltonian
(or equivalently, an eigenvector of the transfer matrix),
while the second, generally speaking, is not (an off-shell Bethe vector).
Attempts to calculate norms and the scalar products directly with the help of the
algebra satisfied by elements of the quantum monodromy matrix
show that this is a difficult combinatorial problem.

The history of the problem is as follows.
The first result is Gaudin's hypothesis which goes back to 1972
\cite{Gau72,Gaudin-book}
(see also the later work \cite{GMW81}) about norms of eigenvectors
of the Hamiltonian of the Bose-gas with point-like interaction.
This hypothesis was proved in 1982 by Korepin \cite{K82} in the framework of the quantum
inverse scattering method for a sufficiently wide class of models. It states that
the squared norm of eigenvectors of the Hamiltonian
is given by determinant of an $m\times m$ matrix ($m$ is the number of
excitations) whose explicit form is restored from
the form of Bethe equations.
In 1989, for models
with the 6-vertex $R$-matrix, a compact determinant formula was
proved \cite{S89} for scalar products of two Bethe vectors one of
which is on-shell and another one
is an arbitrary off-shell Bethe vector. The original method to obtain this result
was a complicated combinatorial analysis of the structure of scalar products and
application of recurrence relations for them.
In 1998, Kitanine, Maillet and Terras \cite{KitMT99} obtained this result by a different method
and showed that
the matrix elements of the matrix participating in the determinant representation
of scalar products are expressed through derivatives of eigenvalues of the transfer matrix.
Later similar results were
obtained for models with the 6-vertex $R$-matrix with non-periodic boundary conditions
\cite{Wan02,KitKMNST07,BelP15,BelP15a}.
In paper \cite{LT13}, a determinant representation
was obtained for scalar products in the elliptic solid-on-solid (SOS) model, which is closely
related to the 8-vertex model.

Recently, a new method was
proposed in \cite{BS19}, which avoids all combinatorial difficulties and
allows one to reduce the calculation of scalar
products of on-shell and off-shell Bethe vectors in models with the 6-vertex $R$-matrix
to solving a system of linear equations. This method explains why the scalar products
of on-shell and off-shell Bethe vectors have determinant representations.

The existence of determinant representations for scalar products allows us to directly
proceed to the calculation of correlation functions. Here we should mention the method
based on the use of auxiliary quantum operators (dual fields) \cite{Kor87,KojKS97,BogIK93L},
as well as the method based on the representation for correlation functions in the form of
multiple integrals \cite{JimMMN92,KitMT00,GohKS04,KitMST05,KitKMST09}. However,
for today, the most powerful and relatively simple approach is the method of form factor expansion.
In particular, a significant successes have been achieved along this path in the study
of the correlation functions of the $XXZ$ chain and the model of one-dimensional
bosons (the Lieb--Liniger model). Both analytical \cite{KitKMST11,KitKMST12} and numerical \cite{CauHM05,PerSCHMWA06,PerSCHMWA07,CauCS07} results were obtained.

The lack of compact determinant representations for scalar products in the
8-vertex model (or the $XYZ$ chain) is a serious obstacle for the study of the correlation
functions  within the framework of the QISM.  At the same time, the method of \cite{BS19}
relies solely on the formula
for the action of the transfer matrix on the Bethe vectors. For the $XYZ$ chain,
this formula was obtained in pioneer work \cite{FT79}. Therefore, this approach is
carried over without significant changes to the case of models with the 8-vertex $R$-matrix.

In this paper, we use the approach suggested in \cite{BS19}. Namely, we obtain a
system of linear equations whose solutions are the scalar products of the on-shell
and off-shell Bethe vectors of the inhomogeneous 8-vertex model. We find these
solutions in terms of determinants (minors of the matrix of the linear system).
At this stage, we basically follow the strategy of
\cite{BS19}. A significant difference appears at the final stage. A feature of this
method is that the above system of linear equations is homogeneous. Therefore, its
solutions contain an ambiguity which must be fixed. In models with the 6-vertex
(trigonometric or rational)
$R$-matrix, this is achieved by considering a special particular case, in which the scalar
product is reduced to the partition function of the 6-vertex model with the domain
wall boundary conditions which has a determinant
representation \cite{K82,I87}. An analogue of this result in
the case of the 8-vertex model is not known (see, however, papers
\cite{PRS08,Ros09}, where the representation of the partition function
of the elliptic SOS model with the domain wall boundary conditions as a finite linear
combination of determinants was obtained). Therefore, the study of this
particular case for the models with the 8-vertex $R$-matrix does not lead to the desired results.
Instead, we fix the freedom in the resulting solution using the quasiperiodic transformation properties
of Bethe vectors under sifts of the variables by
periods. Despite of this method allows one
to fix the ambiguity only partially, the residual arbitrariness disappears in the normalized expressions.

Let us present here the main result. We consider the inhomogeneous 8-vertex model
or the  $XYZ$ spin-$\frac{1}{2}$ chain
on an even number of sites $N=2n$ with a rational anisotropy parameter
$\eta =2P/Q$. Let $\bigl <\Psi_{\nu}(\{v_i\})\bigr |$ be the (dual) eigenvector
of the transfer matrix ${\sf T}(u)$ with the eigenvalue $T_{\nu}(u; \{v_i\})$
explicitly given by (\ref{a4a}) below
(here $\nu =0,1, \ldots , Q-1$, and the parameters $v_1, \ldots , v_n$
satisfy the Bethe equations)
and let $\bigl |\Psi_{\mu}(\{u_i\})\bigr >$ be an off-shell Bethe vector
with arbitrary parameters $u_1, \ldots , u_n$.  Let also define
$$
V=\sum_{i=1}^n v_i, \qquad U=\sum_{i=1}^n u_i,
$$
and $r=V-U$.
Our main result is the following
determinant formula for the scalar product for the on-shell and off-shell Bethe vectors:
%
\begin{multline}\label{int1}
\bigl <\Psi_{\nu}(\{v_i\})\bigr |\Psi_{\mu}(\{u_i\})\bigr >=
\phi_1^{(\nu \mu)}(r)\phi^{(\nu)}_2(\{v_i\})
\\
\times \,
\frac{\prod\limits_{a,b=1}^n
\theta_1(u_a-v_b|\tau)}{\prod\limits_{p<q}\theta_1(u_p-u_q|\tau)\theta_1(v_q-v_p|\tau)}\,
\det_{1\leq j,k\leq n}T_{jk}^{(\nu \mu)}(r),
\end{multline}
where the matrix $T_{jk}^{(\nu \mu)}(r)$ is given by
\beq\label{int2}
T_{jk}^{(\nu \mu)}(r)\! =\frac{\theta_1'(0|\tau)
\theta_1(u_k\! -\! v_j\! +\! r|\tau)}{\theta_1(u_k-v_j|\tau)
\theta_1(r|\tau)}\Bigl (T_{\nu}(u_k; v_1, \ldots , v_n)\! -\! T_{\mu}(u_k,
v_1,\ldots , v_j\! -\! r, \ldots , v_n)\Bigr ).
\eeq
In (\ref{int1}), (\ref{int2}), $\theta_1(z|\tau)$ is the odd
Jacobi theta function with modular parameter
$\tau \in \CC$ with ${\rm Im}\, \tau >0$. The function $\phi_1^{(\nu \mu)}(r)$ in
(\ref{int1}) is known explicitly (see (\ref{fix21}), (\ref{fix22}) below). The function
$\phi^{(\nu)}_2(\{v_i\})$ can not be fixed by our method. However, it does not enter the
expression for specially normalized scalar products
$$
\frac{\bigl <\Psi_{\nu}(\{v_i\})\bigr |\Psi_{\mu}(\{u_i\})\bigr >}
{\bigl <\Psi_{\nu}(\{v_i\})\bigr |\Psi_{\nu}(\{v_i\})\bigr >}.
$$
 We argue that only such expressions are essential for finding correlation functions.

The paper is organized as follows.
In section \ref{section:8vertex} we introduce the main objects of the
8-vertex model and $XYZ$ spin chain: the $R$-matrix
(in the elliptic parametrization), the $L$-operator,
the quantum monodromy matrix and the transfer matrix. After that we study how the
$R$-matrix acts on the tensor products of some specially parameterized vectors
in $\CC^2$ and introduce
vacuum vectors for gauge-transformed $L$-operators. Section \ref{section:gaba}
explains how the generalized algebraic Bethe ansatz works for the construction of
eigenvectors of the transfer matrix. First, the commutation relations for the elements
of the gauge-transformed quantum monodromy matrix are derived and then their algebra
is used to construct right and left (dual) eigenvectors, basically in the same way as in
\cite{FT79}. In section \ref{section:qoperator}
we recall the alternative method of diagonalization of the transfer matrix
based on the $Q$-operator and derive the sum rule for Bethe roots for the inhomogeneous
model. This section is included for completeness and is not directly related
to what follows. The main content of the paper
is contained in section \ref{section:scalarproducts}, where we obtain a
homogeneous system of linear
equations for the scalar products of Bethe vectors, prove its solvability and solve it
in terms of determinants. We also prove that our result implies orthogonality
of on-shell Bethe vectors.

There are also three appendices and a list of notations. In Appendix A we show that
our result implies that some special Bethe vectors are in fact null-vectors.
Appendix B is devoted to a detailed account of the simplest case $N=2$.
In Appendix C we show that in the case $\eta =\frac{1}{2}$ (the case of free fermions)
even more explicit results can be obtained.

\section{Inhomogeneous 8-vertex model}

\label{section:8vertex}

 Since the $XYZ$ spin-$\tfrac12$ chain and the 8-vertex model are equivalent, below we mostly talk about the latter for definiteness.
We stress that we consider an inhomogeneous model only for reasons of generality. All formulas below allow a smooth homogeneous limit.

\subsection{The $R$-matrix}

The matrix of Boltzmann weights of the 8-vertex model (the $R$-matrix)
has a natural elliptic parametrization \cite{Baxter-book}. In order to write
it explicitly, we use the
Jacobi theta functions
\beq\label{8v1}
\begin{array}{l}
\displaystyle{\theta_1(u|\tau )=-i\sum_{k\in \z}
(-1)^k q^{(k+\frac{1}{2})^2}e^{\pi i (2k+1)u},}
\\ \\
\displaystyle{\theta_2(u|\tau )=\sum_{k\in \z}
q^{(k+\frac{1}{2})^2}e^{\pi i (2k+1)u},}
\\ \\
\displaystyle{\theta_3(u|\tau )=\sum_{k\in \z}
q^{k^2}e^{2\pi i ku},}
\\ \\
\displaystyle{\theta_4(u|\tau )=\sum_{k\in \z}
(-1)^kq^{k^2}e^{2\pi i ku},}
\end{array}
\eeq
where $\tau \in \CC$, ${\rm Im}\, \tau >0$, and
$q=e^{\pi i \tau}$. Let us mention the infinite product representation
\beq\label{8v2}
\begin{array}{l}
\displaystyle{\theta_1(u|\tau )=2q^{\frac{1}{4}} \sin \pi u \prod_{n\geq 1}
(1-q^{2n})(1-q^{2n}e^{2\pi i u})(1-q^{2n}e^{-2\pi i u}).}
\end{array}
\eeq
 Similar infinite product representations exist also for the other the\-ta-func\-ti\-ons
which are connected with $\theta_1(u|\tau)$ by the formulas
$$
\begin{array}{c}
\theta_2(u|\tau)=\theta_1(u+\frac{1}{2}|\tau), \! \quad \! \!
\theta_3(u|\tau)=q^{\frac{1}{4}}e^{\pi iu}\theta_1(u+\frac{\tau+1}{2}|\tau), \! \quad \!\!
\theta_4(u|\tau)=-iq^{\frac{1}{4}}e^{\pi iu}\theta_1(u+\frac{\tau}{2}|\tau).
\end{array}
$$
It is seen from here that $\theta_1$ and $\theta_2$ become respectively sin and cos
as $q\to 0$ while $\theta_3$ and $\theta_4$ become constants (equal to 1). The function
$\theta_1$ is odd while the other three are even.
The theta functions satisfy a large number of non-trivial identities which are used below
without comments. Most of these identities with proofs can be found in
\cite{KZ15}.

Let
$$
\sigma_0 =\left (\begin{array}{rr}1&0\\ 0&1 \end{array}\right ),
\quad
\sigma_1 =\left (\begin{array}{cc}0&1\\ 1&0 \end{array}\right ),
\quad
\sigma_2 =\left (\begin{array}{rr}0&-i\\ i&0 \end{array}\right ),
\quad
\sigma_3 =\left (\begin{array}{rr}1&0\\ 0&-1 \end{array}\right )
$$
be Pauli matrices, $\sigma_{\pm}=\frac{1}{2}\, (\sigma_1 \pm i\sigma_2)$.
The Baxter's $R$-matrix of the symmetric 8-vertex model in the elliptic pa\-ra\-met\-ri\-za\-tion
has the form
\beq\label{8v3}
{\sf R}(u)={\sf R}(u; \eta , \tau )=\sum_{a=0}^3W_a(u)\, \sigma_a \otimes \sigma_a ,
\eeq
where
\beq\label{8v4}
W_a(u)= W_a(u; \eta ,\tau )=\theta_1(\eta |\tau )\,
\frac{\theta_{5-a}\Bigl (u+\frac{\eta}{2}\Bigl |\tau
\Bigr )}{2\theta_{5-a}\Bigl (\frac{\eta}{2}\Bigl |\tau \Bigr )}
\eeq
and the index of the theta functions is understood modulo 4
(for example, $\theta_5(u|\tau)=\theta_1(u|\tau)$).
This $R$-matrix acts in the tensor product $V_1\otimes V_2$ ($V_i\cong \CC^2$)
and can be also denoted as $\displaystyle{{\sf R}_{12}(u)=
\sum_{a=0}^3W_a(u)\, \sigma_a^{(1)} \sigma_a^{(2)}}$. (Pictorially, the $R$-matrix
is represented as the vertex {\huge{${+}$}}
formed by intersection of two lines,
the first space
being associated with the horizontal line and the second one with the vertical line.)
In the matrix form we have:
\beq\label{8v5}
\begin{array}{c}
\displaystyle{
{\sf R}(u)=\left (\begin{array}{cccc}
W_0(u)\! + \! W_3(u) & 0&0&W_1(u)\! -\! W_2(u)
\\
0& W_0(u)\! -\! W_3(u)& W_1(u)\! +\! W_2(u)&0
\\
0& W_1(u)\! +\! W_2(u)& W_0(u)\! -\! W_3(u)&0
\\
W_1(u)\! -\! W_2(u)& 0&0& W_0(u)\! +\! W_3(u)
\end{array}
\right )}
\\ \\
\displaystyle{=\left (\begin{array}{cccc}
a^{\rm 8v}(u) & 0&0& d^{\rm 8v}(u)
\\
0& b^{\rm 8v}(u)& c^{\rm 8v}(u)&0
\\
0& c^{\rm 8v}(u)& b^{\rm 8v}(u)&0
\\
d^{\rm 8v}(u)& 0&0& a^{\rm 8v}(u)
\end{array}
\right ),}
\end{array}
\eeq
where
\beq\label{8v6}
\begin{array}{l}
\displaystyle{
a^{\rm 8v}(u)=\frac{2\theta_4(\eta |2\tau ) \, \theta_1(u+\eta |2\tau )\,
\theta_4(u|2\tau )}{\theta_2(0|\tau )\,\theta_4(0|2\tau )}},
\\ \\
\displaystyle{
b^{\rm 8v}(u)=\frac{2\theta_4(\eta |2\tau ) \, \theta_4(u+\eta |2\tau )\,
\theta_1(u|2\tau )}{\theta_2(0|\tau )\,\theta_4(0|2\tau )}},
\\ \\
\displaystyle{
c^{\rm 8v}(u)=\frac{2\theta_1(\eta |2\tau ) \, \theta_4(u+\eta |2\tau )\,
\theta_4(u|2\tau )}{\theta_2(0|\tau )\, \theta_4(0|2\tau )}},
\\ \\
\displaystyle{
d^{\rm 8v}(u)=\frac{2\theta_1(\eta |2\tau ) \, \theta_1(u+\eta |2\tau )\,
\theta_1(u|2\tau )}{\theta_2(0|\tau )\, \theta_4(0|2\tau )}}.
\end{array}
\eeq
It is easy to see that when the spectral parameter $u$ is shifted
by the quasiperiods $1$ and $\tau$, the $R$-matrix transforms as follows:
\beq\label{8v6a}
\begin{array}{l}
{\sf R}_{12}(u+1)=-\sigma_3^{(1)}{\sf R}_{12}(u)\sigma_3^{(1)},
\\ \\
{\sf R}_{12}(u+\tau)=-e^{-\pi i (2u+\eta +\tau)}\sigma_1^{(1)}{\sf R}_{12}(u)\sigma_1^{(1)}.
\end{array}
\eeq
It can be shown that this
$R$-matrix satisfies the Yang-Baxter equation \cite{Baxter72}
\beq\label{yb}
{\sf R}_{12}(u_1-u_2){\sf R}_{13}(u_1){\sf R}_{23}(u_2)=
{\sf R}_{23}(u_2){\sf R}_{13}(u_1){\sf R}_{12}(u_1-u_2)
\eeq
and commutes with $\sigma_a\otimes \sigma_a$:
\beq\label{sigmaa}
\sigma_a\otimes \sigma_a {\sf R}(u)={\sf R}(u)\sigma_a\otimes \sigma_a ,
\quad a=1,2,3.
\eeq
Below we also need the following properties of the $R$-matrix (\ref{8v5}):
\beq\label{prop}
\begin{array}{l}
{\sf R}_{12}(-u;-\eta , \tau )=-{\sf R}_{12}(u; \eta , \tau ),
\\ \\
{\sf R}_{12}^{t_1t_2}(u)={\sf R}_{12}(u),
\\ \\
{\sf R}_{12}(u-\eta ; \eta , \tau )=e^{\pi i (2u-\eta +\tau )}{\sf R}_{12}^{t_1}(
u+\tau +1; -\eta , \tau ),
\end{array}
\eeq
where $t_i$ means transposition in the $i$-th space.

In the limit $\tau \to +i\infty$ ($q\to 0$)
the elliptic $R$-matrix degenerates into the standard
trigonometric $R$-matrix of the 6-vertex model:
$$
{\sf R}(u)\rightarrow 2q^{\frac{1}{4}}\left (\begin{array}{cccc}
\sin \pi (u+\eta )& 0 & 0& 0
\\
0&\sin \pi u & \sin \pi \eta & 0
\\
0& \sin \pi \eta & \sin \pi u & 0
\\
0&0&0&\sin \pi (u+\eta )\end{array} \right )+O(q^{\frac{5}{4}}).
$$

The $R$-matrix (\ref{8v5}) can be also represented in the form of the $L$-operator
\beq\label{8v5a}
{\sf L}(u)=\left (\begin{array}{cc}
W_0(u)\sigma_0 +W_3(u)\sigma_3 & W_1(u)\sigma_1 -iW_2(u)\sigma_2
\\ \\
 W_1(u)\sigma_1 +iW_2(u)\sigma_2 & W_0(u)\sigma_0 -W_3(u)\sigma_3
 \end{array} \right )=
 \left ( \begin{array}{cc}{\sf a}(u)&{\sf b}(u)
 \\ \\ {\sf c}(u)& {\sf d}(u)\end{array} \right ),
 \eeq
 which is the $2\times 2$ matrix whose matrix elements are operators in
 $\CC^2$. Clearly, it is the same $R$-matrix (\ref{8v5}) written as a
 block matrix\footnote{Usually in the literature the $L$-operator differs
 from the $R$-matrix by a shift of the spectral parameter $u\to u-\eta /2$.
 We do not make this shift.}.
 The Yang-Baxter equation for ${\sf R}$ is the $RLL=LLR$ relation for ${\sf L}$
\beq\label{RLL}
{\sf R}_{12}(u-v){\sf L}_1(u){\sf L}_2(v)={\sf L}_2(v){\sf L}_1(u){\sf R}_{12}(u-v),
\eeq
where
${\sf L}_1(u)={\sf L}(u)\otimes 1$, ${\sf L}_2(v)=1\otimes {\sf L}(v)$.

The quantum monodromy matrix of the inhomogeneous 8-vertex model is
\beq\label{m1}
{\cal T}(u)={\sf L}_1(u-\xi_1){\sf L}_2(u-\xi_2)\ldots {\sf L}_N(u-\xi_N)
=\left (\begin{array}{cc}
A(u)& B(u) \\ C(u) & D(u)\end{array}\right ),
\eeq
where complex numbers $\xi_i$
are inhomogeneity parameters. It is an operator in $\CC^2 \otimes {\cal H}$,
$\displaystyle{{\cal H}=\bigotimes_{i=1}^N V_i}$, $V_i\cong \CC^2$.
Here ${\sf L}_j(u)$ is given by the formula (\ref{8v5a}),
where the $\sigma$-matrices $\sigma_{a}^{(j)}$ act in the $j$-th copy of $\CC^2$ associated
with the $j$-th site of the lattice. The operators $A(u), B(u), C(u), D(u)$ act in the
space ${\cal H}$.
We consider the case of even $N=2n$, otherwise
solvability of the model is problematic. Equations (\ref{8v6a}) imply the following
properties of the quantum monodromy matrix:
\beq\label{m1a}
\begin{array}{l}
{\cal T}(u+1)=\sigma_3 {\cal T}(u)\sigma_3 = \left (
\begin{array}{rr} A(u)&-B(u)\\ -C(u)&D(u)\end{array}\right ),
\\ \\
{\cal T}(u+\tau)=e^{-\pi i c(u)}\sigma_1 {\cal T}(u)\sigma_1 = e^{-\pi i c(u)}\left (
\begin{array}{rr} D(u)&C(u)\\ B(u)&A(u)\end{array}\right ),
\end{array}
\eeq
where
\beq\label{c(u)}
c(u)=N(2u+\eta +\tau)-2\sum_{k=1}^N \xi_k.
\eeq
It follows from (\ref{RLL}) and (\ref{m1}) that the quantum monodromy matrix satisfies the
$RTT=TTR$ relation
\beq\label{RTT}
{\sf R}_{12}(u-v){\cal T}_1(u){\cal T}_2(v)={\cal T}_2(v){\cal T}_1(u){\sf R}_{12}(u-v),
\eeq
This relation implies that the transfer matrices
\beq\label{8v7}
\begin{array}{c}
{\sf T}(u)=\mbox{tr}_0\, \Bigl ({\sf R}_{01}(u-\xi_1)
{\sf R}_{02}(u-\xi_2)\ldots {\sf R}_{0N}(u-\xi_N)\Bigr )
\\ \\
=\, \mbox{tr}\Bigl ( {\sf L}_1(u-\xi_1){\sf L}_2(u-\xi_2)\ldots {\sf L}_N(u-\xi_N)\Bigr )
=\mbox{tr}\, {\cal T}(u) \, =A(u)+D(u)
\end{array}
\eeq
commute for any values
of the spectral parameter $u$. The transfer matrix is an operator in the space
${\cal H}$. For the solution of the model one is interested in eigenvectors and
eigenvalues of the transfer matrix.

It follows from (\ref{sigmaa}) that the transfer matrix commutes with the
operators
\beq\label{Ua}
{\sf U}_a=(\sigma_a)^{\otimes N}.
\eeq
Note that the operators ${\sf U}_a$
commute with each other: $${\sf U}_a{\sf U}_b=(\sigma_a \sigma_b)^{\otimes N}=
(-1)^N (\sigma_b \sigma_a)^{\otimes N}={\sf U}_b{\sf U}_a$$ (because $N$ is an even number).
Therefore, the problem is to find common eigenvectors of the transfer matrix and the
operators ${\sf U}_a$.

The homogeneous 8-vertex model (when all $\xi_i$ are equal to 0)
is closely related to the $XYZ$ spin-$\frac{1}{2}$ chain.
The connection goes as follows:
the Hamiltonian $H^{\rm XYZ}$ of the $XYZ$ spin chain is contained in the
commuting family of operators ${\sf T}(u)$ in the following way:
\beq\label{8v8}
\p_u \log {\sf T}(u)\Bigl |_{u=0}=
\frac{\theta_1'(0|\tau )}{2\theta_1(\eta |\tau )}\,
H^{\rm XYZ} +J_0 N \mathbf{1},
\eeq
where $J_0=\frac{1}{2}\,
\theta_1'(\eta |\tau )/\theta_1(\eta |\tau )$, and  $\mathbf{1}$ is the identity operator.
The Hamiltonian of the $XYZ$ chain is given by
\beq\label{HamXYZ}
H^{\rm XYZ}=\sum_{j=1}^{N} \Bigl (J_1 \sigma_1^{(j)}\sigma_1^{(j+1)}+
J_2 \sigma_2^{(j)}\sigma_2^{(j+1)}+J_3 \sigma_3^{(j)}\sigma_3^{(j+1)}\Bigr )
\eeq
with the constants
$$
J_1=\frac{\theta_4(\eta |\tau )}{\theta_4(0|\tau )}\,, \quad
J_2=\frac{\theta_3(\eta |\tau )}{\theta_3(0|\tau )}\,, \quad
J_3=\frac{\theta_2(\eta |\tau )}{\theta_2(0|\tau )}\,.
$$
The transfer matrix of the homogeneous 8-vertex model is a generating function
for conserved quantities of the $XYZ$ spin-$\tfrac12$ chain.

\subsection{Intertwining vectors}

\label{section:int}

The $L$-operator of the 8-vertex model does not have a vacuum vector,
i.e. a vector annihilated by the operator ${\sf c}(u)$, because the matrix
${\sf c}(u)$ is non-degenerate for almost all $u$. This fact makes it impossible
to apply directly the algebraic Bethe ansatz method used for the solution of the 6-vertex model.
Instead, one can apply the so-called generalized algebraic Bethe ansatz \cite{FT79}.
The key ingredient of the generalized algebraic Bethe ansatz for the 8-vertex model
is the rule of the action of the $R$-matrix (\ref{8v5}) to some special vectors.

Let us introduce a family of vectors
\beq\label{i1}
\bigl |\phi (s)\bigr >=\left (\begin{array}{c}
\theta_1(s|2\tau )\\ \theta_4 (s|2\tau )
\end{array}\right ),
\eeq
where $s$ is a complex parameter. They are called intertwining vectors.
The covector orthogonal to $\left |\phi (s)\right >$ is
$$
\bigl <\phi^{\bot}(s)\bigr |=\Bigl (-\theta_4(s|2\tau ), \theta_1(s|2\tau )\Bigr )=
ie^{-\pi i (s+\frac{\tau}{2})}\bigl < \phi (s+\tau +1)\bigr |
$$
and the scalar product $\bigl <\phi^{\bot}(t)\bigl |\phi (s)\bigr >$ is given by
\beq\label{i2}
\begin{array}{c}
\left <\phi^{\bot}(t)\bigl |\phi (s)\right >=
\theta_1 \left (\frac{1}{2}\, (t-s)\bigl |\tau \right )
\theta_2 \left (\frac{1}{2}\, (t+s)\bigl |\tau \right )
\\ \\
\displaystyle{=2\frac{\theta_1(\frac{1}{2}\, (t-s)|2\tau )\theta_4(\frac{1}{2}\, (t-s)|2\tau )
\theta_2(\frac{1}{2}\, (t+s)|2\tau )\theta_3(\frac{1}{2}\, (t+s)|2\tau )}{\theta_2
(0|2\tau )\theta_3(0|2\tau )}}.
\end{array}
\eeq

Using the identities for the theta functions, one can prove the following important identity
for the intertwining vectors:
\beq\label{i3}
{\sf R}(u)\bigl |\phi (s\! +\! \eta )\bigr >\otimes \bigl |\phi (s-u)\bigr >=
\theta_1(u+\eta |\tau ) \, \bigl |\phi (s)\bigr >\otimes
\bigl |\phi (s\! -\! u\! +\! \eta )\bigr >
\eeq
or, indicating explicitly the spaces where the vectors live:
\beq\label{i3a}
{\sf R}_{12}(u)\bigl |\phi (s\! +\! \eta )\bigr >_1\bigl |\phi (s-u)\bigr >_2=
\theta_1(u+\eta |\tau ) \, \bigl |\phi (s)\bigr >_1
\bigl |\phi (s\! -\! u\! +\! \eta )\bigr >_2.
\eeq
Note that when one acts by the $R$-matrix to the tensor product of two vectors,
one in general obtains a linear combination of pure tensor products. The situation when
one gets just one tensor product term as in (\ref{i3}) is exceptional. This property was called
by Baxter ``passing of a pair of vectors through the vertex'' \cite{Baxter-book}
and it played a very important
role in his solution of the 8-vertex model. The vectors that satisfy this property are
parameterized by points of an elliptic curve which is uniformized by the parameter $s$.
This is the origin of the parameter $s$ in (\ref{i3}).

Let us give some other useful versions
of identity (\ref{i3a}).
Changing $u\to -u$, $\eta \to -\eta$ in (\ref{i3a}) and using (\ref{prop}), we arrive at
\beq\label{i3b}
{\sf R}_{12}(u)\bigl |\phi (s\! -\! \eta )\bigr >_1\bigl |\phi (s+u)\bigr >_2=
\theta_1(u+\eta |\tau ) \, \bigl |\phi (s)\bigr >_1
\bigl |\phi (s\! +\! u\! -\! \eta )\bigr >_2.
\eeq
Shifting $s\to s+\tau +1$ and transposing in the both spaces, we also get
the transposed version of equation (\ref{i3a}):
\beq\label{i3c}
\bigl <\phi^{\bot} (s+\eta )\bigr |_1 \bigl <\phi^{\bot} (s-u )\bigr |_2
{\sf R}_{12}(u)=\theta_1(u+\eta |\tau ) \,
\bigl <\phi^{\bot} (s)\bigr |_1 \bigl <\phi^{\bot} (s-u+\eta )\bigr |_2.
\eeq

Shifting $u\to u-\xi$ and then $s\to s+u$ in (\ref{i3a})
and taking the scalar product of both sides with the covector
$\left <\phi^{\bot}(s+u)\right |$, we get:
\beq\label{i4}
\bigl <\phi^{\bot}(s+u)\bigr |_1{\sf R}_{12}(u-\xi )\bigl |\phi
(s\! +\! u\! +\! \eta )\bigr >_1 \, \bigl |\phi (s+\xi )\bigr >_2=0,
\eeq
where $\xi$ is an additional arbitrary parameter.
Here the operator $\bigl <\phi^{\bot}(s+u)\bigr |_1{\sf R}_{12}(u-\xi )\bigl |\phi
(s\! +\! u\! +\! \eta )\bigr >_1$ acts in the vertical space (the space number 2).
Taking the scalar product of
(\ref{i3a}) with the covector $\bigl <\phi^{\bot}(t)\bigr |$, we obtain
$$
\bigl <\phi^{\bot}(t)\bigr |_1{\sf R}_{12}(u)\bigl |\phi (s+\eta )\bigr >_1
\bigl |\phi (s-u)\bigr >_2=
\theta_1(u+\eta |\tau ) \, \bigl <\phi^{\bot}(t)\bigr |\phi (s)\bigr >\,
\bigl |\phi (s-u+\eta )\bigr >_2
$$
or, what is the same but with an additional parameter $\xi$ introduced by the shift
$u\to u-\xi$,
\beq\label{i5}
\frac{\bigl <\phi^{\bot}(t-u)\bigr |_1{\sf R}_{12}(u-\xi )
\bigl |\phi (s\! +\! u\! +\! \eta )\bigr >_1}{\bigl <
\phi^{\bot}(t-u)\bigr |\phi (s+u)\bigr >}
\, \bigl |\phi (s+\xi )\bigr >_2=
\theta_1(u\! -\! \xi \! +\! \eta |\tau ) \, \bigl |\phi (s\! +\! \xi \! +\! \eta \bigr >_2.
\eeq

Shifting the arguments in (\ref{i3a}) and changing
$\eta \to -\eta$, using the property (\ref{prop}) and transposing
in the first space, we obtain the following important corollary:
\beq\label{i6}
\bigl  <\phi^{\bot} (s)\bigr |_1 {\sf R}_{12}(u)\bigl |\phi (s-u)\bigr >_{2}=
\theta_1(u|\tau )\bigl <\phi^{\bot}(s+\eta )\bigr |_1 \bigl |\phi (s-u-\eta )\bigr >_2
\eeq
or, what is the same but with an additional parameter $\xi$,
\beq\label{i6a}
\bigl  <\phi^{\bot} (s+u)\bigr |_1 {\sf R}_{12}(u-\xi )\bigl |\phi (s+\xi )\bigr >_{2}=
\theta_1(u-\xi |\tau )\bigl <\phi^{\bot}(s+u+\eta )\bigr |_1 \bigl |\phi (s+\xi -\eta )\bigr >_2.
\eeq
We stress that
$\bigl <\ldots \bigr |_1\bigl |\ldots \bigr >_2$ here
is not a scalar product but the tensor product
of the vector and covector (which live in difference spaces).
Taking the scalar product with the vector $\bigl |\phi (t-u+\eta )\bigr >_1$
in the first space, we can write
this identity in the following form:
\beq\label{i7}
\frac{\bigl  <\phi^{\bot} (s+u)\bigr |_1 {\sf R}_{12}(u-\xi )
\bigl |\phi (t-u+\eta )\bigr >_{1}}{\bigl <\phi^{\bot} (s+u+\eta )\bigr |
\phi (t-u+\eta )\bigr >}\, \bigl |\phi (s+\xi )\bigr >_2=
\theta_1(u-\xi |\tau )\bigl |\phi (s+\xi -\eta )\bigr >_2.
\eeq

Let us now give a more general identity for the intertwining vectors
which can be proved basically in the same way as (\ref{i3a}):
\beq\label{i8}
\begin{array}{c}
{\sf R}_{12}(u)\bigl |\phi (s+\eta )\bigr >_1 \bigl |\phi (t-u)\bigr >_2
\\ \\
=\, \displaystyle{\frac{\theta_1(\eta |\tau )\theta_2(\frac{1}{2}\, (s+t)-u|\tau )}{\theta_2
(\frac{1}{2}\, (s+t)|\tau )}\,
\bigl |\phi (t)\bigr >_1 \bigl |\phi (s+u+\eta )\bigr >_2}
\\ \\
\phantom{aaaaaaaaaaaaaaaa}+\,
\displaystyle{\frac{\theta_1(u |\tau )\theta_2(\frac{1}{2}\, (s+t)+\eta |\tau )}{\theta_2
(\frac{1}{2}\, (s+t)|\tau )}\,
\bigl |\phi (s)\bigr >_1 \bigl |\phi (t-u-\eta )\bigr >_2}.
\end{array}
\eeq
It provides a rule of how the $R$-matrix acts on the tensor products of two arbitrary
vectors. At $t=s$ (\ref{i8}) coincides with (\ref{i3a}) (this can be seen after using
an identity for the theta functions).
Substituting $u\to -u$, $\eta \to -\eta$, we also obtain:
\beq\label{i8a}
\begin{array}{c}
{\sf R}_{12}(u)\bigl |\phi (s-\eta )\bigr >_1 \bigl |\phi (t+u)\bigr >_2
\\ \\
=\, \displaystyle{\frac{\theta_1(\eta |\tau )\theta_2(\frac{1}{2}\, (s+t)+u|\tau )}{\theta_2
(\frac{1}{2}\, (s+t)|\tau )}\,
\bigl |\phi (t)\bigr >_1 \bigl |\phi (s-u-\eta )\bigr >_2}
\\ \\
\phantom{aaaaaaaaaaaaaaaa}
+\, \displaystyle{\frac{\theta_1(u |\tau )\theta_2(\frac{1}{2}\, (s+t)-\eta |\tau )}{\theta_2
(\frac{1}{2}\, (s+t)|\tau )}\,
\bigl |\phi (s)\bigr >_1 \bigl |\phi (t+u+\eta )\bigr >_2}.
\end{array}
\eeq

Equations (\ref{i3a}), (\ref{i3b}), (\ref{i8}), (\ref{i8a}) can be unified in the
``intertwining relation'' between the $R$-matrix and the collection of Boltzmann weights
of a IRF-type model. Introduce the vectors
\beq\label{i9}
\begin{array}{l}
\bigl |\phi_k^{k+1}(u)\bigr >=\bigl |\phi (s-u+k\eta +\frac{\eta}{2})\bigr >,
\\ \\
\bigl |\phi_{k+1}^{k}(u)\bigr >=\bigl |\phi (s+u+k\eta +\frac{\eta}{2})\bigr >,
\end{array}
\eeq
then the above mentioned equations can be compactly written as
\beq\label{i10}
{\sf R}_{12}(u-v)\bigl |\phi_{k}^{k'}(u)\bigr >_1 \bigl |\phi_{k'}^{k''}(v)\bigr >_2
=\sum_l  \bigl |\phi_{l}^{k''}(u)\bigr >_1
\bigl |\phi_{k}^{l}(v)\bigr >_2
W\left [\begin{array}{cc}k&k' \\ l&k'' \end{array}\right ](u-v),
\eeq
where $W\left [\begin{array}{cc}k&k' \\ l&k'' \end{array}\right ](u)=0$ unless
$|k-k'|=|k'-k''|=|l-k''|=|l-k|=1$. The non-zero weights are:
\beq\label{i11}
\begin{array}{l}
\displaystyle{W\left [\begin{array}{cc}k&k\! \pm \! 1 \\ k\! \pm \! 1&k\! \pm \! 2
\end{array}\right ](u)=
\theta_1(u+\eta |\tau )},
\\ \\
\displaystyle{W\left [\begin{array}{cc}k&k\! \pm \! 1 \\ k\! \pm \! 1&k \end{array}\right ](u)
=\frac{\theta_1(\eta |\tau )\theta_2(s+k\eta \mp u|\tau )}{\theta_2(s+k\eta |\tau )},}
\\ \\
\displaystyle{W\left [\begin{array}{cc}k&k\! \pm \! 1 \\ k\! \mp \! 1&k \end{array}\right ](u)
=\frac{\theta_1(u |\tau )\theta_2(s+(k\pm 1)\eta |\tau )}{\theta_2(s+k\eta |\tau )}.}
\end{array}
\eeq
A similar intertwining relation obtained from (\ref{i10}) by transposition in both spaces
holds for the corresponding covectors.

\subsection{Vacuum vectors}

\label{section:vac}

Let us consider the gauge transformation of the $L$-operator
\beq\label{v1}
{\sf L}'_k (u, \xi_k)=M_{k+l-1}^{-1}(u){\sf L}_k (u-\xi_k)M_{k+l}(u)
=\left ( \begin{array}{cc}{\sf a}_k'(u)&{\sf b}_k'(u)
 \\ \\ {\sf c}_k'(u)& {\sf d}_k'(u)\end{array} \right ),
\eeq
where $l\in \ZZ$ is an integer parameter.
The matrix $M_k(u)$ is given by
\beq\label{v2}
M_k(u)=\left (\begin{array}{ll}
\theta_1(s_{k} +u|2\tau )&\gamma_k\theta_1(t_{k} -u|2\tau )
\\
\theta_4(s_{k} +u|2\tau )&\gamma_k\theta_4(t_{k} -u|2\tau )
\end{array}
\right ),
\eeq
where $s_k=s+k\eta$, $t_k=t+k\eta$,
$s,t\in \CC$ are arbitrary parameters and
\beq\label{gamma}
\gamma_k =\frac{1}{\theta_2(\tau_k |2\tau )\theta_3(\tau_k|2\tau )}, \quad
\tau_k =\frac{1}{2}\, (s_k+t_k).
\eeq
Note that the columns of this matrix are the intertwining vectors.
The inverse matrix is
\beq\label{v2a}
M_k^{-1}(u)=\frac{1}{\det M_k(u)}\left (\begin{array}{ll}
\gamma_k\theta_4(t_{k} -u|2\tau )&-\gamma_k\theta_1(t_{k} -u|2\tau )
\\
-\theta_4(s_{k} +u|2\tau )&\phantom{-\gamma_k} \theta_1(s_{k} +u|2\tau )
\end{array}
\right ),
\eeq
where
\beq\label{v3}
\begin{array}{c}
\det M_k(u)=\, - \gamma_k\bigl <\phi^{\bot}(t_k-u)\bigr |\phi (s_k+u)\bigr >
\\ \\
=\gamma_k\theta_1\bigl (\frac{1}{2}\, (s-t)\! +\! u\bigl |\tau \bigr )
\theta_2 (\tau_{k} |\tau )
\\ \\
=\displaystyle{2\frac{\theta_1\bigl (\frac{1}{2}\, (s-t)\! +\! u\bigl |2\tau \bigr )
\theta_4\bigl (\frac{1}{2}\, (s-t)\! +\! u\bigl |2\tau \bigr )}{\theta_2(0|2\tau )
\theta_3(0|2\tau )}\equiv \mu (u)}.
\end{array}
\eeq
Note that $\det M_k(u)=\mu (u)$ does not depend on $k$.

The gauge-transformed $L$-operator (\ref{v1})
has a local $u$-independent vacuum vector
\beq\label{v4}
\bigl |\omega_k^l\bigr >=\left (\begin{array}{c}
\theta_1(s_{k+l-1}+\xi_k|2\tau )
\\
\theta_4(s_{k+l-1}+\xi_k|2\tau )
\end{array}
\right ) =\bigl |\phi (s_{k+l-1}+\xi_k)\bigr >_k \in V_k
\eeq
which is annihilated by the left lower
element ${\sf c}_k'(u)$:
\beq\label{v4a}
{\sf c}_k'(u)\bigl |\omega_k^l\bigr >=0
\eeq
(recall that ${\sf c}_k'(u)$ depends also on $s$ and $l$). This directly follows from
equation (\ref{i4}) (one should put $s=s_{k+l-1}$ in the latter).
In their turn, equations (\ref{i5}) and (\ref{i7}) (where one should put
$s=s_{k+l-1}$, $t=t_{k+l-1}$) tell us how
the operators ${\sf a}_k'(u)$, ${\sf d}_k'(u)$ act to the vacuum vector:
\beq\label{v5}
\begin{array}{l}
{\sf a}_k'(u)\bigl |\omega_k^l\bigr >=
\theta_1 (u\! - \!\xi_k \! +\! \eta |\tau )\bigl |\omega_k^{l+1}\bigr >,
\\ \\
{\sf d}_k'(u)\bigl |\omega_k^l\bigr >=
\theta_1 (u -\xi_k |\tau )\bigl |\omega_k^{l-1}\bigr >.
\end{array}
\eeq
Unlike the situation in the 6-vertex model, the vacuum vector is not an eigenvector
for these operators but transforms in a simple way.

The gauge-transformed quantum monodromy matrix is
$$
{\cal T}'(u)={\sf L}'_1(u-\xi_1){\sf L}'_2(u-\xi_2)\ldots {\sf L}'_N(u-\xi_N)
$$
$$
=M_{l}^{-1}(u){\cal T}(u)M_{N+l}(u)=\left (\begin{array}{cc}
A^l(u)& B^l(u)\\ C^l(u)&D^l(u)\end{array}\right ).
$$
The global vacuum vectors are defined as
$$
\left |\Omega ^l\right >=\left |\omega_1^l\right >\otimes \left |\omega_2^l\right >\otimes \ldots
\otimes \left |\omega_N^l\right >.
$$
According to (\ref{v4a}), (\ref{v5}), the action of the operators $A^l(u)$, $D^l(u)$ and $C^l(u)$
on the global vacuum vector is given by
\beq\label{v6}
\begin{array}{l}
\displaystyle{C^l(u)\left |\Omega^l\right >=0,}
\\ \\
\displaystyle{A^l(u)\left |\Omega^l\right >=
\prod_{i=1}^N \theta_1 (u\! - \!\xi_i \! +\! \eta |\tau )
\left |\Omega^{l+1}\right >},
\\ \\
\displaystyle{D^l(u)\left |\Omega^l\right >=
\prod_{i=1}^N \theta_1 (u\! - \!\xi_i  |\tau )
\left |\Omega^{l-1}\right >.}
\end{array}
\eeq

The same formulas (\ref{i3a}), (\ref{i3b}) allow one to check that the local
dual vacuum vector
\beq\label{du1}
\bigl <\bar \omega_k^l\bigr |=\bigl <\phi^{\bot}(t_{k+l}-\xi_k)\bigr |
\eeq
satisfies the following properties:
\beq\label{du2}
\begin{array}{l}
\bigl <\bar \omega_k^l\bigr |{\sf b}_k' =0,
\\ \\
\bigl <\bar \omega_k^l\bigr |{\sf a}_k'=\gamma_{k+l-1}\gamma_{k+l}^{-1}\,
\theta_1(u-\xi_k +\eta |\tau )\bigl <\bar \omega_k^{l-1}\bigr |,
\\ \\
\bigl <\bar \omega_k^l\bigr |{\sf d}_k'=\gamma_{k+l}\gamma_{k+l-1}^{-1}\,
\theta_1(u-\xi_k |\tau )\bigl <\bar \omega_k^{l+1}\bigr |.
\end{array}
\eeq
The global dual (left) vacuum vectors are defined as tensor products of the local ones:
\beq\label{du3}
\left <\bar \Omega^l\right |=\left <\bar \omega_1^l\right |\otimes
\left <\bar \omega_2^l\right |\otimes \ldots \otimes
\left <\bar \omega_N^l\right |.
\eeq
The action of the operators $A^l(u)$, $D^l(u)$, $B^l(u)$ to the left vacuum is given by
\beq\label{du4}
\begin{array}{l}
\left <\bar \Omega^l \right |B^l(u)=0,
\\ \\
\displaystyle{\left <\bar \Omega^l \right |A^l(u)=
\gamma_l\gamma_{l+N}^{-1}\prod_{i=1}^N \theta_1(u-\xi_i +\eta |\tau)
\left <\bar \Omega^{l-1} \right |,}
\\ \\
\displaystyle{\left <\bar \Omega^l \right |D^l(u)=
\gamma_{l+N}\gamma_{l}^{-1}\prod_{i=1}^N \theta_1(u-\xi_i|\tau)
\left <\bar \Omega^{l+1} \right |.}
\end{array}
\eeq
These formulas will be used in the generalized algebraic Bethe ansatz.

\section{The generalized algebraic Bethe ansatz}

\label{section:gaba}

In this section, we construct off-shell and on-shell Bethe vectors and describe their properties. We continue to use the ``bra-ket''
notation in order to distinguish between right $\bigl |\Psi_{\nu}(u_1,\dots,u_n)\bigr >$
and left (dual) $\bigl <\Psi_{\nu}(v_1,\dots,v_n)\bigr |$ Bethe vectors.
We draw the reader's attention to the fact that the meaning of this notation is
completely different from that in the previous section.
First of all, in distinction of the two-component vectors $\bigl |\phi (u)\bigr >$
and $\bigl <\phi^{\bot}(u)\bigr |$ the Bethe vectors
$\bigl |\Psi_{\nu}(u_1,\dots,u_n)\bigr >$ belong to the quantum space of
the model, that is, $(\CC^2)^{\otimes N}$. Respectively, the dual vectors
$\bigl <\Psi_{\nu}(v_1,\dots,v_n)\bigr |$ belong to the dual space.
Besides, since the procedures for constructing left and right vectors are slightly different, we
generally do not require that the left and right vectors be connected by transposition or Hermitian conjugation. In particular, the
scalar product $\bigl <\Psi_{\nu}(v_1,\dots,v_n)\bigr |\Psi_{\nu}(v_1,\dots,v_n)\bigr >$,
generally speaking, is not
the square of the vector norm. However, it does become the square of
the norm for the values of parameters that ensure the positivity of
Boltzmann weights \eqref{8v6} or the self-conjugacy of the Hamiltonian
(\ref{HamXYZ}). This treatment of the right and left Bethe vectors
is traditional in the algebraic  Bethe ansatz approach.

\subsection{The permutation relations}

Let us introduce the generalized (gauge-transformed) monodromy matrices
\beq\label{p1}
{\cal T}_{k,l}(u)=M^{-1}_k(u) {\cal T}(u)M_l(u)=
\left (\begin{array}{cc}
A_{k,l}(u)&B_{k,l}(u)
\\
C_{k,l}(u)& D_{k,l}(u)\end{array} \right ).
\eeq
Note that in this new notation ${\cal T}'(u)={\cal T}_{l, l+N}(u)$. We have:
\beq\label{p2}
\begin{array}{l}
A_{k,l}(u)=\displaystyle{\frac{\bigl <\phi^{\bot}(t_k-u)\bigr |{\cal T}(u)
\bigl |\phi (s_l+u)\bigr >}{\bigl <\phi^{\bot}(t_k-u)\bigr |\phi (s_k+u)\bigr >}},
\\ \\
B_{k,l}(u)=\gamma_l\displaystyle{\frac{\bigl <\phi^{\bot}(t_k-u)\bigr |{\cal T}(u)
\bigl |\phi (t_l-u)\bigr >}{\bigl <\phi^{\bot}(t_k-u)\bigr |\phi (s_k+u)\bigr >}},
\\ \\
C_{k,l}(u)=-\displaystyle{\frac{1}{\gamma_k}\,
\frac{\bigl <\phi^{\bot}(s_k+u)\bigr |{\cal T}(u)
\bigl |\phi (s_l+u)\bigr >}{\bigl <\phi^{\bot}(t_k-u)\bigr |\phi (s_k+u)\bigr >}},
\\ \\
D_{k,l}(u)=-\displaystyle{\frac{\gamma_l}{\gamma_k}\,
\frac{\bigl <\phi^{\bot}(s_k+u)\bigr |{\cal T}(u)
\bigl |\phi (t_l-u)\bigr >}{\bigl <\phi^{\bot}(t_k-u)\bigr |\phi (s_k+u)\bigr >}}.
\end{array}
\eeq
It follows from equations (\ref{m1a}) that the generalized monodromy matrix
has the following quasiperiodicity properties:
\beq\label{p2a}
\begin{array}{lll}
{\cal T}_{k,l}(u+1)&=&{\cal T}_{k,l}(u),
\\ &&\\
{\cal T}_{k,l}(u+\tau)&=&e^{-\pi i c(u)}
\left (\begin{array}{cc}e^{\pi i s_k} &0\\ 0& -e^{-\pi it_k}\end{array}\right )
{\cal T}_{k,l}(u)
\left (\begin{array}{cc}e^{-\pi i s_l} &0\\ 0& -e^{\pi it_l}\end{array}\right )
\\ &&\\
&& =\, e^{-\pi i c(u)}\left (\begin{array}{cc}
e^{\pi i (s_k-s_l)}A_{k,l}(u)& -e^{\pi i(s_k+t_l)}B_{k,l}(u)
\\ \\
-e^{-\pi i(s_l+t_k)}C_{k,l}(u)&e^{\pi i (t_l-t_k)}D_{k,l}(u)
\end{array}\right ),
\end{array}
\eeq
where $c(u)$ is defined in (\ref{c(u)}).

For the calculations below it is convenient to introduce a temporary notation
for the vectors and covectors
$$
X^l (u)=\bigl |\phi (s_l+u)\bigr >, \quad Y^l (u)=\bigl |\phi (t_l-u)\bigr >,
$$
$$
\tilde X^k(u)=\bigl <\phi^{\bot}(s_k+u)\bigr |, \quad
\tilde Y^k(u)=\bigl <\phi^{\bot}(t_k-u)\bigr |,
$$
then
$$
\begin{array}{l}
\displaystyle{A_{k,l}(u)=-\frac{\gamma_k}{\mu (u)}\,
\tilde Y^k(u){\cal T}(u) X^l (u),}
\\ \\
\displaystyle{B_{k,l}(u)=-\frac{\gamma_k\gamma_l}{\mu (u)}\,
\tilde Y^k(u){\cal T}(u) Y^l (u),}
\\ \\
\displaystyle{C_{k,l}(u)=\, \frac{1}{\mu (u)}\,
\tilde X^k(u){\cal T}(u) X^l (u),}
\\ \\
\displaystyle{D_{k,l}(u)=\, \frac{\gamma_l}{\mu (u)}\,
\tilde X^k(u){\cal T}(u) Y^l (u).}
\end{array}
$$
In this notation, equations (\ref{i3a}),
(\ref{i3b}), (\ref{i8}), (\ref{i8a}) look as follows:
\beq\label{p3a}
{\sf R}_{12}(u-v)X_1^{l+1}(u)X_2^{l}(v)=\theta_1(u-v+\eta |\tau )
X_1^{l}(u)X_2^{l+1}(v),
\eeq
\beq\label{p3}
{\sf R}_{12}(u-v)Y_1^{l-1}(u)Y_2^{l}(v)=\theta_1(u-v+\eta |\tau )
Y_1^{l}(u)Y_2^{l-1}(v),
\eeq
\beq\label{p4}
{\sf R}_{12}(u-v)Y_1^{l+1}(u)X_2^{l}(v)=f^+_l(u-v)Y_1^{l}(u)X_2^{l-1}(v)+
g_l(v-u)X_1^{l}(u)Y_2^{l+1}(v),
\eeq
\beq\label{p5}
{\sf R}_{12}(u-v)X_1^{k}(u)Y_2^{k-1}(v)=f^-_k(u-v)X_1^{k+1}(u)Y_2^{k}(v)+
g_k(u-v)Y_1^{k-1}(u)X_2^{k}(v),
\eeq
\beq\label{p4a}
{\sf R}_{12}(u-v)Y_1^{l}(u)X_2^{l+1}(v)=f^+_l(u-v)Y_1^{l-1}(u)X_2^{l}(v)+
g_l(v-u)X_1^{l+1}(u)Y_2^{l}(v),
\eeq
\beq\label{p5a}
{\sf R}_{12}(u-v)X_1^{k-1}(u)Y_2^{k}(v)=f^-_k(u-v)X_1^{k}(u)Y_2^{k+1}(v)+
g_k(u-v)Y_1^{k}(u)X_2^{k-1}(v),
\eeq
where
$$
f^{\pm}_k(u)=\frac{\theta_1(u|\tau )\theta_2(\tau_{k\pm 1}|\tau )}{\theta_2(\tau_k|\tau )},
\quad
g_k(u)=\frac{\theta_1(\eta |\tau )\theta_2(\tau_{k}+u|\tau )}{\theta_2(\tau_k|\tau )}.
$$
Similar relations hold for the covectors $\tilde X^l(u)$, $\tilde Y^l(u)$; they are obtained
by transposition in both spaces.

Multiplying both sides of the $RTT=TTR$ relation (\ref{RTT}) by the vectors
$Y_1^{l-1}(u)Y_2^{l}(v)$ from the right and
$\tilde Y_1^{k-1}(u)\tilde Y_2^{k}(v)$ from the left and using (\ref{p3}), one
obtains the permutation relation
\beq\label{p6}
B_{k+1, l}(u)B_{k, l+1}(v)=B_{k+1, l}(v)B_{k, l+1}(u).
\eeq
Similarly, multiplying both sides of (\ref{RTT}) by the vectors
$X_1^{l+1}(u)X_2^{l}(v)$ from the right and
$\tilde X_1^{k+1}(u)\tilde X_2^{k}(v)$ from the left
and using (\ref{p3a}), we get
\beq\label{p6a}
C_{k, l+1}(u)C_{k+1, l}(v)=C_{k, l+1}(v)C_{k+1, l}(u).
\eeq
The commutation relations between $A$- and $B$-operators are obtained by
multiplying both sides of (\ref{RTT}) by the vectors
$Y_1^{l+1}(u)X_2^{l}(v)$ from the right and
$\tilde Y_1^{k-1}(u)\tilde Y_2^{k}(v)$ from the left and using
the transposed version of (\ref{p3}) and (\ref{p4}):
\begin{multline}\label{p7}
\theta_1(u-v+\eta |\tau )B_{k, l+1}(u)A_{k-1, l}(v)
\\
=\theta_1(u-v|\tau )A_{k, l-1}(v)B_{k-1, l}(u)+
g_l(v-u)B_{k, l+1}(v)A_{k-1, l}(u).
\end{multline}
The other commutation relations can be obtained in a similar way.
Multiplying both sides of (\ref{RTT}) by the vectors
$Y_1^{l}(u)Y_2^{l+1}(v)$ from the right and
$\tilde X_1^{k}(u)\tilde Y_2^{k-1}(v)$ from the left and using
the transposed version of (\ref{p5}),
one obtains
\begin{multline}\label{p8}
\theta_1(u-v+\eta |\tau )B_{k-1, l}(v)D_{k, l+1}(u)
\\
=\theta_1(u-v|\tau )D_{k+1, l}(u)B_{k, l+1}(v)+
g_k(u-v)B_{k-1, l}(u)D_{k, l+1}(v).
\end{multline}
Multiplying both sides of the $RTT=TTR$ relation by the vectors
$X_1^{l+1}(u)X_2^{l}(v)$  from the right and
$\tilde X_1^{k-1}(u)\tilde Y_2^{k}(v)$
from the left and using
the transposed version of (\ref{p5a}),
one obtains:
\begin{multline}\label{p7a}
\theta_1(u-v+\eta |\tau )A_{k, l+1}(v)C_{k-1, l}(u)
\\
=\frac{\gamma_k^2}{\gamma_{k+1}\gamma_{k-1}}
\, \theta_1(u-v|\tau )C_{k, l+1}(u)A_{k+1, l}(v)+
g_k(u-v)A_{k, l+1}(u)C_{k-1, l}(v).
\end{multline}
Finally, multiplying both sides of the $RTT=TTR$ relation by the vectors
$Y_1^{l}(u)X_2^{l+1}(v)$  from the right and
$\tilde X_1^{k+1}(u)\tilde X_2^{k}(v)$
from the left and using (\ref{p4a}),
one obtains:
\begin{multline}\label{p8a}
\theta_1(u-v+\eta |\tau )D_{k, l}(u)C_{k+1, l+1}(v)
\\
=\frac{\gamma_l^2}{\gamma_{l+1}\gamma_{l-1}}
\, \theta_1(u-v|\tau )C_{k, l}(v)D_{k+1, l-1}(u)+
g_l(v-u)D_{k, l}(v)C_{k+1, l+1}(u).
\end{multline}
These are the main operator permutation relations used in the generalized
algebraic Bethe ansatz
procedure.

\subsection{Right eigenvectors}
\label{section:right}

Let us consider a vector
\beq\label{a1}
\left |\Psi^l (u_1, \ldots , u_n)\right >=B_{l-1, l+1}(u_1)B_{l-2, l+2}(u_2)\ldots
B_{l-n, l+n}(u_n)
\left |\Omega ^{l-n}\right >.
\eeq
We recall that $n$ is fixed and
is equal to $N/2$. The commutation relation (\ref{p6}) implies that this vector
is a symmetric function of the parameters $u_1, \ldots , u_n$.
We are going to act to this vector by the transfer matrix ${\sf T}(u)=
A_{l,l}(u)+D_{l,l}(u)$. To this end,
let us rewrite equations (\ref{p7}), (\ref{p8}) in a more convenient form
suitable for moving the operators $A$ and $D$ to the right through $B$:
\beq\label{p9}
\begin{array}{l}
A_{k,l}(u)B_{k-1, l+1}(v)
\\ \ \\
\phantom{aaaaa}=\alpha (u-v)B_{k, l+2}(v)A_{k-1, l+1}(u)+
\beta_{l+1}(u-v)B_{k, l+2}(u)A_{k-1, l+1}(v),
\end{array}
\eeq
\beq\label{p10}
\begin{array}{l}
D_{k,l}(u)B_{k-1, l+1}(v)
\\ \ \\
\phantom{aaaaa}=
\alpha (v-u)B_{k-2, l}(v)D_{k-1, l+1}(u)-
\beta_{k-1}(u-v)B_{k-2, l}(u)D_{k-1, l+1}(v),
\end{array}
\eeq
where
\beq\label{p11}
\alpha (u)= \frac{\theta_1(u-\eta |\tau )}{\theta_1(u|\tau )}, \quad
\beta_k(u)=\frac{\theta_1(\eta |\tau )\, \theta_2(\tau_k +u|\tau )}{\theta_1(u|\tau )\,
\theta_2(\tau_k |\tau )}.
\eeq
The action of the operators
$A_{l,l}(u)$, $D_{l,l}(u)$ to the vector (\ref{a1})
can be found, with the help of the standard
algebraic Bethe ansatz argument, using the permutation relations (\ref{p9}),
(\ref{p10}) and the property (\ref{v6}). The result is:
$$
A_{l,l}(u)\left |\Psi^l (u_1, \ldots , u_n)\right >=
T_A(u)\left |\Psi^{l+1} (u_1, \ldots , u_n)\right >
$$
$$
\phantom{aaaaaaaaaaaaaaaa}+\sum_{j=1}^n
\Lambda_{A,j}^{l}(u)\left |\Psi^{l+1} (u_1, \ldots , u_{j-1}, u, u_{j+1}, \ldots , u_n)\right >,
$$
$$
D_{l,l}(u)\left |\Psi^l (u_1, \ldots , u_n)\right >=
T_D(u)\left |\Psi^{l-1} (u_1, \ldots , u_n)\right >
$$
$$
\phantom{aaaaaaaaaaaaaaaa} +\sum_{j=1}^n
\Lambda_{D,j}^{l}(u)\left |\Psi^{l-1} (u_1, \ldots , u_{j-1}, u, u_{j+1}, \ldots , u_n)\right >,
$$
where
$$
T_A(u)=\prod_{i=1}^N \theta_1(u-\xi_i +\eta |\tau )\,
\prod_{k=1}^n \frac{\theta_1(u-u_k-\eta |\tau )}{\theta_1(u-u_k|\tau )},
$$
$$
T_D(u)=\prod_{i=1}^N \theta_1(u-\xi_i |\tau )\,
\prod_{k=1}^n \frac{\theta_1(u-u_k+\eta |\tau )}{\theta_1(u-u_k|\tau )},
$$
$$
\Lambda_{A,j}^{l}(u)=\frac{\theta_1(\eta |\tau )}{\theta_1'(0|\tau )}\,
\Phi \Bigl (u\! -\! u_j, \tau_{l+1}\! +\! \frac{1}{2}\Bigr )
\prod_{i=1}^N \theta_1(u_j-\xi_i +\eta |\tau )\,
\prod_{k=1, \neq j}^n \frac{\theta_1(u_j-u_k-\eta |\tau )}{\theta_1(u_j-u_k|\tau )},
$$
$$
\Lambda_{D,j}^{l}(u)=-
\frac{\theta_1(\eta |\tau )}{\theta_1'(0|\tau )}\,
\Phi \Bigl (u\! -\! u_j, \tau_{l-1}\! +\! \frac{1}{2}\Bigr )
\prod_{i=1}^N \theta_1(u_j-\xi_i |\tau )\,
\prod_{k=1, \neq j}^n \frac{\theta_1(u_j-u_k+\eta |\tau )}{\theta_1(u_j-u_k|\tau )}.
$$
In the last two formulas we have introduced a function
\beq\label{a2}
\Phi (u, v)=\frac{\theta_1'(0|\tau )\, \theta_1(u+v|\tau )}{\theta_1(u|\tau )\,
\theta_1(v|\tau )}.
\eeq
It has a simple pole at $u=0$ with the residue $1$.

Consider now the Fourier transform of the vector $\left |\Psi^l\right >$:
\beq\label{a3}
\bigl |\Psi_{\nu} (u_1, \ldots , u_n)\bigr >=
\sum_{l\in \z}e^{-il\pi \eta \nu}\bigl |\Psi^l (u_1, \ldots , u_n)\bigr >.
\eeq
For arbitrary parameters $u_j$ we call such vectors {\it off-shell Bethe vectors}.
The action of the transfer matrix ${\sf T}(u)=A_{l,l}(u)+D_{l,l}(u)$ on such vector
is given by
\begin{multline}\label{a4}
{\sf T}(u)\bigl |\Psi_{\nu} (u_1, \ldots , u_n)\bigr >=
T_{\nu}(u)\bigl |\Psi_{\nu} (u_1, \ldots , u_n)\bigr >
\\
+\sum_{l\in \z}\sum_{j=1}^n e^{-il\pi \eta \nu}\left (e^{i\pi \eta \nu}\Lambda_{A,j}^{l-1}(u)+
e^{-i\pi \eta \nu}\Lambda_{D,j}^{l+1}(u)\right )
\bigl |\Psi^{l} (u_1, \ldots , u_{j-1}, u, u_{j+1}, \ldots , u_n)\bigr >,
\end{multline}
where
\begin{multline}\label{a4a}
T_{\nu}(u) = e^{i\pi \eta \nu}\! \prod_{i=1}^N \theta_1(u-\xi_i +\eta |\tau )\!
\prod_{k=1}^n \! \frac{\theta_1(u-u_k-\eta |\tau )}{\theta_1(u-u_k|\tau )}
\\
+ e^{-i\pi \eta \nu}\!
\prod_{i=1}^N \theta_1(u-\xi_i |\tau )\!
\prod_{k=1}^n \! \frac{\theta_1(u-u_k+\eta |\tau )}{\theta_1(u-u_k|\tau )}.
\end{multline}
Note that one can rewrite (\ref{a4}) in the form
\begin{multline}\label{a5}
{\sf T}(u)\bigl |\Psi_{\nu} (u_1, \ldots , u_n)\bigr >=
T_{\nu}(u)\bigl |\Psi_{\nu} (u_1, \ldots , u_n)\bigr >
\\
-\sum_{l\in \z}\sum_{j=1}^n e^{-il\pi \eta \nu}
\Phi \Bigl (u-u_j, \tau_l +\frac{1}{2}\Bigr )\!
\, \Bigl (\res_{u=u_j}\! T_{\nu}(u) \Bigr )
\bigl |\Psi^{l} (u_1, \ldots , u_{j-1}, u, u_{j+1}, \ldots , u_n)\bigr >.
\end{multline}
In this form, it is clear that the r.h.s. is regular at $u=u_j$ as it should be.

The eigenvalue of the transfer matrix should be a regular function of $u_j$. The
conditions $\displaystyle{\res_{u=u_j}\! T_{\nu}(u) =0}$ are simultaneously the conditions of
cancellation of the ``unwanted terms'' in (\ref{a5}). These conditions have the form
of the Bethe equations
\beq\label{Bethe}
e^{2i\pi \eta
\nu}\prod_{i=1}^N \frac{\theta_1(u_j-\xi_i +\eta |\tau )}{\theta_1(u_j-\xi_i |\tau )}
=\prod_{k=1, \neq j}^{n}\frac{\theta_1(u_j-u_k +\eta |\tau )}{\theta_1(u_j-u_k -\eta |\tau )}.
\eeq
For $N=2$ there is only one Bethe equation and it can be solved explicitly
(see Appendix B).
If the Bethe equations are satisfied, then the unwanted terms cancel and the
vector $\left |\Psi_{\nu}\right >=\left |\Psi_{\nu}(u_1, \ldots , u_n)\right >$
is an eigenvector of the transfer matrix provided that $\nu$
is such that the series (\ref{a3}) converges and is non-zero. Presumably, this holds
for some particular values of $\nu$ and $\nu =0$ is among them. We call such vectors
{\it on-shell Bethe vectors}.

For what follows we need analytical properties of the Bethe vectors as functions
of the parameters $u_j$. First of all, we note that $\left |\Psi^l\right>$
(and, therefore, $\left |\Psi_{\nu}\right >$ is an entire function of $u_j$.
Indeed, a possible pole could only occur at $u_j=(t-s)/2$ when $\mu(u)$ in the denominator
of the expression for $B_{l-j,l+j}(u_j)$ vanishes. In this case the matrix
$M_k(u_j)$ becomes degenerate and one can immediately see that
$$
\res_{u_j=(t-s)/2}B_{l-j,l+j}(u_j)\propto \res_{u_j=(t-s)/2}C_{l-j,l+j}(u_j).
$$
Using the fact that the Bethe vector is a symmetric function of the $u_i$'s,
one can move $u_j$ to the very right end of the chain of the $B$-operators, where
we have
$$
\res_{u_j=(t-s)/2}B_{l-n,l+n}(u_j)\left |\Omega^{l-n}\right >
=\res_{u_j=(t-s)/2}C_{l-n,l+n}(u_j)\left |\Omega^{l-n}\right >=0.
$$
Therefore, $\res\limits_{u_j=(t-s)/2}\left |\Psi^l(u_1, \ldots , u_n)
\right>=0$ and thus the Bethe vector
is a regular function of each of $u_j$.

Next, let us derive the quasiperiodic properties of the Bethe vector under the shifts
$u_j\to u_j+1$ and $u_j\to u_j+\tau$.
Using (\ref{p2a}), we have:
\beq\label{qua1}
\begin{array}{l}
B_{l-j,l+j}(u+1)=B_{l-j,l+j}(u),
\\ \\
B_{l-j,l+j}(u+\tau)=-e^{\pi i(s+t)+2\pi i l\eta -\pi i c(u)}B_{l-j,l+j}(u),
\end{array}
\eeq
where $c(u)$ is given by (\ref{c(u)}). It then follows that
 \beq\label{qua2}
\begin{array}{l}
\bigl |\Psi_{\nu}(u_1,\ldots,u_{j-1},u_j+1,u_{j+1},\dots,u_n )\bigr > =
\bigl |\Psi_{\nu}(u_1,\ldots,u_n)\bigr >,
\\ \\
\bigl |\Psi_{\nu}(u_1,\ldots,u_{j-1},u_j+\tau,u_{j+1},\dots,u_n )\bigr >=
-e^{\pi i(s+t)-\pi i c(u)}
\bigl |\Psi_{\nu -2}(u_1,\ldots,u_n)\bigr >.
\end{array}
\eeq
In particular, if the vector is on-shell and $\{\nu , u_1, \ldots , u_n\}$ is the
corresponding solution of the Bethe equations, the set
$\{\nu +2 , u_1, \ldots , u_j+\tau , \ldots , u_n\}$ also solves the Bethe equations and
the two eigenvectors are proportional to each other, i.e., correspond to the same physical
state. This fact was mentioned in \cite{T95}.

\subsection{Left eigenvectors}

For the construction of left eigenvectors of the transfer matrix it is convenient to redefine
the operators $A_{kl}$, $B_{k,l}$, $C_{kl}$, $D_{kl}$ in the following way:
\beq\label{b0}
\bar A_{kl}=\gamma_k^{-1}\gamma_l A_{kl}, \quad
\bar B_{kl}=\gamma_k^{-1}\gamma_l^{-1} B_{kl}, \quad
\bar C_{kl}=\gamma_k\gamma_l C_{kl}, \quad
\bar D_{kl}=\gamma_k\gamma_l^{-1} D_{kl}.
\eeq
Note that these operators act to the left vacuum as follows (see (\ref{du4})):
$$
\begin{array}{l}
\displaystyle{\left <\bar \Omega^{l} \right |\bar A_{l,l+N}(u)=
\prod_{i=1}^N \theta_1(u-\xi_i +\eta |\tau)
\left <\bar \Omega^{l-1} \right |,}
\\ \\
\displaystyle{\left <\bar \Omega^{l} \right |\bar D_{l,l+N}(u)=
\prod_{i=1}^N \theta_1(u-\xi_i|\tau)
\left <\bar \Omega^{l+1} \right |}.
\end{array}
$$
The new operators are matrix elements of the gauge-transformed quantum monodromy matrix
$\bar {\cal T}(u)=\bar M_{k}^{-1}(u){\cal T}(u)\bar M_l(u)$ by means of the matrix
\beq\label{v2b}
\bar M_k(u)=\left (\begin{array}{ll}
\gamma_k\theta_1(s_{k} +u|2\tau )&\theta_1(t_{k} -u|2\tau )
\\
\gamma_k \theta_4(s_{k} +u|2\tau )&\theta_4(t_{k} -u|2\tau )
\end{array}
\right )
\eeq
(comparing to the matrix (\ref{v2}), the first rather than second column
is multiplied by $\gamma_k$).

Let us consider the dual vector
\beq\label{b1}
\left <\Psi^l (v_1, \ldots , v_n)\right |=
\left <\bar \Omega ^{l-n}\right |\bar C_{l-n, l+n}(v_n)\ldots
\bar C_{l-2, l+2}(v_2)\bar C_{l-1, l+1}(v_1).
\eeq
The commutation relation (\ref{p6a}) implies that this vector
is a symmetric function of the parameters $v_1, \ldots , v_n$.
We are going to act to this vector by the transfer matrix ${\sf T}(u)=
\bar A_{l,l}(u)+\bar D_{l,l}(u)$ to the left. To this end,
let us rewrite equations (\ref{p7a}), (\ref{p8a}) in a more convenient form
suitable for moving the operators $\bar A$ and $\bar D$ to the left through $\bar C$:
\beq\label{b2}
\begin{array}{l}
\bar C_{k-1,l+1}(v)\bar A_{k,l}(u)
\\ \ \\
\phantom{aaaaa}=
\alpha (u-v)\bar A_{k-1,l+1}(u)\bar C_{k-2,l}(v)
-\beta_{k-1}(v-u)\bar A_{k-1,l+1}(v)\bar C_{k-2,l}(u),
\end{array}
\eeq
\beq\label{b3}
\begin{array}{l}
\bar C_{k-1,l+1}(v)\bar D_{k,l}(u)
\\ \ \\
\phantom{aaaaa}=\alpha (v-u)\bar D_{k-1,l+1}(u)\bar C_{k,l+2}(v)
+\beta_{l+1}(v-u)\bar D_{k-1,l+1}(v)\bar C_{k,l+2}(u),
\end{array}
\eeq
with the same functions $\alpha (u)$, $\beta_k(u)$ as in (\ref{p11}).

Consider now the Fourier transform of the dual vector $\left <\Psi^l\right |$:
\beq\label{a3a}
\bigl <\Psi_{\nu} (v_1, \ldots , v_n)\bigr |=
\sum_{l\in \z}e^{il\pi \eta \nu}\bigl <\Psi^l (v_1, \ldots , v_n)\bigr |.
\eeq
The arguments similar to the ones used in the case of right vectors lead to
the following formula for the action of the transfer matrix to the dual vector:
\begin{multline}\label{a5a}
\bigl <\Psi_{\nu} (v_1, \ldots , v_n)\bigr |{\sf T}(u)=
T_{\nu}(u)\bigl <\Psi_{\nu} (v_1, \ldots , v_n)\bigr |
\\
+\sum_{l\in \z}\sum_{j=1}^n e^{il\pi \eta \nu}
\Phi \Bigl (v_j-u, \tau_l +\frac{1}{2}\Bigr )\!
\, \Bigl (\res_{u=v_j}\! T_{\nu}(u) \Bigr )
\bigl <\Psi^{l} (v_1, \ldots , v_{j-1}, u, v_{j+1}, \ldots , v_n)\bigr |.
\end{multline}
Again, the conditions $\displaystyle{\res\limits_{u=v_j}\! T_{\nu}(u)=0}$ for all $j$
ensure cancellation
of the ``unwanted terms'' in (\ref{a5a}) and are equivalent to the Bethe equations
(\ref{Bethe}) for the parameters $v_j$.

Analytic properties of the left Bethe vectors are similar to those of the right ones.
They are regular functions of the parameters $v_i$.
Using (\ref{p2a}), we have:
\beq\label{qua3}
\begin{array}{l}
\bar C_{l-j,l+j}(u+1)=\bar C_{l-j,l+j}(u),
\\ \\
\bar C_{l-j,l+j}(u+\tau)=-e^{-\pi i(s+t)-2\pi i l\eta -\pi i c(u)}\bar C_{l-j,l+j}(u).
\end{array}
\eeq
It then follows that
 \beq\label{qua4}
\begin{array}{l}
\bigl <\Psi_{\nu}(v_1,\dots,v_{j-1},v_j+1,v_{j+1},\dots,v_n)\bigr | =
\bigl <\Psi_{\nu}(v_{1},\dots,v_n)\bigr |,
\\ \\
\bigl <\Psi_{\nu}(v_1,\dots,v_{j-1},v_j+\tau,v_{j+1},\dots,v_n)\bigr |=
-e^{-\pi i(s+t)-\pi i c(v_j)}
\bigl <\Psi_{\nu -2}(v_{1},\dots,v_n)\bigr |.
\end{array}
\eeq

\subsection{Action of the operators ${\sf U}_a$ to Bethe vectors}

\label{section:action}

The key identity necessary to derive how the operators ${\sf U}_a$ act to
(in general off-shell) Bethe vectors is
\beq\label{ao1}
\sigma_a \otimes {\sf U}_a {\cal T}(u)={\cal T}(u)\, \sigma_a \otimes {\sf U}_a,
\eeq
which follows from (\ref{sigmaa}) (here $\sigma_a$ acts in the auxiliary space
$\CC^2$ and
${\sf U}_a$ acts in the quantum space ${\cal H}$), or, in more detail:
$$
\begin{aligned}
\left (\begin{array}{cc}
A(u){\sf U}_1 & B(u){\sf U}_1 \\
C(u){\sf U}_1 & D(u){\sf U}_1
\end{array} \right )&=
\left (\begin{array}{cc}
{\sf U}_1 D(u) & {\sf U}_1 C(u)\\
{\sf U}_1 B(u) & {\sf U}_1 A(u)
\end{array} \right ),\\
\left (\begin{array}{cc}
A(u){\sf U}_3 & B(u){\sf U}_3 \\
C(u){\sf U}_3 & D(u){\sf U}_3
\end{array} \right )&=
\left (\begin{array}{rr}
{\sf U}_3 A(u) & -{\sf U}_3 B(u)\\
-{\sf U}_3 C(u) & {\sf U}_3 D(u)
\end{array} \right ).
\end{aligned}
$$
It then follows that
\begin{align}\label{ao2}
{\sf U}_1 B_{k,l}(u;s,t)&=e^{-\pi i(2u+(k+l)\eta +2s+\tau)}
B_{k,l}(u; s+\tau, t+\tau){\sf U}_1,\\
\label{ao3}
\bar C_{k,l}(u;s,t){\sf U}_1& =e^{-\pi i(-2u+(k+l)\eta +2t+\tau)}{\sf U}_1
\bar C_{k,l}(u; s+\tau, t+\tau),\\
\label{ao4}
{\sf U}_3 B_{k,l}(u;s,t)&=-B_{k,l}(u; s+1, t+1){\sf U}_3,\\
\label{ao5}
\bar C_{k,l}(u;s,t){\sf U}_3 & =-{\sf U}_3 \bar C_{k,l}(u; s+1, t+1),
\end{align}
and also
\begin{align}\label{ao2a}
{\sf U}_1 B_{k,l}(u;s,t)&=-e^{-\pi i(t_k+t_l+s-t)}
B_{k,l}(u; s+\tau, t-\tau){\sf U}_1,\\
\label{ao3a}
\bar C_{k,l}(u;s,t){\sf U}_1& =-e^{\pi i(s_k+s_l +t-s)}{\sf U}_1
\bar C_{k,l}(u; s+\tau, t-\tau),\\
\label{ao4a}
{\sf U}_3 B_{k,l}(u;s,t)&=B_{k,l}(u; s+1, t-1){\sf U}_3,\\
\label{ao5a}
\bar C_{k,l}(u;s,t){\sf U}_3 & ={\sf U}_3 \bar C_{k,l}(u; s+1, t-1).
\end{align}
It is also straightforward to find how the operators ${\sf U}_a$ act
to the vacua:
\begin{align}\label{ao6}
{\sf U}_1\! \left |\Omega^{l-n}(s)\right >&=(-1)^n e^{\pi i
\bigl (Ns_l -n\eta +\sum_k\xi_k +n\tau \bigr )}
\left |\Omega^{l-n}(s+\tau)\right >,\\
\label{ao7}
\left <\bar \Omega^{l-n}(t)\right |{\sf U}_1&=(-1)^n e^{\pi i
\bigl (Nt_l +n\eta -\sum_k\xi_k +n\tau \bigr )}
\left <\bar \Omega^{l-n}(t+\tau)\right |,\\
\label{ao8}
{\sf U}_3\! \left |\Omega^{l-n}(s)\right >&=
\left |\Omega^{l-n}(s+1)\right >,\\
\label{ao9}
\left <\bar \Omega^{l-n}(t)\right |{\sf U}_3&=
\left <\bar \Omega^{l-n}(t+1)\right |.
\end{align}

Combining equations
(\ref{ao2})--(\ref{ao9}), one can derive the following properties of
the Bethe vectors:
\begin{align}\label{ao10}
{\sf U}_1\! \bigl |\Psi_{\mu}(u_1, \ldots , u_n;s,t)\bigr >&=(-1)^n e^{-2\pi i
\sigma (u_1, \ldots , u_n)}\bigl |\Psi_{\mu}(u_1, \ldots , u_n;s+\tau,t+\tau)\bigr >,
\\
\label{ao11}
\bigl <\Psi_{\nu}(v_1, \ldots , v_n;s,t)\bigr |{\sf U}_1&=(-1)^n e^{2\pi i
\sigma (v_1, \ldots , v_n)}
\bigl <\Psi_{\nu}(v_1, \ldots , v_n;s+\tau,t+\tau)\bigr |,
\\
\label{ao12}
{\sf U}_3\! \bigl |\Psi_{\mu}(u_1, \ldots , u_n;s,t)\bigr >&=(-1)^n
\bigl |\Psi_{\mu}(u_1, \ldots , u_n;s+1,t+1)\bigr >,
\\
\label{ao13}
\bigl <\Psi_{\nu}(v_1, \ldots , v_n;s,t)\bigr |{\sf U}_3&=(-1)^n
\bigl <\Psi_{\nu}(v_1, \ldots , v_n;s+1,t+1)\bigr |,
\end{align}
where
\beq\label{q15}
\sigma (v_1, \ldots , v_n)
= \sum_{i=1}^n v_i -\frac{1}{2}\sum_{k=1}^N \xi_k +\frac{1}{2}\, n\eta,
\eeq
and
\begin{align}\label{ao10a}
{\sf U}_1\! \bigl |\Psi^{l}(u_1, \ldots , u_n;s,t)\bigr >&=e^{\pi in (s-t-\eta)
+\pi i n\tau +\pi i \sum_k\xi_k}\bigl |\Psi^{l}(u_1, \ldots , u_n;s+\tau,t-\tau)\bigr >,
\\
\label{ao11a}
\bigl <\Psi^{l}(v_1, \ldots , v_n;s,t)\bigr |{\sf U}_1&= e^{\pi in(s-t-\eta )
+\pi i n\tau +\pi i \sum_k\xi_k}
\bigl <\Psi^{l}(v_1, \ldots , v_n;s+\tau,t-\tau)\bigr |,
\\
\label{ao12a}
{\sf U}_3\! \bigl |\Psi^{l}(u_1, \ldots , u_n;s,t)\bigr >&=
\bigl |\Psi^{l}(u_1, \ldots , u_n;s+1,t-1)\bigr >,
\\
\label{ao13a}
\bigl <\Psi^{l}(v_1, \ldots , v_n;s,t)\bigr |{\sf U}_3&=
\bigl <\Psi^{l}(v_1, \ldots , v_n;s+1,t-1)\bigr |.
\end{align}

\subsection{The case of rational $\eta$}

For irrational values of $\eta$ the Fourier transform (\ref{a3}), (\ref{a3a}) is rather
formal because convergence of the infinite series is problematic. Formal expressions
of this kind become really meaningful for rational $\eta$,
\beq\label{eta1}
\eta =\frac{2P}{Q},
\eeq
where $P,Q$ are mutually prime integers. In this case all functions in question
become $Q$-periodic in $l$ and the infinite Fourier series (\ref{a3}) can be
substituted by the finite sum
\beq\label{a3b}
\bigl |\Psi_{\nu} (u_1, \ldots , u_n)\bigr >=
\sum_{l\in \z_Q}e^{-2\pi ilP \nu/Q}\bigl |\Psi^l (u_1, \ldots , u_n)\bigr >,
\eeq
where $\ZZ_Q=\{0, 1, \ldots , Q-1\}$. Because of the $Q$-periodicity the admissible
values of $n$ are not restricted by $n=N/2$ anymore but can be found from the condition
\beq\label{eta2}
2n=N \,\,\, (\mbox{mod $Q$}).
\eeq

The case $\eta =1/2$ ($Q=4$) is the case of free fermions when the 8-vertex model
splits into two Ising models. In this case the right hand side of Bethe equations
becomes $(-1)^{n-1}$, so the Bethe equations convert into $n$ uncoupled equations
for each $u_j$ separately.

\subsection{Dependence of the eigenvectors on $s,t$}

\label{section:dependence}

Here we investigate the dependence of the left eigenvectors $\left <\Psi_{\nu}\right |$
on the parameters $s,t$. It is natural to expect that the eigenvectors of the transfer
matrix do not essentially depend on these auxiliary parameters, i.e., all the dependence
is concentrated in the common scalar factor:
$$
\bigl <\Psi_{\nu}\bigr |=\varphi_{\nu}(x,y) \bigl <\Psi_{\nu}^{(0)}\bigr |,
$$
where we have denoted
\beq\label{xy11}
x=\frac{1}{2}\, (s+t+1)\,,\quad y=\frac{1}{2}\, (s-t)\,,
\eeq
 and
the eigenvector $\bigl <\Psi_{\nu}^{(0)}\bigr |$ does not depend on $s,t$.

We will use equations
(\ref{ao11}), (\ref{ao11a}), (\ref{ao13}), (\ref{ao13a}) taking into account
the fact that the eigenvectors of the transfer matrix are simultaneously
eigenvectors of the operators ${\sf U}_a$ ($a=1,3$).
Since ${\sf U}_a$ are involutions, their eigenvalues
are $(-1)^{\nu_a}$, where $\nu_a =0,1$:
 \beq\label{nu11}
\bigl <\Psi_{\nu}\bigr |{\sf U}_a = (-1)^{\nu_a} \bigl
<\Psi_{\nu}\bigr |.
 \eeq
 Then we have from (\ref{ao11}),
(\ref{ao11a}), (\ref{ao13}), (\ref{ao13a}): \beq\label{st1}
\begin{array}{l}
\varphi_{\nu}(x+1,y)=(-1)^{n+\nu_3}\varphi_{\nu}(x,y),
\\ \\
\varphi_{\nu}(x+\tau,y)=(-1)^{n+\nu_1}e^{-2\pi i\sigma}\varphi_{\nu}(x,y)
\end{array}
\eeq
($\sigma$ is defined in (\ref{q15})) and
\beq\label{st2}
\begin{array}{l}
\varphi_{\nu}(x,y+1)=(-1)^{\nu_3}\varphi_{\nu}x,y),
\\ \\
\varphi_{\nu}(x,y+\tau)=(-1)^{\nu_1}e^{-\pi in\tau -2\pi i ny +\pi i n \eta
-\pi i \sum_k \xi_k}\varphi_{\nu}(x,y).
\end{array}
\eeq
Besides,  it straightforwardly follows from the construction of the eigenvectors that
\beq\label{st3}
\varphi_{\nu}(x+\eta, y)=e^{-\pi i\nu\eta} \varphi_{\nu}(x,y).
\eeq

Below in section \ref{section:sumrule} we argue that $n+\nu_3=-\nu$ and
$$
2\sigma = n+\nu_1 +\nu\tau
$$
(the sum rule (\ref{q19})). Using these relations and the definition of $\sigma$,
we can represent equations (\ref{st1}), (\ref{st2}), (\ref{st3}) in the form
\beq\label{st1a}
\begin{array}{l}
\varphi_{\nu}(x+1,y)=(-1)^{\nu}\varphi_{\nu}(x,y),
\\ \\
\varphi_{\nu}(x+\tau,y)=e^{-\pi i\nu\tau}\varphi_{\nu}(x,y),
\\ \\
\varphi_{\nu}(x+\eta, y)=e^{-\pi i\nu\eta} \varphi_{\nu}(x,y)
\end{array}
\eeq
and
\beq\label{st2a}
\begin{array}{l}
\varphi_{\nu}(x,y+1)=(-1)^{n+\nu}\varphi_{\nu}(x,y),
\\ \\
\varphi_{\nu}(x,y+\tau)=(-1)^{n}e^{-\pi in\tau -2\pi i ny
+\pi i\nu\tau -2\pi i \sum_j v_j}\varphi_{\nu}(x,y).
\end{array}
\eeq
We know that $\varphi_{\nu}(x,y)$ is an entire function of $y$ and may have poles
in $x$ at the points $x=-l\eta$ in the fundamental domain, where $l=0,\ldots Q/2$ for
even $Q$ and $l=0, \ldots , Q$ for odd $Q$. It then follows from (\ref{st1a}), (\ref{st2a})
and these properties that the poles in $x$ actually cancel and as a function of $y$
$\varphi_{\nu}(x,y)$ is a theta function of order $n$, i.e. it has $n$ zeros $y_j$
in the fundamental
domain:
\beq\label{st4}
\varphi_{\nu}(x,y)=b_\nu e^{\pi i \nu(y-x)}\prod_{j=1}^n \theta_1(y-y_j|\tau )
\eeq
with the condition
$$
\sum_{j=1}^n y_j=-\sum_{j=1}^nv_j.
$$
We conjecture that $y_j=-v_j$.  This conjecture
is supported by numerical calculations for $N=2,4$.
Remarkably, the dependence on $x$ and $y$
factorizes.

\section{The $Q$-operator and the sum rule}

\label{section:qoperator}

 Here we construct the Baxter's $Q$-operator. This construction
is necessary to obtain a sum rule, which is an additional requirement to the eigenvectors
besides the Bethe equations.
 This section is not directly connected with what follows and is included for
completeness.

\subsection{Construction of the $Q$-operator}

A minor modification of the formulas in sections \ref{section:int} and \ref{section:vac}
leads to the construction of the $Q$-operator. We begin with the
construction of the right $Q$-operator ${\sf Q}_R(u)$. Basically, one should shift
$s\to s-u$, $t\to t+u$ and consider the $L$-operators
\beq\label{q1}
{\sf L}^{\pm}(u)=M_{l}^{-1}{\sf L}(u)M_{l\pm 1} =
\left ( \begin{array}{cc}
{\sf a}^{\pm}(u)& {\sf b}^{\pm}(u) \\
{\sf c}^{\pm}(u)& {\sf d}^{\pm}(u)
\end{array}
\right ),
\eeq
where
\beq\label{q2}
M_l=\left (\begin{array}{ll}
\theta_1(s_l|2\tau ) & \gamma_l \theta_1(t_l|2\tau)\\
\theta_4(s_l|2\tau ) & \gamma_l \theta_4(t_l|2\tau)
\end{array} \right ).
\eeq
The identities from section~\ref{section:int} allow one to prove the following important
relations:
\beq\label{q3}
\begin{aligned}
&{\sf c}^{\pm}(u-\xi )\bigl |\phi (s_l\pm (u-\xi ))\bigr >=0,
\\
&{\sf a}^{\pm}(u-\xi )\bigl |\phi (s_l\pm (u-\xi ))\bigr >=
\theta_1(u-\xi +\eta |\tau )\bigl |\phi (s_l\pm (u-\eta -\xi ))\bigr >,
\\
&{\sf d}^{\pm}(u-\xi )\bigl |\phi (s_l\pm (u-\xi ))\bigr >=
\theta_1(u-\xi |\tau )\bigl |\phi (s_l\pm (u+\eta -\xi ))\bigr >.
\end{aligned}
\eeq

Let $\epsilon_i=\pm 1$, $i=1, \ldots , N$ be such that
$\displaystyle{\sum_{i=1}^N \epsilon _i=0}$ (for even $N$ this is always possible)
and set $\displaystyle{e_m =\sum_{i=1}^m \epsilon _i}$, $e_0=e_N=0$.
Let us introduce a family of vectors
$$
\bigl |\omega (u; \epsilon_1, \ldots , \epsilon_N)\bigr >=
\bigotimes_{i=0, \ldots , N-1}^{\rightarrow} \bigl |\phi (
s_{l+e_i}-\epsilon_{i+1}(u-\xi_i)\bigr >.
$$
Then the relations (\ref{q3}) imply that
\beq\label{q4}
{\sf T}(u)\bigl |\omega (u; \epsilon_1, \ldots , \epsilon_N)\bigr >=
a(u)\bigl |\omega (u-\eta ; \epsilon_1, \ldots , \epsilon_N)\bigr >+
d(u)\bigl |\omega (u+\eta ; \epsilon_1, \ldots , \epsilon_N)\bigr >,
\eeq
where
\beq\label{ad}
a(u)=\prod_{i=1}^N\theta_1(u-\xi_i+\eta),\qquad d(u)=\prod_{i=1}^N\theta_1(u-\xi_i).
\eeq
The vectors $\bigl |\omega (u; \epsilon_1, \ldots , \epsilon_N)\bigr >$
can be regarded as columns of an operator ${\sf Q}_R(u)$
(a ``pre-$Q$-operator) which, therefore, satisfies
the relation
\beq\label{q5}
{\sf T}(u){\sf Q}_R(u) =a(u){\sf Q}_R(u-\eta)+d(u){\sf Q}_R(u+\eta).
\eeq
Since $s$ can be any complex number and $\epsilon_1, \ldots , \epsilon_N$ is any
set of numbers $\pm 1$ (such that their sum is zero), the set of all possible vectors
$\bigl |\omega (u; \epsilon_1, \ldots , \epsilon_N)\bigr >$ spans the total $2^N$-dimensional
quantum space ${\cal H}$ of the model.

The construction of the left $Q$-operator, ${\sf Q}_L(u)$, is similar.
We define
\beq\label{q1a}
\bar {\sf L}^{\pm}(u)=\bar M_{l}^{-1}{\sf L}(u)\bar M_{l\pm 1} =
\left ( \begin{array}{cc}
\bar {\sf a}^{\pm}(u)& \bar {\sf b}^{\pm}(u) \\
\bar {\sf c}^{\pm}(u)& \bar {\sf d}^{\pm}(u)
\end{array}
\right ),
\eeq
where
\beq\label{q2a}
\bar M_l=\left (\begin{array}{ll}
\gamma_l \theta_1(s_l|2\tau ) & \theta_1(t_l|2\tau)\\
\gamma_l \theta_4(s_l|2\tau ) & \theta_4(t_l|2\tau)
\end{array} \right ).
\eeq
The identities from section \ref{section:int} allow one to prove the
relations
\beq\label{q6}
\begin{aligned}
&\bigl <\phi^{\bot}(t_{l\pm 1}\pm (u-\xi))\bigr |\bar {\sf b}^{\pm}(u-\xi)=0,
\\
&\bigl <\phi^{\bot}(t_{l\pm 1}\pm (u-\xi))\bigr |\bar {\sf a}^{\pm}(u-\xi)=
\theta_1(u-\xi +\eta |\tau)
\bigl <\phi^{\bot}(t_{l\pm 1}\pm (u-\eta -\xi))\bigr |,
\\
&\bigl <\phi^{\bot}(t_{l\pm 1}\pm (u-\xi))\bigr |\bar {\sf d}^{\pm}(u-\xi)=
\theta_1(u-\xi |\tau)
\bigl <\phi^{\bot}(t_{l\pm 1}\pm (u+\eta -\xi))\bigr |.
\end{aligned}
\eeq
We introduce a family of dual vectors
\beq\label{q7}
\bigl <\bar \omega (u; \epsilon_1, \ldots , \epsilon_N)\bigr |=
\bigotimes_{i=1,\ldots , N}^{\rightarrow}\bigl <\phi^{\bot} (t_{l+e_i}+
\epsilon_i(u-\xi_i)\bigr |,
\eeq
which satisfy
\beq\label{q8}
\bigl <\bar \omega (u;\epsilon_1, \ldots , \epsilon_N)\bigr |{\sf T}(u)=
a(u)\bigl <\bar \omega (u-\eta ;\epsilon_1, \ldots , \epsilon_N)\bigr |
+d(u) \bigl <\bar \omega (u+\eta ;\epsilon_1, \ldots , \epsilon_N)\bigr |,
\eeq
and regard them as rows of an operator ${\sf Q}_L(u)$ which, therefore, satisfies
the relation
\beq\label{q5a}
{\sf Q}_L(u){\sf T}(u) =a(u){\sf Q}_L(u-\eta)+d(u){\sf Q}_L(u+\eta).
\eeq

Now, using the argument of Baxter's works (see \cite{Baxter-book}), one can prove the
commutation relation
\beq\label{q9}
{\sf Q}_L(u){\sf Q}_R(v)={\sf Q}_L(v){\sf Q}_R(u)
\eeq
and define the $Q$-operator ${\sf Q}(u)={\sf Q}_R(u){\sf Q}^{-1}_R(u_0)=
{\sf Q}^{-1}_L(u_0){\sf Q}_L(u)$, where $u_0$ is some point for which the
left and right $Q$-operators are invertible.
It then follows that $[{\sf Q}(u), {\sf Q}(v)]=[{\sf Q}(u), {\sf T}(v)]=0$ and
the $Q$-operator obeys the $TQ$-relation
\beq\label{q10}
{\sf T}(u){\sf Q}(u) =a(u){\sf Q}(u-\eta)+d(u){\sf Q}(u+\eta)
\eeq
which is the main property of the $Q$-operator.

\subsection{The sum rule}

\label{section:sumrule}

It is straightforward to check that
\beq\label{q11}
\begin{array}{l}
{\sf Q}_R(u+1)={\sf U}_3 {\sf Q}_R(u),
\\ \\
\displaystyle{
{\sf Q}_R(u+\tau )=e^{-\pi i c(u)/2}
{\sf U}_1{\sf Q}_R(u)},
\end{array}
\eeq
\beq\label{q12}
\begin{array}{l}
{\sf Q}_L(u+1)={\sf Q}_L(u){\sf U}_3 ,
\\ \\
\displaystyle{
{\sf Q}_L(u+\tau )=e^{-\pi i c(u)/2}
{\sf Q}_L(u){\sf U}_1},
\end{array}
\eeq
where ${\sf U}_a$ are operators
defined in (\ref{Ua}).
It follows from these relations that $[{\sf Q}(u), {\sf U}_a]=0$ and
\beq\label{q13}
\begin{array}{l}
{\sf Q}(u+1)={\sf U}_3 {\sf Q}(u),
\\ \\
\displaystyle{
{\sf Q}(u+\tau )=e^{-\pi i c(u)/2}
{\sf U}_1{\sf Q}(u)}.
\end{array}
\eeq
Let $Q(u)$ be the eigenvalue of the operator ${\sf Q}(u)$ on a common eigenfunction
with the operators ${\sf U}_a$. As we have seen before, the eigenvalues
of the operators ${\sf U}_a$ are $(-1)^{\nu_a}$, where $\nu_a =0,1$. We can write
\beq\label{q14}
\begin{array}{l}
Q(u+1)=(-1)^{\nu_3}Q(u),
\\ \\
\displaystyle{
Q(u+\tau )=(-1)^{\nu_1} e^{-\pi i c(u)/2}Q(u)}.
\end{array}
\eeq
It follows from (\ref{q14}) that the entire function $Q(u)$ has exactly
$n$ zeros $v_i$ in the fundamental domain. Let us consider a
function
$$
F(u)=\frac{Q(u)}{\prod\limits_{i=1}^n\theta_1(u-v_i|\tau)}.
$$
It is an entire function of $u$, and
equations (\ref{q14}) imply that
$$
F(u+1)=(-1)^{n+\nu_3}F(u), \quad F(u+\tau)=
(-1)^{n+\nu_1}e^{-2\pi i \sigma}F(u),
$$
where $\sigma$ is defined in (\ref{q15}).
It follows from these properties that $F(u)$ is the exponential function
$F(u)=e^{\pi i(n+\nu_3)u}$ and
\beq\label{q16}
2\sigma = n+\nu_1 -(n+\nu_3)\tau .
\eeq
Therefore,
\beq\label{q17}
Q(u)=e^{\pi i(n+\nu_3)u}\prod_{i=1}^n \theta_1(u-v_i|\tau ).
\eeq

It remains to identify zeros of the $Q(u)$ with Bethe roots.
Writing the $TQ$-relation (\ref{q10}) for the eigenvalues,
\beq\label{q18}
T(u; v_1, \ldots , v_n)=a(u)\, \frac{Q(u-\eta)}{Q(u)}+
d(u)\, \frac{Q(u+\eta)}{Q(u)},
\eeq
and using the fact that $T(u; v_1, \ldots , v_n )$ does not have poles at $u=v_i$,
one obtains the Bethe equations
\beq\label{Bethe2}
e^{-2\pi i(n+\nu_3)\eta }\, \frac{a(v_j)}{d(v_j)}=
\prod_{k=1, \neq j}\frac{\theta_1(v_j-v_k+\eta |\tau)}{\theta_1(v_j-v_k-\eta |\tau)}.
\eeq
Comparing with (\ref{Bethe}), one identifies
\beq\label{nu3}
\nu =-(n+\nu_3)\,.
\eeq
 In addition, we conclude
that the Bethe roots have to satisfy the sum rule
\beq\label{q19}
\sum_{i=1}^n v_i =\frac{1}{2}\sum_{k=1}^N \xi_k -\frac{1}{2}\, n\eta
+\frac{1}{2}\, (n+\nu_1 +\nu \tau ),
\eeq
and $\nu$ can only take integer values. Taking this into account,
one may say that the eigenvector of the transfer matrix is determined by a solution
of the {\it extended} system of Bethe equations, i.e., the system (\ref{Bethe})
supplemented by the sum rule (\ref{q19}) for the $n+1$ unknown variables
$\{\nu, v_1 , \ldots , v_n\}$.

It should be noted that
the $Q$-operator has some eigenvectors which are not eigenvectors
of the spin reflection operator (see \cite{FMC03,FMC06}).
For such states, the sum rule is not valid.

\section{Scalar products of Bethe vectors}

\label{section:scalarproducts}

 In this section, we obtain a system of linear equations for
scalar products of the on-shell and off-shell Bethe vectors.
Basically, we follow the method of \cite{BS19}. However, for models with
the 8-vertex $R$-matrix, some generalization is required.

\subsection{The notation}

In this section we denote
$\bu=\{u_1,\dots,u_{n+1}\}$, $\bv=\{v_1,\dots,v_{n}\}$, $\bw=\{w_1,\dots,w_{n+1}\}$. We also denote
$\bu_j=\bu\setminus u_j$, $\bv_k=\bv\setminus v_k$ and so on.
Let us also introduce the functions
\beq\label{gfh}
g(u,v)=\frac{\theta_1(\eta)}{\theta_1(u-v)},
\qquad f(u,v)=\frac{\theta_1(u-v+\eta)}{\theta_1(u-v)},
\qquad h(u,v)=\frac{\theta_1(u-v+\eta)}{\theta_1(\eta)}.
\eeq
Here and below in this section $\theta_1(x)\equiv \theta_1(x|\tau)$. Observe that
\beq\label{prop1}
g(u,v)=-g(v,u),\qquad f(u,v)=g(u,v)h(u,v), \qquad h(u,u)=1.
\eeq
 In order to make the formulas more compact, we use a shorthand notation for products of these functions. Namely, if any of them  depends on a (sub)set
of variables, then one should take a product with respect to the corresponding (sub)set.
For example,
\beq\label{shnprod}
f(u,\bv)=\prod_{j=1}^nf(u,v_j),\quad  h(\bw,u_k)=\prod_{j=1}^{n+1}h(w_j,u_k),
\quad g(u_k,\bu_k)=\prod_{j=1, \neq k}^{n+1}g(u_k,u_j)
\eeq
 and so on.
 We stress that this convention is applied to the functions (\ref{gfh}) only.

Finally, the functions $a(u)$, $d(u)$ are defined in (\ref{ad}).

\subsection{A system of linear equations for scalar products}

 We now proceed directly to the derivation of the system of equations for scalar products.
Using definitions \eqref{gfh} and convention \eqref{shnprod} we  rewrite equation (\ref{a4}) for the action of the operator
${\sf T}(u)=A_{l,l}(u)+D_{l,l}(u)$ as follows:
\beq\label{actADn}
\begin{array}{lll}
{\sf T}(u_j)\bigl |\Psi^{l}(\bu_j)\bigr >&=&
\displaystyle{
\sum_{k=1}^{n+1}\Biggl[a(u_k)f(\bu_k,u_k)\frac{\theta_1(u_j-u_k+
\tau_{l+1}+\frac{1}{2})}{h(u_j,u_k)\theta_1(\tau_{l+1}+\frac{1}{2})}\bigl |\Psi^{l+1}(\bu_k)\bigr >}
\\ &&\\
&&\displaystyle{\phantom{aaaa}
+\, d(u_k)f(u_k,\bu_k)\frac{\theta_1(u_j-u_k+
\tau_{l-1}+\frac{1}{2})}{h(u_k,u_j)\theta_1(\tau_{l-1}+\frac{1}{2})}\bigl |\Psi^{l-1}(\bu_k)
\bigr > \Biggr]}.
\end{array}
\eeq
Set
\beq\label{Xjl} 
X_j^l= \bigl <\Psi_{\nu}(\bv)\bigl |\Psi^{l}(\bu_j)\bigr >,
\eeq
where $\bigl <\Psi_{\nu}(\bv)\bigr |$ is a dual on-shell Bethe vector:
\beq\label{actADl}
\bigl <\Psi_{\nu}(\bv)\bigr |{\sf T}(u_j)=T_{\nu}(u_j,\bv)
\bigl <\Psi_{\nu}(\bv)\bigr | \quad \mbox{for all $u_j\in\CC$}.
\eeq
This means that the parameters $\bv$ are supposed to satisfy the Bethe equations (\ref{Bethe}).
The eigenvalue is given by
\beq\label{Eival}
T_{\nu}(u_j,\bv)=e^{i\pi \eta \nu}a(u_j)f(\bv,u_j)+e^{-i\pi \eta \nu}d(u_j)f(u_j,\bv).
\eeq
Multiplying (\ref{actADn}) from the left by $\langle\Psi_{\nu}(\bv)|$ and using (\ref{actADl}),
we obtain:
\beq\label{SP1}
\begin{array}{l}
\displaystyle{
\sum_{k=1}^{n+1}\Biggl[a(u_k)f(\bu_k,u_k)\frac{\theta_1(u_j-u_k+\tau_{l+1}+
\frac{1}{2})}{h(u_j,u_k)\theta_1(\tau_{l+1}+\frac{1}{2})}X^{l+1}_k}+\\ \\
\displaystyle{
\phantom{aaaaaaa}
+\, d(u_k)f(u_k,\bu_k)\frac{\theta_1(u_j-u_k+\tau_{l-1}+\frac{1}{2})}{h(u_k,u_j)
\theta_1(\tau_{l-1}+\frac{1}{2})}X^{l-1}_k-\delta_{jk}T_{\nu}(u_j,\bv)X^{l}_k\Biggr ]=0}.
\end{array}
\eeq
We recall that $\tau_l =\frac{1}{2}\, (s_l+t_l)=x-\frac{1}{2}+l\eta$.
This is a homogeneous system of linear equations for the scalar products,
similar to the one familiar from the rational and trigonometric cases \cite{BS19}
but with an
additional integer parameter $l$.

In what follows we assume that $\eta$ is a rational number
$\eta =2P/Q$ (\ref{eta1}).
In this case $X_k^l$ and coefficients of the system (\ref{SP1}) are $Q$-periodic
in $l$ and the index $l$ in (\ref{SP1}) should be understood modulo $Q$
($l+Q=l$). Therefore, \eqref{SP1} is a homogeneous system of $(n+1)Q$ linear
equations for $(n+1)Q$ unknown variables $X_k^l$, $k=1, \ldots n+1$,
$l\in \ZZ_Q$, $\ZZ_Q=\{0,\ldots , Q-1\}$. In the next subsection we show that
this system has non-trivial solutions.

Below we consider only the case when the number of Bethe parameters in
the two vectors in the scalar product (\ref{Xjl}) is the same and is equal to
$N/2$ although for rational $\eta$ this number can also take other values
(see \eqref{eta2}).

\subsection{Transformation of the system and solvability}

 Since the obtained system of equations \eqref{SP1}
is homogeneous, its solutions (if any) are ambiguous.
Namely, if $X_k^l$ is a solution to the system, then
$\phi(\bv,\bu)X_k^l$ is also a solution to the system,
where $\phi(\bv,\bu)$ is an arbitrary function of the variables
$\bv$ and $\bu$. In order to minimize possible
arbitrariness, we, following the method of \cite{BS19},
transform the system to a new (equivalent) form and show that the
solutions of the new system are determined up to a function that
depends on the variables $\bv$, but does not
depend on the parameters $\bu$.
As a byproduct, we prove that the rank of the system is less than $(n+1)Q$, and
therefore, it does have nontrivial solutions.

Let us introduce $(n+1)\! \times \! (n+1)$ matrices $W^l$ with the entries
\beq\label{Wjk}
W_{jk}^l=g(u_k,w_j)\frac{g(u_k,\bu_k)}{g(u_k,\bw)}\, \theta_1
\Bigl (u_k-w_j-S-\tau_l-\frac{1}{2}\Bigr ),\quad j,k=1,\ldots,n+1, \quad l\in\ZZ_Q ,
\eeq
where $\bw=\{w_1,\dots,w_{n+1}\}$ are arbitrary pairwise distinct complex numbers and
\beq\label{S}
S=\sum_{j=1}^{n+1}(u_j-w_j).
\eeq
The matrix $W^l$ is nothing else than an elliptic Cauchy matrix multiplied by a diagonal
matrix from the right. The determinant of the elliptic Cauchy matrix is given by
\beq\label{int8}
\det\limits_{1\leq i,j\leq n+1}\Phi (u_i-w_j, \lambda )
=\frac{(\theta_{1}'(0))^{n+1}
\theta_1 \Bigl (\lambda +\sum\limits_{i=1}^{n+1} (u_i-w_j)\Bigr )}{\theta_1 (\lambda )}\,
\frac{\prod\limits_{p<q}\theta_1 (u_p-u_q)\theta_1
(w_q-w_p)}{\prod\limits_{r,s}\theta_1 (u_r-w_s)},
\eeq
where $\Phi$ is the function (\ref{a2}). It is seen from this formula that $\det W^l \neq 0$
if all $u_j$ and $w_j$ are distinct and $\tau_l +\frac{1}{2}\neq 0$,
$S+\tau_l +\frac{1}{2}\neq 0$ modulo the lattice spanned by $1$ and $\tau$.

Multiplying the system (\ref{SP1}) from the left by $W^l$ we obtain:
\begin{equation}\label{SP2}
\sum_{k=1}^{n+1}\Biggl[\frac{a(u_k)f(\bu_k,u_k)}{\theta_1(\tau_{l+1}+\frac{1}{2})}
E^+_{jk} X^{l+1}_k
+\frac{d(u_k)f(u_k,\bu_k)}{\theta_1(\tau_{l-1}+\frac{1}{2})}
E^-_{jk}X^{l-1}_k -W_{jk}^lT_{\nu}(u_k,\bv)X^{l}_k\Biggr]=0.
\end{equation}
Here
\beq\label{Epm}
E^\pm_{jk}=\theta_1(\pm\eta)\sum_{m=1}^{n+1}W^l_{jm}
\frac{\theta_1(u_m-u_k+\tau_{l\pm1}+\frac{1}{2})}{\theta_1(u_m-u_k\pm\eta)}.
\eeq
As soon as the matrix $W^l$ is non-degenerate, the new system is equivalent to
the previous one.

The sum (\ref{Epm}) can be calculated via  an auxiliary contour integral. Let
%
$$
I^\pm=\frac{\theta_1(\pm\eta)\theta'_1(0)}{2\pi i}\times
\oint
\frac{\theta_1(z-u_k+\tau_{l}\pm\eta+\frac{1}{2})}{\theta_1(z-u_k\pm\eta)}
\frac{\theta_1(z-w_j-S-\tau_{l}-\frac{1}{2})}{\theta_1(z-w_j)}\prod_{p=1}^{n+1}
\frac{\theta_1(z-w_p)}{\theta_1(z-u_p)}\,dz,
$$
%
where the integration goes along the boundary of the
fundamental parallelogram. Then $I^\pm=0$ due to the periodicity of the
 integrand.
On the other hand, this integral can be calculated
as sum of the residues in the interior of the contour.
It is easy to see that the sum of the residues at the
points $z=u_m$ gives exactly $E^\pm_{jk}$.
One more contribution comes from the pole at $z-u_k\pm\eta=0$. Thus we arrive at
\beq\label{0Epm}
0=E^\pm_{jk}+\theta_1(\pm\eta)\theta_1(\tau_{l}+\frac{1}{2})
\frac{\theta_1(u_k-w_j-S-\tau_{l\pm1}-\frac{1}{2})}{\theta_1(u_k-w_j\mp\eta)}
\prod_{p=1}^{n+1}\frac{\theta_1(u_k-w_p\mp\eta)}{\theta_1(u_k-u_p\mp\eta)},
\eeq
leading to
\beq\label{Epmres}
\begin{array}{l}
\displaystyle{E^+_{jk}=\theta_1 \Bigl (u_k-w_j-S-
\tau_{l+1}-\frac{1}{2}\Bigr )\frac{\theta_1(\tau_{l}+\frac{1}{2})}{h(w_j,u_k)}
\frac{h(\bw,u_k)}{h(\bu,u_k)}},\\ \\
\displaystyle{
E^-_{jk}=\theta_1 \Bigl (u_k-w_j-S-\tau_{l-1}-\frac{1}{2}\Bigr )\frac{\theta_1
(\tau_{l}+\frac{1}{2})}{h(u_k,w_j)}\frac{h(u_k,\bw)}{h(u_k,\bu)}}.
\end{array}
\eeq
Substituting these expressions into (\ref{SP2}), we obtain:
\beq\label{SP3}
\begin{array}{c}
\displaystyle{\sum_{k=1}^{n+1}g(u_k,\bu_k)\Biggl[(-1)^n a(u_k)h(\bw,u_k)
\frac{\theta_1(u_k-w_j-S_{l+1})}{\theta_1(\tau_{l+1} +\frac{1}{2})h(w_j,u_k)}X^{l+1}_k}\\ \\
\displaystyle{
+\, d(u_k)h(u_k,\bw)\frac{\theta_1(u_k-w_j-S_{l-1})}{\theta_1(\tau_{l-1} +
\frac{1}{2})h(u_k,w_j)}X^{l-1}_k}
\end{array}
\eeq
$$
\displaystyle{
-\, \frac{\theta_1(u_k-w_j-S_{l})}{\theta_1(\tau_{l} +\frac{1}{2})}\,
\frac{g(u_k,w_j)}{g(u_k,\bw)}\Bigl(e^{i\pi \eta \nu}a(u_k)f(\bv,u_k)
+e^{-i\pi \eta \nu}d(u_k)f(u_k,\bv)\Bigr)X^{l}_k\Biggr]=0,}
$$
where
\beq\label{sl}
S_l=S+\tau_l+\frac{1}{2}.
\eeq
This is a new system of equations which contains the set of arbitrary complex parameters $\bw$.
Note that they may depend on $l$.

Let us set $\bw_{n+1}=\bv$, while the parameter $w_{n+1}$ remains free.  Set
\beq\label{P}
P_l=\sum_{j=1}^{n+1}u_j- \sum_{j=1}^{n}v_j+\tau_l+\frac{1}{2}.
\eeq
Consider equations (\ref{SP3}) for $j=n+1$. We have $Q$ equations of the form
\begin{multline}\label{SPvn1}
\sum_{k=1}^{n+1}\Biggl[(-1)^na(u_k)h(\bv,u_k)\Biggl(
\frac{\theta_1(u_k-P_{l+1})}{\theta_1(\tau_{l+1}+\frac{1}{2})}\, X^{l+1}_k-
e^{i\pi \eta \nu}\frac{\theta_1(u_k-P_{l})}{\theta_1(\tau_l+\frac{1}{2})}\, X^{l}_k\Biggr)\\
+\, d(u_k)h(u_k,\bv)\Biggl(
\frac{\theta_1(u_k-P_{l-1})}{\theta_1(\tau_{l-1}+\frac{1}{2})}\, X^{l-1}_k-
e^{-i\pi \eta \nu}\frac{\theta_1(u_k-P_{l})}{\theta_1(\tau_l+\frac{1}{2})}\,
X^{l}_k\Biggr)\Biggr]g(u_k,\bu_k)=0.
\end{multline}
It is easy to see that these equations are linearly dependent. Indeed,
if we multiply (\ref{SPvn1}) by $e^{-il\pi \eta \mu}$, where
$\mu \in \ZZ_Q$, and sum over $l\in\ZZ_Q$, we
immediately see that the l.h.s. vanishes for $\mu =\nu$
(and does not vanish for $\mu\neq\nu$).  Hence, the system does have
non-trivial solutions.

Consider the remaining system for $j<n+1$ and
$\bw_{n+1}=\bv$.  The parameter $w_{n+1}$ is still free and we denote
it by $w$. Then we have $nQ$ equations of the form
\beq\label{SPvj}
\begin{array}{c}
\displaystyle{
\sum_{k=1}^{n+1}g(u_k,\bu_k)\Biggl[\mathcal{A}_k\Bigl(\frac{h(w,u_k)}{h(v_j,u_k)}
\frac{\theta_1(u_k\! -\! v_j\! +\! w\! -\! P_{l+1})}{\theta_1
(\tau_{l+1}+\frac{1}{2})}\,X^{l+1}_k}
\\ \\
\displaystyle{
\phantom{aaaaaaaaaaaaaaaaaaaaaa}-e^{i\pi \eta \nu}\frac{g(v_j,u_k)}{g(w,u_k)}
\frac{\theta_1(u_k\! -\! v_j\! +\! w\! -\! P_{l})}{\theta_1(\tau_l+\frac{1}{2})}\,
X^{l}_k\Bigr)}
\\ \\
\displaystyle{
+\mathcal{D}_k\Bigl(\frac{h(u_k,w)}{h(u_k,v_j)}
\frac{\theta_1(u_k\! -\! v_j\! +\! w\! -\! P_{l-1})}{\theta_1
(\tau_{l-1}+\frac{1}{2})}\,  X^{l-1}_k}\phantom{aaaaaaaaaaaaaaaa}
\\ \\
\displaystyle{\phantom{aaaaaaaaaaaaaaaa}
-e^{-i\pi \eta \nu}\frac{g(u_k,v_j)}{g(u_k,w)}
\frac{\theta_1(u_k\! -\! v_j\! +\! w\! -\! P_{l})}{\theta_1
(\tau_{l}+\frac{1}{2})}\,X^{l}_k
\Bigr)\Biggr]=0},
\end{array}
\eeq
where $j=1,\ldots,n$, $l=0,1, \ldots , Q-1$ and
\beq\label{calAD}
\mathcal{A}_k=  (-1)^na(u_k)h(\bv,u_k),\qquad \mathcal{D}_k=d(u_k)h(u_k,\bv).
\eeq
Now let us transform the system (\ref{SPvj}) further. We will
take advantage of the fact that the parameter $w$ may depend on $l$ and take it to be
\beq\label{w1}
w=l\eta +w_0 +\check U,
\quad \check U=\sum_{j=1}^{n+1}u_j,
\eeq
where $w_0$ is an $l$-independent free parameter.
Then the system (\ref{SPvj}) becomes:
\beq\label{SPvj1}
\hspace{-2cm}\begin{array}{c}
\displaystyle{
\sum_{k=1}^{n+1}g(u_k,\bu_k)\Biggl[\mathcal{A}_k\Bigl(
\frac{\theta_1(u_k\! -\! \check U\! -\! w_0\!-\! (l+1)\eta )}{\theta_1(\tau_{l+1}+\frac{1}{2})}\,
\frac{\theta_1(u_k\! -\! v_j\! +\! r\! -\! \eta)}{\theta_1 (u_k-v_j-\eta )}\,
X^{l+1}_k} \\ \\
\displaystyle{\phantom{aaaaaaaaaaaaaaaaaaaaaa}
-e^{i\pi \eta
\nu}\frac{\theta_1(u_k\! -\! \check U\! -\!w_0\!-\! l\eta )}{\theta_1(\tau_l+\frac{1}{2})}\,
\frac{\theta_1(u_k\! -\! v_j\! +\! r)}{\theta_1(u_k-v_j)}\,
X^{l}_k\Bigr)}\\ \\
\displaystyle{
+\mathcal{D}_k\Bigl(
\frac{\theta_1(u_k\! -\! \check U-\!w_0\!-\! (l-1)\eta )}{\theta_1(\tau_{l-1}+\frac{1}{2})}\,
\frac{\theta_1(u_k\! -\! v_j\! +\! r\! +\! \eta)}{\theta_1 (u_k-v_j+\eta )}\,
X^{l-1}_k}\\ \\
\displaystyle{\phantom{aaaaaaaaaaaaaaaaaaaaaaaaaaa}
-e^{-i\pi \eta
\nu}\frac{\theta_1(u_k\! -\! \check U\! -\!w_0\!-\! l\eta )}{\theta_1(\tau_l+\frac{1}{2})}\,
\frac{\theta_1(u_k\! -\! v_j\! +\! r)}{\theta_1(u_k-v_j)}\,
X^{l}_k
\Bigr)\Biggr]=0},
\end{array}
\eeq
where
\beq\label{w4}
r=\sum_{p=1}^n v_p -\frac{1}{2}\, (s\! +\! t\! +\! 1)+w_0.
\eeq
Multiplying these equations by $e^{-il \pi \eta \mu}$ and summing over $l\in\ZZ_Q$, we obtain the
following system of $nQ$ equations
\beq\label{w2}
\begin{array}{l}
\displaystyle{
\sum_{k=1}^{n+1}g(u_k,\bu_k)\Biggl[
\mathcal{A}_k\left (e^{i\pi \eta \mu}\frac{\theta_1(u_k-v_j -\eta +r)}{\theta_1(u_k-v_j-\eta )}-
 e^{i\pi \eta \nu} \frac{\theta_1(u_k-v_j +r)}{\theta_1(u_k-v_j )}\right )}
\\ \\
\phantom{aaaaaaa}
\displaystyle{
+\mathcal{D}_k\left (e^{-i\pi \eta \mu}\frac{\theta_1(u_k-v_j +
\eta +r)}{\theta_1(u_k-v_j+\eta )}-
 e^{-i\pi \eta \nu}\frac{\theta_1(u_k-v_j +r)}{\theta_1(u_k-v_j )}\right )\Biggr ]Y^{(\mu)}_k=0}
 \end{array}
 \eeq
for $(n+1)Q$ variables
\beq\label{w3}
\begin{array}{lll}
Y^{(\mu)}_k&=&\displaystyle{
\sum_{l\in \z_Q}\frac{\theta_1 (l\eta +w_0 +\check U -u_k )}{\theta_1(\tau_l+\frac{1}{2})}\,
e^{-il \pi \eta \mu }X_k^l}
\\ &&\\
&&\phantom{aaaaaaaaa}=\, \displaystyle{
\sum_{l\in \z_Q}\frac{\theta_1 (l\eta +r+x +U_k -V)}{\theta_1(l\eta +x)}\,
e^{-il \pi \eta \mu }X_k^l}.
\end{array}
\eeq
Here $\mu \in \ZZ_Q$, $x =\frac{1}{2}\,(s+t+1)$ and
\beq\label{Uk}
U_k=\sum_{a=1, \neq k}u_a =\check U-u_k, \qquad V=\sum_{a=1}^n v_a.
\eeq
Comparing with (\ref{SPvj1}), the system (\ref{w2}) is block-diagonal; the $Q$ diagonal blocks
are numbered by $\mu \in \ZZ_Q$ and each block is a system of $n$ equations for
$n+1$ variables $Y^{(\mu)}_k$.
Note that $Y^{(\mu)}_k$ does not depend on $u_k$ but does depend on $r$
(although $X_k^l$ does not depend on it), so we can
denote it as $Y^{(\mu)}_k=Y^{(\mu)}_k(r)$.
We can represent the system (\ref{w2}) in the form
\beq\label{w5}
\sum_{k=1}^{n+1}T^{(\nu \mu )}_{ik}(r)G_kY^{(\mu)}_k(r)=0, \quad i=1, \ldots , n,
\eeq
where
$$
G_k= \frac{g(u_k, \bar u_k)}{g(u_k, \bar v)}
$$
and the matrix $T^{(\nu \mu )}_{ik}(r)$ is given by
\beq\label{w6}
T^{(\nu \mu )}_{ik}(r)=\Phi (u_k-v_i, r)
\Bigl (T_{\nu}(u_k, \bar v)-T_{\mu}(u_k, \bar v_i \cup (v_i-r))\Bigr ).
\eeq
Here we have used the function $\Phi$ which is defined in (\ref{a2}).
Another convenient representation of the matrix $T^{(\nu \mu )}_{ik}(r)$ is
$$
\displaystyle{T^{(\nu \mu)}_{ik}(r)=\frac{\theta_1'(0)}{\theta_1(r)}\left [
a(u_k)f(\bar v, u_k)\left (e^{i\pi \eta \nu}
\frac{\theta_1(u_k-v_i+r)}{\theta_1(u_k-v_i)}-
e^{i\pi \eta \mu}\frac{\theta_1(u_k-v_i-\eta +r)}{\theta_1(u_k-v_i-\eta )}\right )\right.}
$$
\beq\label{Tnumu}
\begin{array}{c}
\displaystyle{\left. +\, d(u_k)f(u_k, \bar v)\left (e^{-i\pi \eta \nu}
\frac{\theta_1(u_k-v_i+r)}{\theta_1(u_k-v_i)}-
e^{-i\pi \eta \mu}\frac{\theta_1(u_k-v_i+\eta +r)}{\theta_1(u_k-v_i+\eta )}\right )\right ]}.
\end{array}
\eeq
Note that
\beq\label{w7}
T^{(\nu \nu)}_{ik}(0)=\frac{\p T_{\nu}(u_k, \bar v)}{\p v_i}.
\eeq
It is interesting to note that this form of the matrix is the same as in
  models with the 6-vertex $R$-matrix (see \cite{S89,KitMT99}).

Let us show that the remaining equations (\ref{SPvn1}) of the original system
follow from (\ref{w5}). Indeed, making the Fourier transform of the system
(\ref{SPvn1}), we get $Q$ equations
\beq\label{w5b}
\sum_{k=1}^{n+1}(T_{\nu}(u_k)-T_{\mu}(u_k))G_kY^{(\mu)}_k(0)=0, \quad \mu\in \ZZ_Q
\eeq
(one of them, at $\mu=\nu$, is trivial). The system (\ref{w5}) should be satisfied
for any $r$, including $r=0$ (at this point the system degenerates but the limit of the
solution as $r\to 0$ is still a solution). At the same time
$$
\lim_{r\to 0}\Bigl ( \theta_1(r)T_{ik}^{(\nu\mu)}(r)\Bigr )=T_{\nu}(u_k)-T_{\mu}(u_k),
$$
hence we see that equations (\ref{w5b}) are satisfied for solutions of the system
(\ref{w5}).

\subsection{Solution of the system}

Fixing some $k\in \{1, \ldots , n+1\}$ and writing the system (\ref{w5}) as
$$
\sum_{j=1, \neq k}^{n+1}T^{(\nu \mu )}_{ij}(r)G_jY^{(\mu)}_j(r)=
-T^{(\nu \mu )}_{ik}(r)G_kY^{(\mu)}_k(r),
$$
we can use the Cramer's rule to write
down the solution in the form
\beq\label{sol}
G_mY^{(\mu)}_m=(-1)^{k-m}G_kY_k^{(\mu)}
\frac{\det\limits_{j\neq m}T_{ij}^{(\nu\mu)}(r)}{\det\limits_{j\neq k}T_{ij}^{(\nu\mu)}(r)}.
\eeq
It is not difficult to verify that
$$
\frac{G_k}{G_m}=(-1)^{k-m}\frac{W_n(\bar u_k, \bar v)}{W_n(\bar u_m, \bar v)},
$$
where
\beq\label{w9}
\displaystyle{
W_n(\bar u_k, \bar v)=\frac{\prod\limits_{a<b, \neq k}\theta_1(u_a-u_b)\prod\limits_{a'>b'}
\theta_1(v_{a'}-v_{b'})}{\prod\limits_{p=1, \neq k}^{n+1}
\prod\limits_{p'=1}^n\theta_1 (u_p-v_{p'})}.}
\eeq
It then follows from (\ref{sol}) that
\beq\label{sol1}
Y^{(\mu)}_k \frac{W_n(\bar u_k,\bar v)}{\det\limits_{j\neq k}T^{(\nu \mu)}_{ij}(r)}=
Y^{(\mu)}_m \frac{W_n(\bar u_m,\bar v)}{\det\limits_{j\neq m}T^{(\nu \mu)}_{ij}(r)}
\eeq
for any $k,m=1, \ldots , n+1$.
The left hand side does not depend on $u_k$ (but may depend on the other variables)
while the right hand side does not depend on $u_m$ (but may depend on the other variables).
Since this is true for any $k,m$,
the both sides in fact do not depend on the variables $\bar u$ at all and
we conclude that
\beq\label{w8}
Y^{(\mu)}_k(r)=\phi^{(\nu ,\mu)}(\bar v,r)
\frac{\det\limits_{j\neq k}T^{(\nu \mu)}_{ij}(r)}{W_n(\bar u_k, \bar v)},
\eeq
where
$\phi^{(\nu ,\mu)} (\bar v,r)$ is some symmetric function of the variables $\bar v$
and a function of $r, \mu , \nu$ (and of $x,y$). This is the solution of the system (\ref{w2}).
Since $Y^{(\mu)}_k(r)$ depends also on $\nu$, below we will sometimes denote it as
$Y_k^{(\nu, \mu)}(r)$.

\subsection{Trying to fix the ambiguity}

In the models with 6-vertex $R$-matrix, the way to fix the unknown function
$\phi$ in (\ref{w8}) is to compare the result (\ref{w8}) with a very particular case
of the scalar product
when $u_k=\xi_k$, $k=1, \ldots , n$ (and so $d(u_k)=0$ for all $k$). In this case
the scalar product is known independently and is expressed through the partition function
of the  6-vertex model with domain wall boundary conditions. The latter is known to
have a determinant representation \cite{I87} which is to be compared with (\ref{w8})
in this particular case \cite{BS19}. Unfortunately, this method does not work for the  8-vertex model,
because the scalar products even in the particular case $u_k=\xi_k$ are not available
in the explicit form.

In order to fix the function $\phi^{(\nu ,\mu)} (\bar v,r)$,
we are going to analyze transformation properties of both sides of
equation (\ref{w8}) under shifts of the variables. This will allow us to fix
the function $\phi^{(\nu ,\mu)} (\bar v,r)$ only partially but, as we will see later,
this is enough for specially normalized scalar products.

First let us analyze transformation properties under the shifts
$r\to r+1$, $r\to r+\tau$.
In this section $r$ is regarded as an independent free parameter.
Clearly, $X_k^l$ does not depend on $r$. Therefore, from (\ref{w3})
we conclude that
\beq\label{fix1}
\begin{array}{l}
Y_k^{(\nu, \mu)}(r+1)=-Y_k^{(\nu, \mu)}(r),
\\ \\
Y_k^{(\nu, \mu)}(r+\tau)= -e^{-\pi i \tau -2\pi i(r+x +U_k-V)}
Y_k^{(\nu, \mu+2)}(r).
\end{array}
\eeq
It is straightforward to check that
\beq\label{fix3}
\begin{array}{l}
\det\limits_{j\neq k}T_{ij}^{(\nu , \mu)}(r+1)=\det\limits_{j\neq k}T_{ij}^{(\nu , \mu)}(r),
\\ \\
\det\limits_{j\neq k}T_{ij}^{(\nu , \mu)}(r+\tau)=e^{2\pi i(V-U_k)}
\det\limits_{j\neq k}T_{ij}^{(\nu , \mu +2)}(r).
\end{array}
\eeq
Substituting (\ref{fix1}), (\ref{fix3}) into the solution (\ref{w8}), one obtains:
\beq\label{fix4}
\begin{array}{l}
\phi^{(\nu, \mu)}(\bar v, r+1)=-\phi^{(\nu, \mu)}(\bar v, r),
\\ \\
\phi^{(\nu, \mu -2)}(\bar v, r+\tau)=-e^{-\pi i\tau -2\pi i r -2\pi i x}
\phi^{(\nu, \mu)}(\bar v, r).
\end{array}
\eeq

Some more information about the function $\phi^{(\nu,\mu)}(\bar v, r)$
can be
obtained by analyzing properties of the solution (\ref{w8})
under shifts of the parameters $s,t$.
First let us consider the shifts
$s\to s+1,\, t\to t+1$
and $s\to s+\tau,\, t\to t+\tau$ (i.e., $x\to x+1$ and $x\to x+\tau$).
For this, we use the results of section \ref{section:action}.
Note that since $r=V+w_0-x$, the shift of $x$ at constant $w_0$
should be accompanied by
the corresponding shift of $r$.
We have:
\beq\label{fix12}
\begin{array}{l}
X_k^l(x+1)=X_k^l(x),
\\ \\
X_k^l(x+\tau)=e^{2\pi i(U_k-V)}X_k^l(x)
\end{array}
\eeq
and thus
\beq\label{fix13}
\begin{array}{l}
Y_k^{(\mu)}(r-1,x+1)=-Y_k^{(\mu)}(r,x),
\\ \\
Y_k^{(\mu)}(r-\tau, x+\tau)=-e^{\pi i \tau +2\pi ix+2\pi i(U_k-V)}Y_k^{(\mu-2)}(r,x).
\end{array}
\eeq
We also have
$$
\det_{j\neq k}T_{ij}^{(\nu\mu)}(r-\tau)=e^{2\pi i(U_k-V)}
\det_{j\neq k}T_{ij}^{(\nu ,\mu -2)}(r).
$$
Substituting this into the solution (\ref{w8}), we find:
\beq\label{fix14}
\begin{array}{l}
\phi^{(\nu, \mu)}(r-1,x+1)=-\phi^{(\nu, \mu)}(r,x),
\\ \\
\phi^{(\nu, \mu +2)}(r-\tau ,x+\tau)=-e^{\pi i\tau +2\pi i x}\phi^{(\nu, \mu)}(r,x).
\end{array}
\eeq
For brevity, the dependence on $\bar v$ is omitted in the notation.
The other possibility is to shift $w_0$ simultaneously with $x$, so that
$r$ remains constant. In this way we get:
\beq\label{fix13a}
\begin{array}{l}
Y_k^{(\mu)}(r,x+1)=Y_k^{(\mu)}(r,x),
\\ \\
Y_k^{(\mu)}(r, x+\tau)=e^{-2\pi i r}Y_k^{(\mu)}(r,x),
\\ \\
Y_k^{(\mu)}(r, x+\eta)=e^{i\pi \eta (\mu-\nu)}Y_k^{(\mu)}(r,x).
\end{array}
\eeq
These properties imply the following properties of the function
$\phi^{(\nu,\mu)}(r,x)$ under shifts of $x$:
\beq\label{fix13b}
\begin{array}{l}
\phi^{(\nu, \mu)}(r,x+1)=\phi^{(\nu, \mu)}(r,x),
\\ \\
\phi^{(\nu, \mu)}(r, x+\tau)=e^{-2\pi i r}\phi^{(\nu, \mu)}(r,x),
\\ \\
\phi^{(\nu, \mu)}(r, x+\eta)=e^{i\pi \eta (\mu-\nu)}\phi^{(\nu, \mu)}(r,x).
\end{array}
\eeq

In order to fix the dependence of $\phi^{(\nu, \mu)}$ on $y=(s-t)/2$, we note that it follows from
(\ref{ao10a})--(\ref{ao13a})  that
$$
\begin{array}{l}
X_k^l(s+1, t-1)=X_k^l(s, t),
\\ \\
X_k^l(s+\tau, t-\tau )=
e^{-2\pi in\tau -2\pi i n(s-t)+2\pi in\eta -2\pi i\sum_k\xi_k}
X_k^l(s, t),
\end{array}
$$
and, therefore,
\beq\label{fix18}
\begin{array}{l}
Y_k^{(\mu)}(y+1)=Y_k^{(\mu)}(y),
\\ \\
Y_k^{(\mu)}(y+\tau)=e^{-\pi iN\tau -2\pi iNy +\pi i N\eta -2\pi i \sum_k\xi_k}
Y_k^{(\mu)}(y).
\end{array}
\eeq
This implies the following properties:
\beq\label{fix19}
\begin{array}{l}
\phi^{(\nu, \mu)}(r,x,y+1)=\phi^{(\nu, \mu)}(r,x,y),
\\ \\
\phi^{(\nu, \mu)}(r,x,y+\tau )=e^{-\pi iN\tau -2\pi iNy +\pi i N\eta -2\pi i \sum_k\xi_k}
\phi^{(\nu, \mu)}(r,x,y).
\end{array}
\eeq
We also know that $\phi^{(\nu, \mu)}(r,x,y)$ is an entire function of $y$.
Therefore, we conclude that
it is a theta function of order $N$, i.e. it has $N$ zeros in the fundamental
domain.

The above properties imply that the dependence on
$x$ and $y$ factorizes and the function $\phi^{(\nu , \mu )}(\bar v, r, x,y)$
can be represented in the form
\beq\label{fix20}
\phi^{(\nu , \mu )}(\bar v, r, x,y)=
\phi_1^{(\nu , \mu )}(r, x)\phi_2^{(\nu )}(\bar v,y),
\eeq
where the function $\phi_1^{(\nu , \mu )}(r, x)$ can be fixed by
the following argument.

As it is seen from (\ref{Tnumu}), the matrix elements of
the matrix $T_{ij}^{(\nu\mu)}(r)$ are non-singular
if $r=m\tau$ and
\beq\label{m1b}
\mu-\nu+2m=Qk, \qquad k\in \ZZ
\eeq
(and the matrix is non-degenerate) while
for other integer values of $m$ the matrix elements have a simple pole at
$r=m\tau$ and the matrix has the form of a matrix of rank $1$ times a singular multiplier.
The determinant of such matrix has a simple pole at $r=m\tau$.
For even $Q$, the values of $m$ satisfying (\ref{m1b}) are
$m=\frac{1}{2}\, (Qk-\mu+\nu)$ ($\mu-\nu$ is always even for even $Q$, see below) and
we conjecture the following form of the function
$\phi_1^{(\nu , \mu)}(r,x)$ which satisfies all the above properties:
\beq\label{fix21}
\phi_1^{(\nu , \mu )}(r, x)=
\delta_{\mu, \nu \, (\mbox{\scriptsize mod $2$})}
e^{\pi i(\mu-\nu)x}
\frac{\theta_1(r|\tau)\,
\theta_1(r+Qx/2 +
(\mu \! -\! \nu)\tau/2|Q\tau/2)}{\theta_1(Qx/2|Q\tau/2)\,
\theta_1(r+(\mu-\nu)\tau/2|Q\tau/2)}.
\eeq
In fact the above transformation
properties still hold if one multiplies $\phi_1^{(\nu , \mu)}(r,x)$ by any function of
$r+(\mu \! -\! \nu)\tau/2$. The choice of this function in (\ref{fix21}) makes
$Y_k^{(\nu, \mu)}(r,x)$ free of poles in $r$ and also free of zeros (except the zero
at $r=-Qx/2 -(\mu-\nu)\tau/2$ modulo the lattice spanned by $1, \tau$).
The delta-symbol $\delta_{\mu, \nu \, (\mbox{\scriptsize mod $2$})}$ comes from the
selection rule (see section \ref{section:selection} below).
Strictly speaking, the above transformation
properties remain the same if one multiplies the right hand side of
(\ref{fix21}) by an elliptic function $f(x)$ of $x$ with periods $1$, $\tau$. However,
it follows from the explicit form of $Y_k^{(\mu)}$ that the only poles of
$\phi_1^{(\nu , \mu)}(r,x)$
are at the points $l\eta$, $l=0, 1, \ldots , Q/2-1$. But the right hand side of
(\ref{fix21}) already has poles at these points, so $f(x)$ must be free of poles
and thus be a constant.
For odd $Q$, the values of $m$ satisfying (\ref{m1b}) are
$m=Qk-(\mu-\nu)(Q+1)/2$, $k\in \ZZ$ and
we conjecture the following form of the function
$\phi_1^{(\nu , \mu)}(r,x)$ which satisfies all the above properties:
\beq\label{fix22}
\phi_1^{(\nu , \mu )}(r, x)=
e^{\pi i \ell_{\mu\nu}x}
\frac{\theta_1(r|\tau)\,
\theta_1(r+Qx +
\ell _{\mu\nu}\tau/2|Q\tau)}{\theta_1(Qx|Q\tau)\,
\theta_1(r+\ell_{\mu\nu}\tau/2|Q\tau)}.
\eeq
Here
\beq\label{fix17}
\ell_{\mu\nu}=\mu-\nu -\frac{1}{2}\, \Bigl (1-(-1)^{\mu-\nu}\Bigr )Q.
\eeq
The formulas (\ref{fix21}), (\ref{fix22}) are also justified by computer
calculations for $N=2$ and $N=4$.
The function $\phi_2^{(\nu )}(\bar v,y)$ is not fixed.
Finally, we note that numerical studies suggest
the following dependence of $\phi_2^{(\nu )}(\bar v,y)$ on $y$:
 \beq\label{fix171}
 \phi_2^{(\nu )}(\bar v,y)=\tilde{\phi}_2^{(\nu )}(\bar v)\, e^{2\pi i \nu y}
 \prod\limits_{k=1}^n\theta_1^2(y+v_k|\tau)\,.
\eeq

\subsection{The result for scalar products}

Now let us consider
$$
Y^{(\mu)}_{n+1}(r)=\sum_{l\in \z_Q}\frac{\theta_1(l\eta +U_{n+1} +w_0)}{\theta_1(l\eta +x )}
\, e^{-il \pi \eta \mu }\bigl < \Psi_{\nu} (\bar v)\bigl |\Psi^l(\bar u)\bigr >,
$$
where $\bar u =\{u_1, \ldots , u_n\}$ are arbitrary
parameters\footnote{ We draw the reader's attention to the fact that now the
set $\bar u$ consists only of $n$ parameters:
$\bar u =\{u_1, \ldots , u_n\}$. The auxiliary parameter
$u_{n+1}$ is no longer required, and we excluded it for the sake of simplification of notation.}.
We see that for the special choice
of $w_0$,
$$
w_0=x - U_{n+1},
$$
$Y^{(\mu)}_{n+1}(r)$ equals the scalar product of the on-shell dual Bethe vector
$\bigl <\Psi_{\nu} (\bar v)\bigr |$ and the off-shell Bethe vector
$$
\bigl |\Psi_{\mu}(\bar u)\bigr >=\sum_{l\in \z_Q}e^{-il \pi \eta \mu}\bigl |\Psi^l
(\bar u)\bigr >.
$$
Note that if we choose $w_0$ in this way, then
\beq\label{r1}
r=\sum_{i=1}^n (v_i-u_i).
\eeq
According to (\ref{w8}), the scalar product has the form
\beq\label{w10}
\bigl <\Psi_{\nu} (\bar v)\bigr | \Psi_{\mu} (\bar u)\bigr >=
\phi_1^{(\nu , \mu )}(r, x)\phi_2^{(\nu )}(\bar v,y)
\frac{\det\limits_{1\leq i,j\leq n} T^{(\nu \mu)}_{ij}(r)}{W_n(\bar u, \bar v)},
\eeq
where
\beq\label{W1}
W_n(\bar u, \bar v)=\frac{\prod\limits^n_{a<b}\theta_1(u_a-u_b)
\theta_1(v_b-v_a)}{\prod\limits^n_{p,q}\theta_1(u_p-v_q)},
\eeq
the matrix $T_{ij}^{(\nu \mu)}(r)$ is given by (\ref{w6}) and the function
$\phi_1^{(\nu , \mu )}(r, x)$ by (\ref{fix21}), (\ref{fix22}).
The function $\phi_2^{(\nu )}(\bar v, y)$ is still unknown.

\subsection{The selection rule}

\label{section:selection}

We are going to show that if $\eta =2P/Q$ with
even $Q$ and odd $P$, then off-shell Bethe vectors $\bigl <\Psi_{\nu}(\bar v)\bigr |$ and
$\bigl |\Psi_{\mu}(\bar u)\bigr >$ with $\mu-\nu=1 \,\, (\mbox{mod $2$})$ are
orthogonal: $\bigl <\Psi_{\nu}(\bar v)\bigr |\Psi_{\mu}(\bar u)\bigr >=0$.
This follows from the fact that the off-shell Bethe vectors are eigenvectors
of the operator ${\sf U}_3$:
\beq\label{sel1}
{\sf U}_3 \bigl |\Psi_{\mu}(\bar u)\bigr >=(-1)^{\mu+n}
\bigl |\Psi_{\mu}(\bar u)\bigr >, \quad
\bigl <\Psi_{\nu}(\bar v)\bigr |{\sf U}_3 =(-1)^{\nu+n}
\bigl <\Psi_{\nu}(\bar v)\bigr |.
\eeq
Indeed, computing the matrix element
$\bigl <\Psi_{\nu}(\bar v)\bigr |{\sf U}_3 \bigl |\Psi_{\mu}(\bar u)\bigr >$
in two ways (acting by ${\sf U}_3$ to the right and to the left), we get:
$$
\Bigl ((-1)^{\mu+n}-(-1)^{\nu+n}\Bigr )
\bigl <\Psi_{\nu}(\bar v)\bigr |\Psi_{\mu}(\bar u)\bigr >=0
$$
which means that $\bigl <\Psi_{\nu}(\bar v)\bigr |\Psi_{\mu}(\bar u)\bigr >=0$
if $\mu-\nu=1 \,\, (\mbox{mod $2$})$.

To prove (\ref{sel1}), we use the result of section
\ref{section:action} that the action of ${\sf U}_3$ to the vectors
$\bigl |\Psi^l\bigr >$ and $\bigl <\Psi^l\bigr |$ is equivalent to the
shift of $s,t$ by $1$ and the fact that
$\bigl |\Psi_{\mu}(s+2,t+2)\bigr > =\bigl |\Psi_{\mu}(s,t)\bigr >$. We have:
\begin{multline}
\bigl |\Psi_{\mu}(s,t)\bigr >=\sum_{l\in \z_Q}e^{-2\pi i lP\mu/Q}
\bigl |\Psi^{l}(s,t)\bigr >=
\sum_{l=0}^{Q -1}e^{-2\pi i lP\mu/Q}\bigl |\Psi^0 \Bigl (
s+\tfrac{2Pl}{Q}, t+\tfrac{2Pl}{Q}\Bigr )\bigr >\\
=\sum_{l=0}^{Q/2 -1}e^{-2\pi i lP\mu/Q}\bigl |\Psi^0 \Bigl (
s+\tfrac{2Pl}{Q}, t+\tfrac{2Pl}{Q}\Bigr )\bigr >
+\sum_{l=Q/2}^{Q -1}e^{-2\pi i lP\mu/Q}\bigl |\Psi^0 \Bigl (
s+\tfrac{2Pl}{Q}, t+\tfrac{2Pl}{Q}\Bigr )\bigr >\\
=\! \! \sum_{l=0}^{Q/2 -1}e^{-2\pi i lP\mu/Q}\bigl |\Psi^0 \Bigl (
s+\tfrac{2Pl}{Q}, t+\tfrac{2Pl}{Q}\Bigr )\bigr > +(-1)^{\mu}\!\!\!
\sum_{l=0}^{Q/2 -1}\!\!\! e^{-2\pi i lP\mu/Q}\bigl |\Psi^0 \Bigl (
s+\tfrac{2Pl}{Q}+1, t+\tfrac{2Pl}{Q}+1\Bigr )\bigr >\\
=\Bigl (1+(-1)^{\mu +n}{\sf U}_3\Bigr )
\sum_{l=0}^{Q/2 -1}e^{-2\pi i lP\mu/Q}\bigl |\Psi^0 \Bigl (
s+\tfrac{2Pl}{Q}, t+\tfrac{2Pl}{Q}\Bigr )\bigr >,\nonumber
\end{multline}
from which the first relation in (\ref{sel1}) follows (recall that ${\sf U}_3^2=1$).
The argument for the dual vector $\bigl <\Psi_{\nu}\bigr |$ is the same.
Note that the selection rule is valid also for $Y_k^{(\nu  , \mu)}$: it vanishes
for $\mu -\nu=1 \,\, (\mbox{mod $2$})$ (the proof is similar).

However, if $Q$ is odd, then (\ref{sel1}) does not hold in general and
the selection rule does not work.

\subsection{Orthogonality and norm of on-shell Bethe vectors}

Let us show that if the vector $\bigl |\Psi_{\mu}(\bar u)\bigr >$ is on-shell and the sets
$\bar u$ and $\bar v$ do not coincide, the scalar product (\ref{w10}) vanishes.
We should show that the matrix $T^{(\nu \mu)}_{ik}(r)$ (\ref{Tnumu})
becomes degenerate if the parameters $\bar u$ satisfy the Bethe equations (\ref{Bethe}),
which we write here in the form
\beq\label{Bethe1}
e^{2i\pi \eta \mu}\frac{a(u_j)}{d(u_j)}=(-1)^{n-1}\frac{h(u_j, \bar u)}{h(\bar u, u_j)}.
\eeq

We are going to show that the rows of the matrix $T^{(\nu \mu)}_{ik}(r)$ are linearly dependent.
Set
\beq\label{or1}
x_j=\frac{g(v_j, \bar v_j)}{g(v_j, \bar u)}.
\eeq
Note that since the sets $\bar u$, $\bar v$ do not coincide, there is at least one
non-vanishing $x_j$. Consider a linear combination
\beq\label{or2}
X=\sum_i x_i T^{(\nu \mu)}_{ik}(r)=a(u_k)f(\bar v, u_k)E^+ +
d(u_k)f(u_k, \bar v)E^-,
\eeq
where
$$
E^{\pm}=\frac{\theta_1'(0)}{\theta_1(r)}\sum_i x_i \left (e^{\pm i\pi \eta \nu}
\frac{\theta_1(v_i-u_k-r)}{\theta_1(v_i-u_k)}-e^{\pm i\pi \eta \mu}
\frac{\theta_1(v_i-u_k\pm \eta -r)}{\theta_1(v_i-u_k \pm \eta )}\right ).
$$
In order to compute $E^{\pm}$, we consider an auxiliary contour integral
$$
I^{\pm}=\frac{\theta_1'(0)}{\theta_1(r)}\oint \frac{dz}{2\pi i}
\left (e^{\pm i\pi \eta \nu}
\frac{\theta_1(z-u_k-r)}{\theta_1(z-u_k)}-e^{\pm i\pi \eta \mu}
\frac{\theta_1(z-u_k\pm \eta -r)}{\theta_1(z-u_k \pm \eta )}\right )
\prod_{a=1}^n \frac{\theta_1(z-u_a)}{\theta_1(z-v_a)}.
$$
The integral is taken along the boundary of the fundamental parallelogram spanned by $1, \tau$.
Since $\displaystyle{r=\sum_i v_i-\sum_i u_i}$ (see (\ref{r1})), it is easy to see that
the  integrand is double-periodic with periods $1$, $\tau$. Therefore,
$I^{\pm}=0$. On the other hand, the integral can be calculated as sum of the residues
inside the fundamental parallelogram. The sum of the residues at the poles at $z=v_j$
gives $E^{\pm}$. One more contribution comes from the simple pole at
$z-u_k\pm \eta =0$. Thus we arrive at
$$
E^{\pm}=-e^{\pm i\pi \eta \mu}\theta_1'(0)\prod_{a=1}^n
\frac{\theta_1(u_k-u_a\mp \eta )}{\theta_1(u_k-v_a\mp \eta )}
$$
or
$$
E^+=-e^{i\pi \eta \mu}\theta_1'(0)\frac{h(\bar u, u_k)}{h(\bar v, u_k)}, \quad
E^-=-e^{-i\pi \eta \mu}\theta_1'(0)\frac{h(u_k, \bar u)}{h(u_k, \bar v)}.
$$
Substituting these results into (\ref{or2}), we obtain:
$$
X=-\theta_1'(0)g(u_k, \bar v)\left [
(-1)^n e^{i \pi \eta\mu}a(u_k)h(\bar u, u_k)+e^{-i\pi \eta \mu}d(u_k)h(u_k, \bar u)\right ]
$$
which is equal to 0 due to the Bethe equations (\ref{Bethe1}). Thus the rows of the
matrix $T^{(\nu \mu)}_{ik}(r)$ are linearly dependent and hence $\det T^{(\nu \mu)}_{ik}(r)=0$.

Note that we did not use the fact that the set $\bar v$ satisfies the Bethe equations.
Therefore, $\det T^{(\nu \mu)}_{ik}(r)$ vanishes when the following
weaker conditions are fulfilled:
i) the set $\bar u$ satisfies the Bethe equations, ii) $\displaystyle{r=\sum_i(v_i-u_i)}$,
iii) the sets $\bar u , \bar v$ do not coincide.

The square of the norm of the vector
$\bigl |\Psi_{\nu} (\bar v)\bigr >$ can be obtained
in the limit\footnote{See remark in the beginning of
section~\ref{section:gaba}.} $u_i\to v_i$. We set $u_i=v_i +\varepsilon$ in equation
(\ref{w10}), so that $r=-n\varepsilon$, and tend $\varepsilon \to 0$. The matrix
$T^{(\nu \nu)}_{ik}(r)$ becomes singular but the factor $W(\bar u, \bar v)$ in the
denominator brings the multiplier $\theta_1^n(\varepsilon )$ which cancels the singularity.
The limiting procedure is straightforward and the result is\footnote{The same result is reproduced
 for the limit $u_i=v_i +\varepsilon_i$, $\varepsilon_i\to 0$.}
$$
\lim_{\varepsilon \to 0}\Bigl (\theta_1(\varepsilon )T^{(\nu \nu)}_{ik}(-n\varepsilon)\Bigr )=
\theta_1(\eta) e^{-i\pi \eta \nu}d(v_k)f(v_k, \bar v_k)K(v_i-v_k), \quad i\neq k,
$$
$$
\lim_{\varepsilon \to 0}\Bigl (\theta_1(\varepsilon )T^{(\nu \nu)}_{ii}(-n\varepsilon)\Bigr )=-
\theta_1(\eta) e^{-i\pi \eta \nu}d(v_i)f(v_i, \bar v_i)\left (
\p_{v_i}\log \frac{a(v_i)}{d(v_i)} +\sum_{j\neq i}K(v_i-v_j)\right ),
$$
where
\beq\label{or3}
K(u)=\frac{\theta_1'(u-\eta)}{\theta_1(u-\eta)}-\frac{\theta_1'(u+\eta)}{\theta_1(u+\eta)}.
\eeq
Therefore,
\beq\label{or4}
\bigl <\Psi_{\nu} (\bar v)\bigr | \Psi_{\nu} (\bar v)\bigr >=
\phi_1^{(\nu, \nu)}(0,x)
\phi_2^{(\nu)} (\bar v,y)\theta_1^n(\eta) e^{-in\pi \eta \nu}d(\bar v)\prod^n_{a\neq b}f(v_a, v_b)
\det_{1\leq i,k\leq n}G_{ik},
\eeq
where
\beq\label{or5}
G_{ik}=-\delta_{ik}\left (\p_{v_i}\log \frac{a(v_i)}{d(v_i)}+\sum_{j=1}^n
K(v_i-v_j)\right )+K(v_i-v_k)\,,
\eeq
\beq\label{or6}
\phi_1^{(\nu, \nu)}(0,x)=\left \{
\begin{array}{l}
\displaystyle{\frac{\theta_1'(0|\tau)}{\theta_1'(0|Q\tau/2)}} \quad \mbox{for $Q$ even},
\\ \\
\displaystyle{\frac{\theta_1'(0|\tau)}{\theta_1'(0|Q\tau)}}    \qquad \mbox{for $Q$ odd}.
\end{array}\right.
\eeq
 and similarly to (\ref{shnprod})
 \beq\label{dbar}
 d(\bar v)=\prod_{i=1}^n d(v_i)\,.
\eeq
This is the elliptic version of the Gaudin's formula \cite{GMW81} and $G_{ik}$ is
the elliptic analogue of the Gaudin's matrix. If we take logarithm of the
Bethe equations, denote
$$
B_j=-\log \frac{a(v_j)}{d(v_j)}+\log \frac{f(v_j, \bar v)}{f(\bar v, v_j)}-2\pi i \eta \nu
$$
(so that the Bethe equations read $B_j =2\pi i n_j$, $n_j\in \ZZ$), then
$G_{jk}=\p B_j/\p v_k$.

However, in the case of the norm the result (\ref{or4})
is somewhat meaningless because it is multiplied by
an unknown function $\phi_2$ of $\bar v$.

\subsection{Normalized scalar products}

The unknown function $\phi_2$ does not enter the {\it specially normalized} scalar products
\beq\label{norm1}
S_{\nu\mu}(\bar v, \bar u)=\frac{\bigl <\Psi_{\nu}(\bar v)
\bigr |\Psi_{\mu}(\bar u)\bigr >}{\bigl <\Psi_{\nu}(\bar v)
\bigr |\Psi_{\nu}(\bar v)\bigr >},
\eeq
where the vector $\bigl <\Psi_{\nu}(\bar v)\bigr |$ is an on-shell
vector and $\bigr |\Psi_{\mu}(\bar u)\bigr >$ is an arbitrary
of-shell Bethe vector.  It is easy to see that \eqref{norm1}
differs from the usual normalized scalar product,
in which the denominator would contain the norms of both vectors.
Nevertheless, in the models with the 6-vertex $R$-matrix,
it is normalization (\ref{norm1}) that is sufficient to
calculate the form factors of local operators and correlation functions
(see the more detailed discussion of this issue in section~\ref{section:conc}).
It is for this reason that we are interested in such a special normalization.

Collecting the formulas obtained above together, we arrive at
the following result for these scalar products:
\beq\label{norm2}
\displaystyle{
S_{\nu\mu}(\bar v, \bar u)=e^{\pi i\eta n\nu}\prod_{p,q}
\frac{\theta_1(u_p-v_q)}{\theta_1(v_p-v_q+\eta)}\prod_{a<b}
\frac{\theta_1(v_a-v_b)}{\theta_1(u_a-u_b)}\,
\frac{\phi_1^{(\nu,\mu)}(r,x)}{\phi_1^{(\nu,\nu)}(0,x)}\,
\frac{\det\limits_{n\times n}T_{jk}^{(\nu\mu)}(r)}{d(\bar v)\det\limits_{n\times n}G_{jk}},}
\eeq
where $\displaystyle{r=\sum_i v_i -\sum_iu_i}$, the matrices $T^{(\nu\mu)}_{jk}(r)$,
$G_{jk}$ are given by formulas (\ref{Tnumu}), (\ref{or5}) respectively, $d(\bar v)$ by (\ref{dbar}) and the functions
$\phi_1^{(\nu,\mu)}(r,x)$, $\phi_1^{(\nu,\nu)}(0,x)$ by (\ref{fix21}), (\ref{fix22}),
(\ref{or6}). Note that at $\mu=\nu$ and $\displaystyle{\sum_i u_i =\sum_iv_i}$
this formula becomes
\beq\label{norm3}
\displaystyle{
S_{\nu\nu}(\bar v, \bar u)=e^{\pi i\eta n\nu}\prod_{p,q}
\frac{\theta_1(u_p-v_q)}{\theta_1(v_p-v_q+\eta)}\prod_{a<b}
\frac{\theta_1(v_a-v_b)}{\theta_1(u_a-u_b)}\,
\frac{\det\limits_{n\times n}\Bigl (
\p T_{\nu}(u_k, \bar v)/\p v_j\Bigr )}{d(\bar v)\det\limits_{n\times n}G_{jk}},}
\eeq
which resembles the result for the $XXZ$ case \cite{KitMT99}.

\section{Concluding remarks\label{section:conc}}

We have obtained the determinant representation (\ref{norm2})
for the specially normalized scalar products
of Bethe vectors (\ref{norm1}) in the inhomogeneous 8-vertex model
 (or equivalently, in the inhomogeneous
$XYZ$ spin-$\frac{1}{2}$ chain)
in the case when the anisotropy parameter  $\eta$ is a rational number
$\eta = 2P/Q$.  Recall, however, that one can take the homogeneous limit in all our formulas.
The matrix $T_{jk}^{(\nu\mu)}(r)$ in (\ref{norm2})
is given by (\ref{w6}) or (\ref{Tnumu}). Note that this matrix is essentially the same
as the matrix entering the determinant representation for scalar products of Bethe
vectors in the elliptic cyclic SOS model obtained in \cite{LT13}.

A more general case when the Bethe vectors are well-defined is the
case when $\eta$ is a point of finite order on the elliptic curve, i.e.,
$Q\eta = 2P_1+P_2\tau$ with some integer $Q, P_1,P_2$. We hope that
it is not too difficult to
extend our results to this case. Other possible generalizations of our results
are related to the scalar products of Bethe vectors in the 8-vertex model with
twisted boundary conditions \cite{NT15} (in the 8-vertex case the only
possible twist matrices are
the Pauli matrices) and in the $XYZ$ spin chain with higher spin \cite{T95,T15}.

 We also did not consider the case of scalar products in which the
right and left vectors depend on a different number of parameters.
Note that the derivation of the system of linear equations for this case remains exactly the same.
However, presumably, in this case, the resulting system has only trivial solutions.

A comment on the limit to the $XXZ$ case (the 6-vertex model) is in order.
Formally, this is the limit $\tau \to +i\infty$ when the Jacobi theta functions
tend to trigonometric functions. However, as is seen from (\ref{norm2}),
(\ref{Tnumu}), the limit of our result does not coincide with the well known answer
for the $XXZ$ case \cite{S89}. The reason is in the different structure of the off-shell
Bethe vectors: the trigonometric limit of the off-shell Bethe vectors constructed in the
framework of the generalized algebraic Bethe ansatz method differs from off-shell Bethe vectors
usually considered in the $XXZ$ type models. It is enough to say that our off-shell
vectors essentially depend on the auxiliary parameters $s,t$ which are absent in the
standard algebraic Bethe ansatz approach.

One of the most attractive areas is the application of the obtained
result (\ref{norm2}) to the calculation of form factors and correlation functions.
However, for this it is necessary to obtain formulas for the action of the
monodromy matrix entries on Bethe vectors.
In models with the 6-vertex $R$-matrix, such formulas are well known. Schematically, they can be represented in the form
\beq\label{ActB}
\mathcal{T}_{ab}(z)\bigl |\Psi(\bv)\bigr > =
\sum_{\bv'} t_{ab}(\bv') \bigl |\Psi(\bv')\bigr >,
\eeq
where $\bv'$ is a subset of $\bv\cup z$, and $t_{ab}(\bv')$ are some
numerical coefficients. The sum in \eqref{ActB} is taken with respect to all possible subsets of fixed cardinality.
The latter depends on the concrete matrix element $\mathcal{T}_{ab}$.

If we assume that similar action formulas also exist
in the 8-vertex model, then we immediately get access to the form factors of local spin operators.
Indeed, the latter can be expressed through the elements of
the quantum monodromy matrix using the formulas of the
inverse scattering problem \cite{GohK00,MaiT00}\footnote{Relation (\ref{ISP-sol}) slightly differs from its custom form since we use different ordering of the $R$-matrices in the definition of the monodromy matrix.}:
\begin{equation}\label{ISP-sol}
E^{ij}_m= \left(\prod_{k=1}^m{\sf T}(\xi_k)\right)^{-1}\; \mathcal{T}_{ji}(\xi_m)
\;\left(\prod_{k=1}^{m-1}{\sf T}(\xi_k)\right).
\end{equation}
Here $E^{ij}_m$ is an elementary unit  matrix acting in the $m$th local quantum space $V_m$.
Therefore, all form factors of local operators reduce to
form factors of the entries of the quantum monodromy matrix. For instance,
magnetization $\Bigl <\frac{1}{2}(1-\sigma_3^{(m)})\Bigr >$ is given by
\beq\label{FF-mag}
\frac{\bigl <\Psi_\nu(\bv)|\frac{1}{2}(1-\sigma_3^{(m)})\bigl |
\Psi_\nu(\bv)\bigr >}{\bigl <\Psi_\nu(\bv)|\Psi_\nu(\bv)\bigr >}=
\frac{\bigl <\Psi_\nu(\bv)|D(\xi_m)\bigl
|\Psi_\nu(\bv)\bigr >}{T_{\nu}(\xi_m|\bv)\bigl <\Psi_\nu(\bv)\bigl |\Psi_\nu(\bv)\bigr >},
\eeq
where $T_{\nu}(\xi_m|\bv)$ is the eigenvalue of ${\sf T}(\xi_m)$ on
$\bigl |\Psi_\nu(\bv)\bigr >$.

If the action of the operator $D(\xi_m)$ on the vector $\bigl |\Psi_\nu(\bv)\bigr >$
is given by a linear combination of off-shell vectors similar to (\ref{ActB}),
then this form factor reduces to scalar products (\ref{norm2}). Other form factors are calculated similarly.

However, we would like to emphasize that the assumption of the
existence of the action formula similar to (\ref{ActB}) in the models with the
8-vertex $R$-matrix is rather strong assumption. For the moment, there is no evidence that such a formula does exist. We are going to highlight this issue in our
forthcoming publications.

\section*{Acknowledgments} We thank A. Liashyk for useful comments and discussions.
This work is supported by the Russian Science Foundation under grant 19-11-00062
and performed in Steklov Mathematical Institute of Russian Academy of Sciences.

\addcontentsline{toc}{section}{\hspace{6mm}Acknowledgments}

\section*{List of notations}

\addcontentsline{toc}{section}{\hspace{6mm} List of notations}

Here we list some notations for convenience as they appear in the
text:

$\tau$ -- modular parameter of elliptic functions;

$\eta$ -- anisotropy parameter (or Planck constant) in the $R$-matrix;

$P,Q\in{\ZZ}_+$ -- coprime numbers entering $\eta=2P/Q$;

$u$
-- spectral parameter in of the
$R$-matrix;

$W_a(u)$, $a=0,...,3$; $a^{\rm 8v}(u)$, $b^{\rm 8v}(u)$,
$c^{\rm 8v}(u)$, $d^{\rm 8v}(u)$ -- matrix elements of $R$-matrix (\ref{8v5});

$N$ -- number of sites of the lattice row (an even integer);

$n=N/2$ -- number of Bethe roots, see also (\ref{eta2});

${\sf a}(u)$, ${\sf b}(u)$, ${\sf c}(u)$, ${\sf d}(u)$ -- elements
of the $L$-operator (\ref{8v5a});

$\xi_1,...,\xi_N$ -- inhomogeneity parameters, (\ref{m1}),
(\ref{8v7});

$A(u)$, $B(u)$, $C(u)$, $D(u)$ -- operator matrix elements of the quantum
monodromy matrix
(\ref{m1a});

$\displaystyle{c(u)=N(2u+\eta +\tau)-2\sum_{k=1}^N \xi_k}$, (\ref{c(u)});

${\sf U}_a=(\sigma_a)^{\otimes N}$, (\ref{Ua});

$s,t$ -- the parameters of intertwining (co)vectors
 $\bigl |\phi(s)\bigr >$
 (\ref{i1});

${\sf a}'_k(u)$, ${\sf b}'_k(u)$, ${\sf c}'_k(u)$, ${\sf d}'_k(u)$
-- elements of the gauged transformed $L$-operator (\ref{v1});

$s_k=s+k\eta$, $t_k=t+k\eta$, $\tau_k=(s_k+t_k)/2$;

$\gamma_k$ -- functions entering the (gauge transformation) matrix $M_k(u)$
(\ref{v2})-(\ref{gamma});

$\bigl |\omega_k^l\bigr >$ -- local vacuum vectors (\ref{v4});

$\left |\Omega ^l\right >=\left |\omega_1^l\right >\otimes \left
|\omega_2^l\right >\otimes \ldots \otimes \left |\omega_N^l\right >$
-- global vacuum vectors (\ref{v6});

$A_{k,l}(u)$, $B_{k,l}(u)$, $C_{k,l}(u)$, $D_{k,l}(u)$ -- elements
of ${\cal T}_{k,l}(u)$ (\ref{p1})-(\ref{p2});

$\bigl |\Psi^l (u_1, \ldots , u_n)\bigr >$ -- vectors (\ref{a1});

$\Phi(u,v)$ -- (Kronecker) function (\ref{a2});

$\bigl |\Psi_\nu (u_1, \ldots , u_n)\bigr >$ -- right Bethe vector, Fourier transform
of $\bigl |\Psi^l \bigr >$ (\ref{a3});


${\bar A}_{k,l}(u)$, ${\bar B}_{k,l}(u)$, ${\bar C}_{k,l}(u)$,
${\bar D}_{k,l}(u)$ -- operators (\ref{b0});

$\bigl <\Psi^l (v_1, \ldots , v_n)\bigr |$ -- left vector
(\ref{b1});

$\bigl <\Psi_\nu (v_1, \ldots , v_n)\bigr |$ -- left Bethe vector, Fourier transform
of $\bigl <\Psi^l \bigr |$ (\ref{a3a});

$\displaystyle{\sigma(v_1,...,v_n)=\sum_{i=1}^n v_i -\frac{1}{2}\sum_{k=1}^N \xi_k
+\frac{1}{2}\, n\eta}$, the function (\ref{q15});

$x=(s+t+1)/2$, $y=(s-t)/2$ -- variables (\ref{xy11});

$\nu_a=0,1$ -- the numbers from (\ref{nu11});

$\nu=-n-\nu_3$;

$\displaystyle{a(u)=\prod_{i=1}^N\theta_1(u-\xi_i+\eta)}$,
$\displaystyle{d(u)=\prod_{i=1}^N\theta_1(u-\xi_i)}$ -- the functions (\ref{ad});

Bethe equations -- (\ref{Bethe}), (\ref{Bethe2});

sum rule  -- (\ref{q19});

$v_1,...,v_n$ -- Bethe roots, solutions of (\ref{Bethe2}) and
(\ref{q19});

$g(u,v)$, $f(u,v)$, $h(u,v)$ -- functions
(\ref{gfh})-(\ref{shnprod});

$X_j^l$ -- the scalar products
$\bigl <\Psi_{\nu}(\bv)\bigl |\Psi^{l}(\bu_j)\bigr >$ (\ref{Xjl});

$T_\nu(u)$ -- function (\ref{a4a}), (\ref{Eival});


$w=w_{n+1}=l\eta +\check U+w_0$, (\ref{w1});

$\displaystyle{\check U =\sum_{j=1}^{n+1}u_j}$;

$\displaystyle{r=\sum_{p=1}^n v_p -\frac{1}{2}\, (s\! +\! t\! +\! 1)+w_0}$,
(\ref{w4});

$Y^{(\mu)}_k$ -- weighted Fourier transform of $X_j^l$ (\ref{w3});

$\displaystyle{U_{n+1}=U=\sum_{p=1}^{n}u_p}$,
$\displaystyle{V=\sum_{p=1}^{n}v_p}$, (\ref{Uk});

$T^{(\nu \mu )}_{ik}(r)$ -- matrix (\ref{w6})-(\ref{Tnumu});

$W_n$ -- function (\ref{w9}), (\ref{W1});

$\phi^{(\nu ,\mu)}(\bar v,r)$ -- function (\ref{w8}), (\ref{fix20});

$\phi_1^{(\nu , \mu )}(r, x)$, $\phi_2^{(\nu )}(\bar v,y)$ --
functions (\ref{fix20}), (\ref{fix21})-(\ref{fix17}),
(\ref{fix171});

$S_{\nu\mu}(\bar v, \bar u)$ -- specially normalized scalar products
(\ref{norm1}).

\section*{Appendix A: null-vector}

\def\theequation{A\arabic{equation}}
\setcounter{equation}{0}

\addcontentsline{toc}{section}{\hspace{6mm}Appendix A: null-vector}

As one of the applications of the obtained formulas,  we show that the scalar products $\bigl <\Psi_{\nu}(\bar v)\bigr |\Psi_{\mu}(\bar u)\bigr >$
vanish if the set $\bar u$ contains $\xi_p$ and $\xi_p -\eta$ for some $p=1, \ldots , N$.
(We denote them as $u_1=\xi_p$ and $u_2=\xi_p-\eta$, so that $d(u_1)=a(u_2)=0$.)
To see this, we will show that in this case $\det T_{jk}^{(\nu \mu)}=0$. Indeed,
from the formula (\ref{Tnumu}) we conclude that the first column is
$$
T_{j1}^{(\nu \mu)}=\frac{\theta_1'(0)}{\theta_1(r)}\, a(\xi_p)f(\bar v, \xi_p)
\left (e^{i\pi \eta \nu}\frac{\theta_1(\xi_p-v_j+r)}{\theta_1(\xi_p-v_j)}-
e^{i\pi \eta \mu}\frac{\theta_1(\xi_p-v_j-\eta +r)}{\theta_1(\xi_p-v_j-\eta )}\right )
$$
and the second column is
$$
T_{j2}^{(\nu \mu)}=\frac{\theta_1'(0)}{\theta_1(r)}\, d(\xi_p-\eta)f(\xi_p-\eta, \bar v)
\left (e^{-i\pi \eta \nu}\frac{\theta_1(\xi_p-\eta -v_j+r)}{\theta_1(\xi_p-\eta -v_j)}-
e^{-i\pi \eta \mu}\frac{\theta_1(\xi_p-v_j +r)}{\theta_1(\xi_p-v_j)}\right )
$$
$$
=\frac{\theta_1'(0)}{\theta_1(r)}\, e^{-\pi i \eta (\mu +\nu)}\,
d(\xi_p-\eta)f(\xi_p-\eta, \bar v)
\left (e^{i\pi \eta \mu}\frac{\theta_1(\xi_p-\eta -v_j+r)}{\theta_1(\xi_p-\eta -v_j)}-
e^{i\pi \eta \nu}\frac{\theta_1(\xi_p-v_j +r)}{\theta_1(\xi_p-v_j)}\right ).
$$
We see that the two columns are proportional to each other and hence the determinant
vanishes.

If the set of dual eigenvectors $\bigl <\Psi_{\nu}(\bar v)\bigr |$ is complete (which is
usually believed), this result means that the vector $\bigl |\Psi_{\mu}(\bar u)\bigr >$
in which $u_1=\xi_p$, $u_2=\xi_p-\eta$ is a null-vector.

\section*{Appendix B: the case $N=2$}

\def\theequation{B\arabic{equation}}
\setcounter{equation}{0}

\addcontentsline{toc}{section}{\hspace{6mm}Appendix B: the case $N=2$}

The case $N=2$ is relatively simple but instructive.
In this case there is only one Bethe equation for the Bethe root $v$ of the form
\beq\label{B1}
e^{2\pi i \eta \nu}\frac{\theta_1(v-\xi_1+\eta)
\theta_1(v-\xi_2+\eta)}{\theta_1(v-\xi_1)\, \theta_1(v-\xi_2)}=1.
\eeq
This equation has 4 solutions in the fundamental domain
(the number of solutions is equal to the dimension of the quantum space):
$$
v=\frac{1}{2}\, (\xi_1+\xi_2-\eta )+\omega,
$$
where $\omega$ is a half-period: $\omega = 0, 1/2$ (in these cases $\nu =0$) and
$\omega = \tau/2 , (\tau +1)/2$ (in these cases $\nu =1$). Note that this agrees
with the sum rule (\ref{q19}).

\subsection*{Diagonalization of the transfer matrix}

For brevity, we denote $a_1=a^{\rm 8v}(u-\xi_1)$, $a_2=a^{\rm 8v}(u-\xi_2)$,
$b_1=b^{\rm 8v}(u-\xi_1)$, etc. In the natural basis
$\left |++\right >$, $\left |+-\right >$, $\left |-+\right >$,
$\left |--\right >$ the transfer matrix of the model is given by
$$
{\sf T}(u)=\left (\begin{array}{cccc}
a_1a_2+b_1b_2& 0& 0& c_1d_2+d_1c_2
\\
0& a_1b_2+b_1a_2& c_1c_2+d_1d_2& 0
\\
0& c_1c_2+d_1d_2&a_1b_2+b_1a_2&0
\\
c_1d_2+d_1c_2& 0& 0& a_1a_2+b_1b_2
\end{array}
\right ).
$$
This matrix can be easily diagonalized by hands. The result is:

\vspace{0.5cm}

\begin{tabular}{|c|c|c|c|}
\hline
&Eigenvector & Eigenvalue & Bethe root \\
\hline
1) & $\vphantom{\sum\limits_k^l}\left |+-\right >-
\left |-+\right >$& $a_1b_2+b_1a_2-c_1c_2-d_1d_2$& $v_1=
\frac{1}{2}\, (\xi_1\! +\! \xi_2\! -\! \eta), \, \nu=0\phantom{aaa)}$\\
\hline
2) & $\vphantom{\sum\limits_k^l}
\left |+-\right >+\left |-+\right >$& $a_1b_2+b_1a_2+c_1c_2+d_1d_2$& $v_2=
\frac{1}{2}\, (\xi_1\! +\! \xi_2\! -\! \eta)+\frac{1}{2}, \, \nu=0$\\
\hline
3) & $\vphantom{\sum\limits_k^l}\left |++\right >+
\left |--\right >$& $a_1a_2+b_1b_2+c_1d_2+d_1c_2$& $v_3\! =\!
\frac{1}{2}\, (\xi_1\! +\! \xi_2\! -\! \eta)\! +\! \frac{\tau+1}{2}, \, \nu=1$\\
\hline
4) & $\vphantom{\sum\limits_k^l}\left |++\right >-
\left |--\right >$& $a_1a_2+b_1b_2-c_1d_2-d_1c_2$& $v_4=
\frac{1}{2}\, (\xi_1\! +\! \xi_2\! -\! \eta)+\frac{\tau}{2}, \, \nu=1$\\
\hline
\end{tabular}

\vspace{0.5cm}

\noindent
Similar results hold for left eigenvectors. After
some transformations the eigenvalues can be brought to the form
$$
T_0(u; v_1)=\frac{2}{\theta_2^2(0)\theta_3(0)\theta_4(0)}\left [
\theta_4(0)\theta_3(\eta)\theta_3\Bigl (\frac{\xi_1\! -\! \xi_2\! +\! \eta}{2}\Bigr )
\theta_3\Bigl (\frac{\xi_1\! -\! \xi_2\! -\! \eta}{2}\Bigr )
\theta_4^2\Bigl (u-\frac{\xi_1\! +\! \xi_2\! -\! \eta}{2}\Bigr )\right.
$$
$$
\left. -\,
\theta_3(0)\theta_4(\eta)\theta_4\Bigl (\frac{\xi_1\! -\! \xi_2\! +\! \eta}{2}\Bigr )
\theta_4\Bigl (\frac{\xi_1\! -\! \xi_2\! -\! \eta}{2}\Bigr )
\theta_3^2\Bigl (u-\frac{\xi_1\! +\! \xi_2\! -\! \eta}{2}\Bigr )\right ],
$$
$$
T_0(u; v_2)=\frac{2}{\theta_2^2(0)\theta_3(0)\theta_4(0)}\left [
\theta_3(0)\theta_4(\eta)\theta_3\Bigl (\frac{\xi_1\! -\! \xi_2\! +\! \eta}{2}\Bigr )
\theta_3\Bigl (\frac{\xi_1\! -\! \xi_2\! -\! \eta}{2}\Bigr )
\theta_4^2\Bigl (u-\frac{\xi_1\! +\! \xi_2\! -\! \eta}{2}\Bigr )\right.
$$
$$
\left. -\,
\theta_4(0)\theta_3(\eta)\theta_4\Bigl (\frac{\xi_1\! -\! \xi_2\! +\! \eta}{2}\Bigr )
\theta_4\Bigl (\frac{\xi_1\! -\! \xi_2\! -\! \eta}{2}\Bigr )
\theta_3^2\Bigl (u-\frac{\xi_1\! +\! \xi_2\! -\! \eta}{2}\Bigr )\right ],
$$
$$
T_1(u; v_3)=\frac{2}{\theta_2^2(0)\theta_3(0)\theta_4(0)}\left [
\theta_3(0)\theta_4(\eta)\theta_2\Bigl (\frac{\xi_1\! -\! \xi_2\! +\! \eta}{2}\Bigr )
\theta_2\Bigl (\frac{\xi_1\! -\! \xi_2\! -\! \eta}{2}\Bigr )
\theta_1^2\Bigl (u-\frac{\xi_1\! +\! \xi_2\! -\! \eta}{2}\Bigr )\right.
$$
$$
\left. -\,
\theta_4(0)\theta_3(\eta)\theta_1\Bigl (\frac{\xi_1\! -\! \xi_2\! +\! \eta}{2}\Bigr )
\theta_1\Bigl (\frac{\xi_1\! -\! \xi_2\! -\! \eta}{2}\Bigr )
\theta_2^2\Bigl (u-\frac{\xi_1\! +\! \xi_2\! -\! \eta}{2}\Bigr )\right ],
$$
$$
T_1(u; v_4)=\frac{2}{\theta_2^2(0)\theta_3(0)\theta_4(0)}\left [
\theta_4(0)\theta_3(\eta)\theta_2\Bigl (\frac{\xi_1\! -\! \xi_2\! +\! \eta}{2}\Bigr )
\theta_2\Bigl (\frac{\xi_1\! -\! \xi_2\! -\! \eta}{2}\Bigr )
\theta_1^2\Bigl (u-\frac{\xi_1\! +\! \xi_2\! -\! \eta}{2}\Bigr )\right.
$$
$$
\left. -\,
\theta_3(0)\theta_4(\eta)\theta_1\Bigl (\frac{\xi_1\! -\! \xi_2\! +\! \eta}{2}\Bigr )
\theta_1\Bigl (\frac{\xi_1\! -\! \xi_2\! -\! \eta}{2}\Bigr )
\theta_2^2\Bigl (u-\frac{\xi_1\! +\! \xi_2\! -\! \eta}{2}\Bigr )\right ].
$$
It can be shown by a straightforward
calculation that this form is the same as the expression (\ref{Eival}):
$$
T_{\nu}(u; v_a)= e^{\pi i\eta \nu}\theta_1(u-\xi_1+\eta)\theta_1(u-\xi_2+\eta)\,
\frac{\theta_1(u\! -\! v_a\! -\! \eta)}{\theta_1(u-v_a)}
$$
$$
\phantom{aaaaaaaaaaaaaaaaaaaaaaaaaaaaa}+\, e^{-\pi i\eta \nu}
\theta_1(u-\xi_1)\theta_1(u-\xi_2)\,
\frac{\theta_1(u\! -\! v_a\! +\! \eta)}{\theta_1(u-v_a)}
$$
for $a=1, \ldots , 4$.

\subsection*{The Bethe vectors}

For illustrative purposes, we give here the explicit expression for the (off-shell)
Bethe vectors. We recall that
$$
\begin{array}{l}
\mu (u;s,t)=\theta_1(\frac{1}{2}\, (s-t)+u|\tau),
\end{array}
$$
$$
\gamma_k=\frac{1}{\theta_2(\tau_k|\tau)}, \quad
\tau_k=\frac{1}{2}(s_k+t_k), \quad s_k =s+k\eta , \quad t_k=t+k\eta .
$$

The right vacuum is
\beq\label{rvac}
\bigl |\Omega^{l-1}\bigr >=\left (\begin{array}{c}
\theta_1(s_{l-1}+\xi_1|2\tau) \\ \theta_4(s_{l-1}+\xi_1|2\tau) \end{array}\right )
\otimes
\left (\begin{array}{c}
\theta_1(s_{l}+\xi_2|2\tau) \\ \theta_4(s_{l}+\xi_2|2\tau) \end{array}\right ).
\eeq
Let us denote the basis vectors as
$$
\bigl |e_1\bigr >=\frac{1}{2}\Bigl (\bigl |+-\bigr >-\bigl |-+\bigr >\Bigr ), \quad
\bigl |e_2\bigr >=\frac{1}{2}\Bigl (\bigl |+-\bigr >+\bigl |-+\bigr >\Bigr ),
$$
$$
\bigl |e_3\bigr >=\frac{1}{2}\Bigl (\bigl |++\bigr >-\bigl |--\bigr >\Bigr ), \quad
\bigl |e_4\bigr >=\frac{1}{2}\Bigl (\bigl |++\bigr >+\bigl |--\bigr >\Bigr ).
$$
For the vector
$$
\bigl |\Psi^l(u)\bigr >=B_{l-1, l+1}(u)\bigl |\Omega^{l-1}\bigr >
$$
we have
$$
\bigl |\Psi^l(u)\bigr >=A_1^{(l)} \bigl |e_1\bigr >+
A_2^{(l)} \bigl |e_2\bigr >+A_3^{(l)} \bigl |e_3\bigr >+
A_4^{(l)} \bigl |e_4\bigr >,
$$
where the coefficients $A_i^{(l)}$ read:
{\small
$$
A_1^{(l)}=\frac{\gamma_{l-1}\gamma_{l+1}}{\mu(u;s,t)}\Biggl [
\Bigl (\theta_4(t_{l-1}-u|2\tau)\theta_1(t_{l+1}-u|2\tau)
\theta_1(s_{l-1}+\xi_1|2\tau)\theta_4(s_{l}+\xi_2|2\tau)
$$
$$
+\theta_1(t_{l-1}-u|2\tau)\theta_4(t_{l+1}-u|2\tau)
\theta_4(s_{l-1}+\xi_1|2\tau)\theta_1(s_{l}+\xi_2|2\tau)\Bigr )(a_1b_2 -c_1c_2)
$$
$$
+\Bigl (\theta_4(t_{l-1}-u|2\tau)\theta_1(t_{l+1}-u|2\tau)
\theta_4(s_{l-1}+\xi_1|2\tau)\theta_1(s_{l}+\xi_2|2\tau)
$$
$$\phantom{aaaaaaaaaa}
+\theta_1(t_{l-1}-u|2\tau)\theta_4(t_{l+1}-u|2\tau)
\theta_1(s_{l-1}+\xi_1|2\tau)\theta_4(s_{l}+\xi_2|2\tau)\Bigr )(d_1d_2 -b_1a_2)
$$
$$
+\Bigl (\theta_4(t_{l-1}-u|2\tau)\theta_4(t_{l+1}-u|2\tau)
\theta_1(s_{l-1}+\xi_1|2\tau)\theta_1(s_{l}+\xi_2|2\tau)\phantom{aaaaaaaaa}
$$
$$\phantom{aaaaaaaaaa}
+\theta_1(t_{l-1}-u|2\tau)\theta_1(t_{l+1}-u|2\tau)
\theta_4(s_{l-1}+\xi_1|2\tau)\theta_4(s_{l}+\xi_2|2\tau)\Bigr )(a_1c_2 -c_1b_2)
$$
$$
+\Bigl (\theta_4(t_{l-1}-u|2\tau)\theta_4(t_{l+1}-u|2\tau)
\theta_4(s_{l-1}+\xi_1|2\tau)\theta_4(s_{l}+\xi_2|2\tau)\phantom{aaaaaaaaa}
$$
$$\phantom{aaaaaaaaaa}
+\theta_1(t_{l-1}-u|2\tau)\theta_1(t_{l+1}-u|2\tau)
\theta_1(s_{l-1}+\xi_1|2\tau)\theta_1(s_{l}+\xi_2|2\tau)\Bigr )(d_1a_2 -b_1d_2)\Biggr ],
$$

$$
A_2^{(l)}=\frac{\gamma_{l-1}\gamma_{l+1}}{\mu(u;s,t)}\Biggl [
\Bigl (\theta_4(t_{l-1}-u|2\tau)\theta_1(t_{l+1}-u|2\tau)
\theta_1(s_{l-1}+\xi_1|2\tau)\theta_4(s_{l}+\xi_2|2\tau)
$$
$$
-\theta_1(t_{l-1}-u|2\tau)\theta_4(t_{l+1}-u|2\tau)
\theta_4(s_{l-1}+\xi_1|2\tau)\theta_1(s_{l}+\xi_2|2\tau)\Bigr )(a_1b_2 +c_1c_2)
$$
$$
+\Bigl (\theta_4(t_{l-1}-u|2\tau)\theta_1(t_{l+1}-u|2\tau)
\theta_4(s_{l-1}+\xi_1|2\tau)\theta_1(s_{l}+\xi_2|2\tau)
$$
$$\phantom{aaaaaaaaaa}
-\theta_1(t_{l-1}-u|2\tau)\theta_4(t_{l+1}-u|2\tau)
\theta_1(s_{l-1}+\xi_1|2\tau)\theta_4(s_{l}+\xi_2|2\tau)\Bigr )(d_1d_2 +b_1a_2)
$$
$$
+\Bigl (\theta_4(t_{l-1}-u|2\tau)\theta_4(t_{l+1}-u|2\tau)
\theta_1(s_{l-1}+\xi_1|2\tau)\theta_1(s_{l}+\xi_2|2\tau)\phantom{aaaaaaaaa}
$$
$$\phantom{aaaaaaaaaa}
-\theta_1(t_{l-1}-u|2\tau)\theta_1(t_{l+1}-u|2\tau)
\theta_4(s_{l-1}+\xi_1|2\tau)\theta_4(s_{l}+\xi_2|2\tau)\Bigr )(a_1c_2 +c_1b_2)
$$
$$
+\Bigl (\theta_4(t_{l-1}-u|2\tau)\theta_4(t_{l+1}-u|2\tau)
\theta_4(s_{l-1}+\xi_1|2\tau)\theta_4(s_{l}+\xi_2|2\tau)\phantom{aaaaaaaaa}
$$
$$\phantom{aaaaaaaaaa}
-\theta_1(t_{l-1}-u|2\tau)\theta_1(t_{l+1}-u|2\tau)
\theta_1(s_{l-1}+\xi_1|2\tau)\theta_1(s_{l}+\xi_2|2\tau)\Bigr )(d_1a_2 +b_1d_2)\Biggr ],
$$

$$
A_3^{(l)}=\frac{\gamma_{l-1}\gamma_{l+1}}{\mu(u;s,t)}\Biggl [
\Bigl (\theta_4(t_{l-1}-u|2\tau)\theta_1(t_{l+1}-u|2\tau)
\theta_1(s_{l-1}+\xi_1|2\tau)\theta_1(s_{l}+\xi_2|2\tau)
$$
$$
+\theta_1(t_{l-1}-u|2\tau)\theta_4(t_{l+1}-u|2\tau)
\theta_4(s_{l-1}+\xi_1|2\tau)\theta_4(s_{l}+\xi_2|2\tau)\Bigr )(a_1a_2 -c_1d_2)
$$
$$
+\Bigl (\theta_4(t_{l-1}-u|2\tau)\theta_1(t_{l+1}-u|2\tau)
\theta_4(s_{l-1}+\xi_1|2\tau)\theta_4(s_{l}+\xi_2|2\tau)
$$
$$
+\theta_1(t_{l-1}-u|2\tau)\theta_4(t_{l+1}-u|2\tau)
\theta_1(s_{l-1}+\xi_1|2\tau)\theta_1(s_{l}+\xi_2|2\tau)\Bigr )(d_1c_2 -b_1b_2)
$$
$$
+\Bigl (\theta_4(t_{l-1}-u|2\tau)\theta_4(t_{l+1}-u|2\tau)
\theta_1(s_{l-1}+\xi_1|2\tau)\theta_4(s_{l}+\xi_2|2\tau)
$$
$$
+\theta_1(t_{l-1}-u|2\tau)\theta_1(t_{l+1}-u|2\tau)
\theta_4(s_{l-1}+\xi_1|2\tau)\theta_1(s_{l}+\xi_2|2\tau)\Bigr )(a_1d_2 -c_1a_2)
$$
$$
+ \Bigl (\theta_4(t_{l-1}-u|2\tau)\theta_4(t_{l+1}-u|2\tau)
\theta_4(s_{l-1}+\xi_1|2\tau)\theta_1(s_{l}+\xi_2|2\tau)
$$
$$
+\theta_1(t_{l-1}-u|2\tau)\theta_1(t_{l+1}-u|2\tau)
\theta_1(s_{l-1}+\xi_1|2\tau)\theta_4(s_{l}+\xi_2|2\tau)\Bigr )(d_1b_2-b_1c_2)\Biggr ],
$$

$$
A_4^{(l)}=\frac{\gamma_{l-1}\gamma_{l+1}}{\mu(u;s,t)}\Biggl [
\Bigl (\theta_4(t_{l-1}-u|2\tau)\theta_1(t_{l+1}-u|2\tau)
\theta_1(s_{l-1}+\xi_1|2\tau)\theta_1(s_{l}+\xi_2|2\tau)
$$
$$
-\theta_1(t_{l-1}-u|2\tau)\theta_4(t_{l+1}-u|2\tau)
\theta_4(s_{l-1}+\xi_1|2\tau)\theta_4(s_{l}+\xi_2|2\tau)\Bigr )(a_1a_2 +c_1d_2)
$$
$$
+\Bigl (\theta_4(t_{l-1}-u|2\tau)\theta_1(t_{l+1}-u|2\tau)
\theta_4(s_{l-1}+\xi_1|2\tau)\theta_4(s_{l}+\xi_2|2\tau)
$$
$$
-\theta_1(t_{l-1}-u|2\tau)\theta_4(t_{l+1}-u|2\tau)
\theta_1(s_{l-1}+\xi_1|2\tau)\theta_1(s_{l}+\xi_2|2\tau)\Bigr )(d_1c_2 +b_1b_2)
$$
$$
+\Bigl (\theta_4(t_{l-1}-u|2\tau)\theta_4(t_{l+1}-u|2\tau)
\theta_1(s_{l-1}+\xi_1|2\tau)\theta_4(s_{l}+\xi_2|2\tau)
$$
$$
-\theta_1(t_{l-1}-u|2\tau)\theta_1(t_{l+1}-u|2\tau)
\theta_4(s_{l-1}+\xi_1|2\tau)\theta_1(s_{l}+\xi_2|2\tau)\Bigr )(a_1d_2 +c_1a_2)
$$
$$
+ \Bigl (\theta_4(t_{l-1}-u|2\tau)\theta_4(t_{l+1}-u|2\tau)
\theta_4(s_{l-1}+\xi_1|2\tau)\theta_1(s_{l}+\xi_2|2\tau)
$$
$$
-\theta_1(t_{l-1}-u|2\tau)\theta_1(t_{l+1}-u|2\tau)
\theta_1(s_{l-1}+\xi_1|2\tau)\theta_4(s_{l}+\xi_2|2\tau)\Bigr )(d_1b_2+b_1c_2)\Biggr ].
$$
}

For rational $\eta =2P/Q$ the vector $\bigl |\Psi^l(u)\bigr >$ is
$Q$-periodic in $l$.
The (off-shell) Bethe vectors $\bigl |\Psi_{\nu}(u)\bigr >$
($\nu =0,1, \ldots , Q-1$)
are defined as finite Fourier transforms
of $\bigl |\Psi^l(u)\bigr >$:
\beq\label{F}
\bigl |\Psi_{\nu}(u)\bigr >=\sum_{l=0}^{Q-1}e^{-2\pi i Pl\nu/Q}\bigl |\Psi^l(u)\bigr >.
\eeq
This example illustrates that any direct calculations with the Bethe vectors,
even in the simplest case $N=2$, are hardly possible.

\subsection*{The result for scalar products}

Combining transformation properties under shifts of the variables with results
of computer simulations, one can suggest the following formula for the scalar
products at $N=2$:
\beq\label{spN2}
\bigl <\Psi_{\nu}(v)\bigl |\Psi_{\mu}(u)\bigr >=\phi_1^{(\nu \mu)}(v-u,x)
\phi_2^{(\nu)}(v,y)\theta_1(u-v)T_{11}^{(\nu \mu)}(v-u).
\eeq
The function $\phi_1^{(\nu \mu)}(r,x)$ is given by (\ref{fix21}) for even $Q$
and by (\ref{fix22}) for odd $Q$. The determinant of the matrix $T_{jk}^{(\nu \mu)}$
reduces to the element $T_{11}^{(\nu \mu)}$ (see (\ref{Tnumu})). Plugging $r=v-u$
into (\ref{Tnumu}) we get the $\nu$-independent expression
\beq\label{spN2a1}
 \displaystyle{
 T_{11}^{(\nu \mu)}=-\frac{\theta_1'(0)\theta_1(\eta)}{\theta_1^2(v-u)}
 \Big( a(u)e^{i\pi\mu\eta}-d(u)e^{-i\pi\mu\eta} \Big)\,.
 }
\eeq
The function $\phi_2^{(\nu)}(v,y)$ is
\beq\label{spN2a}
\phi_2^{(\nu)}(v,y)=
C(\eta, \tau)\chi^{(\nu)}(v)
d(v)e^{2\pi i\nu y}\theta_1^2(y+v),
\eeq
where
\beq\label{spN2b}
\chi^{(\nu)}(v)=e^{2\pi i\nu (v-\eta)}\theta_1(v-\xi_1)\theta_1(v-\xi_2)
\, \p_z \log \frac{a(z)}{d(z)}\Biggr |_{z=v}
\eeq
and
\beq\label{spN2c}
C(\eta, \tau)=\left \{ \begin{array}{l}
\displaystyle{\frac{2Q^2\theta_1'(0|Q\tau/2)}{(\theta_1'(0))^3 \theta_2^2(0)\theta_1(\eta)}
\quad \mbox{for even $Q$}},
\\ \\
\displaystyle{\frac{2Q^2\theta_1'(0|Q\tau)}{(\theta_1'(0))^3 \theta_2^2(0)\theta_1(\eta)}
\quad \mbox{for odd $Q$}}.
\end{array}
\right.
\eeq

\section*{Appendix C: free fermions}

\def\theequation{C\arabic{equation}}
\setcounter{equation}{0}

\addcontentsline{toc}{section}{\hspace{6mm}Appendix C: free fermions}

The case $\eta =1/2$ ($Q=4$)
corresponds to free fermions. The Bethe equations drastically simplify
in this case; they have the form
\beq\label{Bethe101}
e^{\pi i\nu}\, \frac{a(v_j)}{d(v_j)}=(-1)^{n-1}.
\eeq
For the case of free fermions it is possible
to obtain more explicit expressions for the scalar products.
The matrix $T_{jk}^{(\nu\mu)}(r)$ is given by
\beq\label{Tik1}
\begin{array}{l}
\displaystyle{
T_{jk}^{(\nu\mu)}(r)=\frac{\theta_1'(0)}{\theta_1(r)}\left(\prod_{p=1}^n
\frac{\theta_2(u_k-v_p)}{\theta_1(u_k-v_p)}\right)}
\\ \\
\displaystyle{ \phantom{aaaaaa}\times \, \left [(-1)^n a(u_k)\left (
e^{i\pi\nu/2}\frac{\theta_1(u_k-v_j+r)}{\theta_1(u_k-v_j)}-
e^{i\pi\mu/2}\frac{\theta_2(u_k-v_j +r)}{\theta_2(u_k-v_j )}\right )\right.}
\\ \\
\displaystyle{\phantom{aaaaaaaaaaaaa}\left. +d(u_k)\left (e^{-i\pi\nu/2}
\frac{\theta_1(u_k-v_j+r)}{\theta_1(u_k-v_j)}-
e^{-i\pi\mu/2}\frac{\theta_2(u_k-v_j+r)}{\theta_2(u_k-v_j)}\right )\right ].}
\end{array}
\eeq

\subsection*{Case $\mu =\nu$}

Set $\mu=\nu$. Then
\begin{equation}\label{Tik12}
 \begin{array}{l}
 \displaystyle{
T_{jk}^{(\nu\nu)}(r)=\frac{\theta_1'(0)}{\theta_1(r)}
\left(\prod_{p=1}^n\frac{\theta_2(u_k-v_p)}{\theta_1(u_k-v_p)}\right)
}
\\ \ \\
 \displaystyle{
 \phantom{aaaaaaaaaaaaaaaaaa}\times\left((-1)^n e^{i\pi\nu/2} a(u_k)+e^{-i\pi\nu/2}d(u_k)\right)
H^{(-)}(u_k,v_j)\,,
}
 \end{array}
\end{equation}
where
$$
\begin{array}{lll}
H^{(-)}(u_k,v_j)&=&\displaystyle{
\frac{\theta_1(u_k-v_j+r)}{\theta_1(u_k-v_j)}-
\frac{\theta_2(u_k-v_j+r)}{\theta_2(u_k-v_j )}}
\\ &&\\
&=&\displaystyle{
\frac{2\theta_1(r|2\tau)}{\theta_4(0|2\tau)}\;
\frac{\theta_4(2u_k-2v_j+r|2\tau)}{\theta_1(2u_k-2v_j|2\tau)},}
\end{array}
$$
and we arrive at
\beq\label{Tik13}
\begin{array}{l}
\displaystyle{
T_{jk}^{(\nu\nu)}(r)=
\frac{2\theta_1(r|2\tau)\theta_1'(0|\tau)}{\theta_1(r|\tau)\theta_4(0|2\tau)}
\left(\prod_{p=1}^n\frac{\theta_2(u_k-v_p)}{\theta_1(u_k-v_p)}\right)}
\\ \\
\displaystyle{\phantom{aaaaaaaaaaaaaa}
\times \, \left((-1)^n e^{i\pi\nu/2} a(u_k)+e^{-i\pi\nu/2}d(u_k)\right)
\frac{\theta_4(2u_k-2v_j+r|2\tau)}{\theta_1(2u_k-2v_j|2\tau)},}
\end{array}
\eeq
i.e. it is the elliptic Cauchy matrix (multiplied by a diagonal matrix).

\subsection*{Case $\mu\neq \nu$}

If $\mu\neq \nu$, then generically we cannot obtain the Cauchy matrix.
However, if $\mu =\nu+2$, then $e^{i\pi\mu/2}=-e^{i\pi\nu/2}$, and the matrix elements
$T_{jk}^{(\nu, \nu+2)}(r)$ take the form
\begin{equation}\label{Tik12p}
\begin{array}{l}
\displaystyle{
T_{jk}^{(\nu, \nu+2)}(r)=
}
\\ \\
\displaystyle{
\phantom{aaaaa}\frac{\theta_1'(0)}{\theta_1(r)}\left(\prod_{p=1}^n
\frac{\theta_2(u_k-v_p)}{\theta_1(u_k-v_p)}\right)
{ \left((-1)^n e^{i\pi\nu/2} a(u_k)+e^{-i\pi\nu/2}d(u_k)\right)}
H^{(+)}(u_k,v_j),
 }
 \end{array}
\end{equation}
where
\beq\label{hjkp}
\begin{array}{lll}
H^{(+)}(u_k,v_j)&=&\displaystyle{
\frac{\theta_1(u_k-v_j+r)}{\theta_1(u_k-v_j)}+\frac{\theta_2(u_k-v_j+r)}{\theta_2(u_k-v_j )}}
\\ && \\
&=&\displaystyle{
\frac{2\theta_4(r|2\tau)}{\theta_4(0|2\tau)}\;
\frac{\theta_1(2u_k-2v_j+r|2\tau)}{\theta_1(2u_k-2v_j|2\tau)},}
\end{array}
\eeq
and finally we arrive at
\beq\label{Tik13p}
\begin{array}{l}
\displaystyle{
T_{jk}^{(\nu ,\nu+2)}
(r)=\frac{2\theta_4(r|2\tau)\theta_1'(0|\tau)}{\theta_1(r|\tau)
\theta_4(0|2\tau)}\left(\prod_{p=1}^n\frac{\theta_2(u_k-v_p)}{\theta_1(u_k-v_p)}\right)}
\\ \\
\displaystyle{\phantom{aaaaaaaaaaaaaa}
\times \, {\left((-1)^n e^{i\pi\nu/2} a(u_k)+e^{-i\pi\nu/2}d(u_k)\right)}
\frac{\theta_1(2u_k-2v_j+r|2\tau)}{\theta_1(2u_k-2v_j|2\tau)}.}
\end{array}
\eeq

\subsection*{Elliptic Cauchy determinants}

Below we use the notation
\beq\label{C-main}
W_n(\bu,\bv|\tau)=\frac{\prod\limits_{1\le a<b\le n}\theta_1(u_a-u_b|\tau)\theta_1(v_b-v_a|\tau)}{\prod\limits_{1\le a,b\le n}\theta_1(u_a-v_b|\tau)}.
\eeq
Then the explicit formula for the elliptic Cauchy determinant tells us that
\beq\label{det1}
\det_{1\leq j,k\leq n}
\left(\frac{\theta_1(2u_k-2v_j+r|2\tau)}{\theta_1(2u_k-2v_j|2\tau)}\right)=
\theta_1^{n-1}(r|2\tau)
\theta_1(r+2U-2V|2\tau)W_n(2\bu,2\bv|2\tau),
\eeq
and
\beq\label{det2}
\det_{1\leq j,k\leq n}
\left(\frac{\theta_4(2u_k-2v_j+r|2\tau)}{\theta_1(2u_k-2v_j|2\tau)}\right)=\theta_4^{n-1}(r|2\tau)
\theta_4(r+2U-2V|2\tau)W_n(2\bu,2\bv|2\tau).
\eeq
Here
\beq\label{UV}
U=\sum_{k=1}^n u_k, \qquad V=\sum_{k=1}^n v_k.
\eeq
Equation (\ref{det2}) follows from (\ref{det1}) if we use the relation
$$\theta_4(x|2\tau) = -ie^{i\pi(x+\tau/2)}\theta_1(x+\tau|2\tau).$$

\subsection*{Determinant of the matrix $T_{jk}^{(\nu\mu)}$}

\paragraph{Case $\mu=\nu$.}
The above results yield
$$
\begin{array}{l}\displaystyle{
\det T_{jk}^{(\nu \nu)}(r)=
\left(\frac{2\theta_1(r|2\tau)\theta_4(r|2\tau)
\theta_1'(0|\tau)}{\theta_1(r|\tau)\theta_4(0|2\tau)}\right)^n
\left(\prod_{a,b=1}^n\frac{\theta_2(u_a-v_b)}{\theta_1(u_a-v_b)}\right)}
\\ \\
\displaystyle{\phantom{aa}
\times \, \prod_{p=1}^n\Bigl((-1)^n e^{i\pi\nu/2} a(u_p)+e^{-i\pi\nu/2}d(u_p)\Bigr)
\frac{\theta_4(r+2U-2V|2\tau)}{\theta_4(r|2\tau)}\, W_n(2\bu,2\bv|2\tau).}
\end{array}
$$
This expression can be simplified using the identity
$\theta_1(x|\tau)\theta_2(0|\tau)=2\theta_1(x|2\tau)\theta_4(x|2\tau)$ and its derivative
at $x=0$:
\beq\label{detT+1}
\begin{array}{l}\displaystyle{
\det T_{jk}^{(\nu \nu)}(r)=
\bigl(2\theta'_1(0|2\tau)\bigr)^n \frac{\theta_4(r+2U-2V|2\tau)}{\theta_4(r|2\tau)}
\left(\prod_{a,b=1}^n\frac{\theta_2(u_a-v_b)}{\theta_1(u_a-v_b)}\right)}
\\ \\
\displaystyle{\phantom{aaaaa}
\times \, W_n(2\bu,2\bv|2\tau)\prod_{p=1}^n\Bigl((-1)^n
e^{i\pi\nu/2} a(u_p)+e^{-i\pi\nu/2}d(u_p)\Bigr).}
\end{array}
\eeq
Observe that in the case of the scalar product
we should set $r=V-U$, and then the ratio of the $\theta_4$-functions disappears:
$$
\frac{\theta_4(r+2U-2V|2\tau)}{\theta_4(r|2\tau)}=\frac{\theta_4(-r|2\tau)}{\theta_4(r|2\tau)}=1.
$$

\paragraph{Case $\mu\ne\nu$.}
In a similar way, we obtain for $\mu=\nu+2$:
\beq\label{detT-1}
\begin{array}{l}\displaystyle{
\det T_{jk}^{(\nu, \nu+2)}(r)=
\bigl(2\theta'_1(0|2\tau)\bigr)^n \frac{\theta_1(r+2U-2V|2\tau)}{\theta_1(r|2\tau)}
\left(\prod_{a,b=1}^n\frac{\theta_2(u_a-v_b)}{\theta_1(u_a-v_b)}\right)}
\\ \\
\displaystyle{\phantom{aaaaa}
\times \, W_n(2\bu,2\bv|2\tau)\prod_{p=1}^n{ \Bigl((-1)^n e^{i\pi\nu/2}a(u_p)+e^{-i\pi\nu/2}d(u_p)\Bigr)}.}
\end{array}
\eeq
Again, in the case of the scalar product the ratio of
the $\theta_1$-functions simplifies, but now it gives the minus sign:
$$
\frac{\theta_1(r+2U-2V|2\tau)}{\theta_1(r|2\tau)}=\frac{\theta_1(-r|2\tau)}{\theta_1(r|2\tau)}=-1.
$$

In the case when $\mu-\nu$ is odd
no simple result for $\det T_{jk}^{(\nu \mu)}(r)$ is available. However, in this case
the scalar product vanishes
due to the selection rule: $\left <\Psi_{\nu}(\bar v)\right |\Psi_{\mu}(\bar u)\bigr >=0$
if $\mu-\nu$ is odd (see section \ref{section:selection}).

\subsection*{Scalar products}

Summarizing the above results, we obtain an explicit
representation for the scalar product
$$X^{(n)}_{\nu \mu}(\bar v, \bar u)=\bigl <\Psi_{\nu}(\bar v)\bigr |\Psi_{\mu}(\bar u)\bigr >.
$$
The result is
 \begin{multline}\label{SP-res}
X^{(n)}_{\nu \mu}(\bar v, \bar u)
=\pm\bigl(2\theta'_1(0|2\tau)\bigr)^n \phi_{1}^{(\nu \mu)}(r,x)\phi_{2}^{(\nu)}(\bar v, y)
\left(\prod_{a,b=1}^n\frac{\theta_2(u_a-v_b)}{\theta_1(u_a-v_b)}\right)
\frac{W_n(2\bu,2\bv|2\tau)}{W_n(\bu,\bv|\tau)}\\
\times \, \prod_{p=1}^n {\Bigl((-1)^n e^{i\pi\nu/2} a(u_p)+ e^{-i\pi\nu/2}d(u_p)\Bigr)}.
\end{multline}
The plus sign corresponds to $\nu=\mu$, while the minus sign  corresponds to $\mu-\nu=\pm2$. For $\mu-\nu=1({\rm mod}\ 2)$ the scalar product vanishes.
The function
$\phi_{1}^{(\nu \mu)}(r,x)$ is
$$
\phi_{1}^{(\nu \mu)}(r,x)=\delta_{\mu, \nu \, \mbox{\scriptsize{(mod 2})}}
e^{\pi i (\mu-\nu)x}\frac{\theta_1(r|\tau)\,
\theta_1(r+2x +(\mu-\nu)\tau/2|2\tau)}{\theta_1(2x|2\tau)\,
\theta_1(r +(\mu-\nu)\tau/2|2\tau)}
$$
(see (\ref{fix21})) and we recall that $\displaystyle{r=\sum_j v_j -\sum_j u_j}$.
The function $\phi_{2}^{(\nu)}(\bar v, y)$ is unknown.

{ A complete analog of (\ref{SP-res}) exists in the $XXZ$ case.
Actually, one can take the limit $\tau\to+i\infty$ in (\ref{SP-res}) and
reproduce the $XXZ$ result up to a common factor. }

The squared norm of the on-shell Bethe vector is
\beq\label{sp1}
\begin{array}{l}
\displaystyle{
\bigl <\Psi_{\nu}(\bar v)\bigr |\Psi_{\nu}(\bar v)\bigr >=
(-1)^n \theta_2^n (0|\tau)\frac{\theta_1'(0|\tau)}{\theta_1'(0|2\tau)}\,
\phi_{2}^{(\nu)}(\bar v, y)e^{-\pi i \nu n /2}\prod^n_{a\neq b}
\frac{\theta_2(v_a-v_b|\tau)}{\theta_1(v_a-v_b|\tau)}}
\\ \\
\displaystyle{ \phantom{aaaaaaaaaaaaaaaaaaaaa}\times \,
\prod_{p=1}^n \Bigl (d(v_p)\p_v \log \frac{a(v)}{d(v)}\Bigr |_{v=v_p} \Bigr ).}
\end{array}
\eeq
The result for the specially normalized scalar product is
\beq\label{sp2}
\begin{array}{r}
\displaystyle{\frac{\bigl <\Psi_{\nu}(\bar v)
\bigr |\Psi_{\mu}(\bar u)\bigr >}{\bigl <\Psi_{\nu}(\bar v)
\bigr |\Psi_{\nu}(\bar v)\bigr >}=
\pm (-1)^n \frac{\theta_1'(0|2\tau)}{\theta_1'(0|\tau)}
\left (\frac{2\theta_1'(0|2\tau)}{\theta_2(0|\tau)}\right )^n
\frac{W_n (2\bar u, 2\bar v |2\tau)}{W_n (\bar u, \bar v |\tau)}
\, \phi_{1}^{(\nu \mu)}(r,x)
}
\\ \ \\
\displaystyle{
\times
\prod^n_{a,b}\frac{\theta_2(u_a\! -\! v_b|\tau)}{\theta_1(u_a\! -\! v_b|\tau)}
\prod^n_{a'\neq b'}\frac{\theta_1(v_{a'}\! -\! v_{b'}|\tau)}{\theta_2(v_{a'}\! -\! v_{b'}|\tau)}
\prod_{p=1}^n
\frac{\!\! \!\!
{ (-1)^n e^{\pi i\nu}a(u_p)+ d(u_p)}}{\,\,\,\, d(v_p)\p_v \log \Bigl
(a(v)/d(v)\Bigr )\Bigr |_{v=v_p}}.
}
\end{array}
\eeq

\end{document}